\documentclass[aps,preprintnumbers, twocolumn, showpacs]{revtex4}
\usepackage{eurosym}
\usepackage{graphicx,epstopdf}
\usepackage{amsmath,verbatim}
\usepackage{bm}
\usepackage{color,multirow}

\def\be{\begin{equation}}
\def\ee{\end{equation}}
\def\bea{\begin{eqnarray}}
\def\eea{\end{eqnarray}}

\begin{document}

\title{Electromagnetic radiation of charged particles in stochastic motion}
\author{Tiberiu Harko$^1$,$^2$}
\email{t.harko@ucl.ac.uk}
\author{Gabriela Mocanu$^3$}
\email{gabriela.mocanu@ubbcluj.ro}
\affiliation{$^1$Department of Physics, Babes-Bolyai University, Kogalniceanu Street,
Cluj-Napoca 400084, Romania,}
\affiliation{$^2$Department of Mathematics, University College London, Gower Street,
London WC1E 6BT, United Kingdom,}
\affiliation{$^3$Astronomical Institute of the Romanian Academy, 19 Cire\c{s}ilor Street, Cluj-Napoca 400487, Romania}
\date{\today}

\begin{abstract}
The study of the Brownian motion of a charged particle in electric and magnetic fields
fields has many important  applications in plasma and heavy ions physics, as well as in astrophysics. In the present paper we consider the electromagnetic radiation properties of a charged non-relativistic particle in the presence of electric and magnetic fields, of an exterior non-electromagnetic potential, and of a friction and stochastic force, respectively. We describe the motion of the charged particle by a
Langevin and generalized Langevin type stochastic differential equation. We investigate in detail the cases of the Brownian motion with or without memory in a constant electric field, in the presence of an external harmonic potential, and of a constant magnetic field. In all cases the corresponding Langevin equations are solved numerically, and a full description of the spectrum of the emitted radiation and of the physical properties of the motion is obtained. The Power Spectral Density (PSD) of the emitted power is also obtained for each case, and, for all considered oscillating systems, it shows the presence of peaks, corresponding to certain intervals of the frequency.
\end{abstract}

\pacs{04.50.Kd,04.20.Cv}
\date{\today}
\maketitle

\section{Introduction}
The stochastic motion of particles in different physical systems, and under the influence of various forces, is a fundamental area of research in plasma physics, astronomy, condensed matter physics, and biology \cite{rev1,rev2}. In particular, the motion and the radiation of charged particles play a key role in the understanding of the nuclear fusion processes in the tokamak plasmas. In the presence of an external electric field, in a fully ionized plasma, electrons with energies higher
than certain critical values of the energy are continuously accelerated at very high-energies. These electrons are called runaway electrons \cite{San1}.

The study of runaway electrons, and in particular of their radiation,  represents a field of great importance in different areas of research such as astronomy, accelerators,
or nuclear fusion.

An important physical problem in plasma and fusion physics is the anomalous transport in a magnetically confined plasma
in a region where the magnetic surfaces are destroyed \cite{Bal1}. One of the basic methods in the study of the anomalous transport is based on the analogy between the transport problem
and the random walk or Brownian motion theory \cite{Bal1,4}. For this
case, the starting point is the equation of motion of a charged
test particle, feeling the action of a magnetic field and of
interparticle collisions. The latter are represented by a random
force and the equation of motion becomes a stochastic
differential equation, the Langevin
equation \cite{Bal1,4, Bal2,Bal3, Coff}
\begin{equation}\label{lang}
\frac{d^2\vec{r}}{dt^2}=\vec{F}\left[\vec{r}(t),\dot{\vec{r}}(t),t\right]+\vec{\eta}(t),
\end{equation}
where $\vec{F}\left[\vec{r}(t),\dot{\vec{r}}(t),t\right]$ is the systematic ("average")
force acting on the article, and $\vec{\eta}(t)$ is a random force modeling the effects of the interparticle collisions. The anomalous transport in plasmas is usually attributed  to the magnetic
fluctuations in  a very strong "basic" magnetic field $\vec{B}_o$, which undergoes
small fluctuations in a perpendicular direction.

The Langevin equation Eq. (\ref{lang}) gives a correct physical
and statistical description of the random Brownian
motion only in the large time limit. This requires that the considered time intervals must be large enough as compared
to the characteristic relaxation time of the velocity
autocorrelation function \cite{Coff}. The  description of  the
dynamics of a homogeneous system without restriction
on a time scale can be realized by generalizing the Langevin equation.
This generalization implies the introduction
of a systematic force term with an integral kernel,  which substitutes the simple friction term \cite{Kubo}.
From a physical point of view the convolution term describes the memory, or the retardation
effects. From an astrophysical point of view such a term can be used to model the stochastic oscillations of the accretion disks \cite{disk1} in the presence of colored noise \cite{disk2,disk3}.

The study of the Brownian motion of charged particles in magnetic fields based on the Langevin equations with the dynamic friction proportional
to the particle velocity was initiated in \cite{b1}, where the diffusion of ions in plasma across the magnetic field, with
the stochasticity arising from the fluctuations of the electric field was considered.  In this work the mean square displacement  was found in the
limit of long times. These studies were further extended in \cite{b2,b3}, where it was shown that for a special symmetry of the dynamical friction matrix for Larmor periods of the order of the relaxation time across the magnetic field the diffusion takes place as an ordinary Brownian motion, uninfluenced by the presence of the external magnetic field. The investigation of the Brownian motion of charged particles in stochastic magnetic fields via the use of the Langevin and generalized Langevin equation has become an active field of research with many potential applications in astrophysics, plasma physics, and condensed matter \cite{b4,b5,b6,b7,b8,b9}. The electromagnetic radiation in small-scale random magnetic fields (jitter radiation) and its spectrum,  as well as the supplementary acceleration produced by the random variations of the Lorentz force were considered in \cite{Nik79}-\cite{Fl07b}.

An interesting phenomenon, that was studied recently, indicates the presence of
 long-time tails and resonant peaks, in the equilibrium and nonequilibrium
correlation functions for the velocity of the Brownian particle as described by the generalized Langevin equation \cite{Hij}. In \cite{sr3,sr4,sr5} it was pointed out that resonant stochastic behavior with a single peak and multi-peaks can be found in PSD curves of the stochastically oscillating accretion disks, described by both standard and generalized Langevin equations.

It is the purpose of the present paper to consider the electromagnetic
radiation of charged, non-relativistic particles, in stochastic motion under the effect of some
random exterior forces, in the presence of electric and magnetic fields, of an external non-electromagnetic potential, and interacting with
the environment by means of interparticle collisions, described as a friction force. To describe the motion of the particle we use the
Langevin equation, describing the non-relativistic motion of a Brownian
particle.  Due to the friction force, the particles in Brownian
motion lose energy to the medium, but simultaneously gain energy from the
random kicks of the external environment, modeled by the random force, as well as from the external non-electromagnetic, electric and magnetic fields. The electromagnetic power emitted by the particles is proportional to the square of its acceleration, which can be computed directly from the Langevin equation.

In our study we compute the electromagnetic radiation for four physical models. The first case we investigate is the Brownian motion of a charged particle in a constant electric field (case I). Therefore the  Langevin equation contain three force terms, the friction force, the electric force, and an external stochastic force. We obtain the stochastic power emitted by the particle, as well as the corresponding power spectral density. As a second case we consider the Brownian motion of a particle in a harmonic potential (case II). The emitted electromagnetic power as well the power spectral density is obtained. For the case of a charged particle with memory in a harmonic potential, described by the generalized Langevin equation, we present the displacement, velocity distribution, the electromagnetic power, and the power spectral density (case III). Finally, a full analysis of the Brownian motion of a charged particle in a constant magnetic field is presented (case IV), including the computation of the electromagnetic emitted power, and of the power spectral density. In all cases we obtain the total emitted electromagnetic power, as well as its statistical properties, from the numerical solution of the Langevin equation.

The present paper is organized as follows. In Section~\ref{sect:A} we introduce the Langevin equation describing the motion of a charged particle in electric and magnetic fields, and in an external, non-electromagnetic potential, and we present the basic relation for the stochastic power emitted during the motion. In Section~\ref{sect:B} we consider in detail the numerical algorithms and the solutions of the Langevin and generalized Langevin equations describing the Brownian motion of a charged particle in a constant electric field, in an exterior harmonic potential, and in a constant magnetic field, respectively. We discuss and conclude our results in Section~\ref{sect:C}.

\section{Stochastic equation of motion of charged particles and the
numerical algorithm for solving the Langevin equations}\label{sect:A}

The equation of motion of a charged particle with mass $m$, charge $Ze$, and
velocity $\vec{v}$ in electric $\vec{E}(t)$ and magnetic field $\vec{B}(t)$ fields, experiencing damping and
random acceleration is given by \cite{Bal1}-\cite{Coff}
\begin{equation}
\frac{d\vec{v}}{dt}=Ze\vec{E}(t)+\frac{Ze}{mc}\left[ \vec{v}(t)\times \vec{B}(t)\right]
-\nu \vec{v}(t)+\nabla V+\vec{\eta}(t),  \label{eq1}
\end{equation}%
where $\nu $ is an effective collisions frequency, $-\nu \vec{v}(t)$ is the
damping term, and $V$ is an external potential, corresponding to the
presence of non-electromagnetic forces. Eq.~(\ref{eq1}) is a stochastic
differential equation, known as the A-Langevin equation \cite{Bal1}-\cite{Bal3}. Generally, there
are two stochastic functions in the A-Langevin equation: the stochastic
electric and magnetic fields $\vec{E}(t)$ and $\vec{B}(t)$, and the random acceleration $\eta (t)$. To
close the system of equations, the stochastic properties of these functions
should be defined. Generally, one assumes that both are Gaussian
processes, which means that the first and the second order correlation
functions provide a complete statistical description of these functions \cite{Bal1}. However, in the following we consider the case in which all the stochastic effects, coming from the electric or magnetic fields, or from the random medium, can be encoded in the stochastic acceleration term $\eta (t)$. Therefore in the present analysis we keep only one stochastic term in the Langevin equations.

For the random acceleration $\vec{\eta }(t)$ we chose first the white noise
approximation, i.e.,
\begin{equation}
\left \langle \eta (t)^i\right \rangle = 0, \left \langle \eta
\left(t_1\right)^i \eta \left(t_2\right)^j \right \rangle= A\delta _{ij}
\delta \left(t_1 - t_2\right) .
\end{equation}

The equilibrium thermal velocity $v_{th}$ is related to the collisions
frequency $\nu $ and value $A$ as $v_{th}^{2}=\left( A/2\right) \nu $. This
relation is valid for charged particles in a magnetic field as well, since
equilibrium thermal velocity is not affected by the Lorenz force.

An extension of the Langevin equation Eq. (\ref{eq1}) was proposed by Kubo \cite{Kubo},
with the dynamical friction becoming frequency dependent. The generalized
Langevin equation in the presence of electric and  magnetic fields is given by
\bea \label{gLang}
\frac{d\vec{v}}{dt}&=&Ze\vec{E}(t)+\frac{Ze}{mc}\left[ \vec{v}(t)\times \vec{B}(t)\right]
- \nonumber\\
&&\int_{0}^{t}\gamma \left( t-t^{\prime }\right) \vec{v}(t^{\prime
})dt^{\prime }+
\nabla V+\vec{\eta}(t),
\eea
where the friction function $\gamma \left( t-t^{\prime }\right) $ represents
now the retarded effect of the frictional force.

The total electromagnetic power $P$ emitted by a moving charge is, in the
non-relativistic limit \cite{LaLi},
\begin{equation}
P=\frac{2}{3}\frac{\left( Ze\right) ^{2}}{c^{3}}\vec{a}^{2},
\end{equation}%
where $\vec{a}=d\vec{v}/dt$ is the acceleration of the particle. By taking
into account that for a charged particle moving in a magnetic
field the acceleration is given by Eq.~(\ref{eq1}), we obtain for the total electromagnetic power
emitted by the stochastically moving particle the expression
\bea
P&=&\frac{2}{3}\frac{\left( Ze\right) ^{2}}{c^{3}}\Bigg\{ Ze\vec{E}(t)+\frac{Ze}{mc}\left[
\vec{v}(t)\times \vec{B}(t)\right] -\nonumber\\
&&\nu \vec{v}(t)+\nabla V+\vec{\eta}%
(t)\Bigg\} ^{2},
\eea
where $\vec{v}(t)$ is obtained as a solution of Eq.~(\ref{eq1}). If the
motion of the charged particle in stochastic motion can be described by the
generalized Langevin equation Eq.~(\ref{gLang}), then the electromagnetic
power emitted by the particle is given by
\begin{eqnarray}
P&=&\frac{2}{3}\frac{\left( Ze\right) ^{2}}{c^{3}}\Bigg\{ Ze\vec{E}(t)+\frac{Ze}{mc}\left[
\vec{v}(t)\times \vec{B}(t)\right] -  \notag \\
&&\int_{0}^{t}\gamma \left( t-t^{\prime }\right) \vec{v}\left( t^{\prime
}\right) dt^{\prime }+\nabla V+\vec{\eta}(t)\Bigg\} ^{2}.
\end{eqnarray}

An important physical and statistical parameter is the steady-state mean
autocorrelation function of the emitted electromagnetic power by the particle in Brownian motion, and which is defined as
\begin{equation}
C_{PP}\left(\tilde{t}  \right) =\lim_{t\rightarrow \infty }\left\langle
P(t)P\left( t+\tilde{t} \right) \right\rangle .
\end{equation}

The Fourier transform of the mean autocorrelation function is called the Power
Spectral Density (PSD) of the power.

In the following we will consider the numerical solutions of the Langevin equations Eqs.~(\ref{eq1}) and (\ref{gLang}), respectively \cite{ermak1980, hershkowitz1998}. In order to obtain the numerical solutions we introduce a set of dimensionless coordinates $\left(\theta ,q,V,W,\bar{\Omega},L\right)$, where $\theta $ represents the dimensionless time, $q$ the dimensionless displacement, $V$ the dimensionless velocity,  $W$ is the dimensionless frequency, $\bar{\Omega}$ is the dimensionless Larmor frequency, and $L$ is the dimensionless power, respectively.

Due to their definition (detailed in the following sections for each case), these dimensionless parameters quantify the relative influence of deterministic parameters versus the amplitude of the noise in the system. Results will generally be discussed as a function of these parameters, thus one should keep in mind that the influence of the noise amplitude is embedded in them.

The results and discussions of the numerical simulations are grouped into three categories:
\begin{enumerate}

\item{}direct results, which are the dimensionless displacement $q (\theta)$ and the velocity $V(\theta)$,

\item{}the luminosity (emitted power) $L(\theta)$, representing an indirect result but which can be at least qualitatively directly compared with observations, and

\item{}statistical characteristics, used to compare between different results and between different parameter sets characterizing the physics of the system. We considered methods and approaches commonly used by astronomers to analyse observational data ~\cite{vau13,tem04}:
    \begin{enumerate}
    \item{}the statistical characteristics of the (simulated) light curve; the mean $\mu$ provides information about the injected energy, the standard deviation $\sigma$ provides information about the dispersion of data with respect to the mean value, the skewness $s$ is an indicator of the lack of symmetry of the distribution of values, with a positive skewness indicating a distribution with a long right tail, while a negative skewness indicates a distribution with a long left tail, and the kurtosis $\kappa$ measures the concentration of data around the peak and in the tails versus the concentration in the flanks, with a normal distribution having $\kappa = 3$;
    \item{}what is called first order statistics, i.e., based on autocorrelation of the data; we will discuss the PSD, obtained by taking the Fourier transform of the ensemble averaged correlation function of $L(\theta)$; in cases where it is appropriate, we discuss also the possible appearance of quasi periodic oscillations (QPOs) which we characterize in terms of their quality factor $Q$ (also sometimes called coherence of oscillation). $Q$ is defined as the ratio between the frequency $f_0$ at which the maximum of the peak occurs and the width of the peak at half-maximum intensity; generally, in astronomical nomenclature, if $Q$ is very large, the oscillation is stricly periodic, if $Q$ decreases towards the value $2$, the feature represents a $QPO$ and if $Q<2$ the signal is usually labelled as noise~\cite{vau13}. There is not one single number or even set of numbers that can be used to fully characterize a PSD of a light curve coming from a source with non-periodic, non-deterministic behaviour; thus a very important component of PSD analysis is the \textit{visual} inspection of the PSD curve.
\end{enumerate}

\end{enumerate}

All results are obtained {\it after mediation over $10^3$ stochastic realisations, and all PSDs were computed starting from a baseline of $10^3$ timesteps}.

\section{Electromagnetic radiation of charged particles in stochastic motion}
\label{sect:B}

In the present Section we consider the properties of the electromagnetic
radiation emitted by charged particles in random media, whose equations of
motion are given by Langevin or generalized Langevin type equations. In
particular, we analyze the total emitted power, and the power spectrum for
charged particles in Brownian motion in the presence of an electric field, in Brownian motion in the presence
of a harmonic potential, for charged particles obeying a
generalized Langevin type equation with memory, and for the stochastic motion of
particles in constant magnetic fields.

\subsection{Radiation of a charged particle in  Brownian motion in the presence of an electric field}

\subsubsection{Equations and physics}

The simplest possible stochastic motion of a charged particle with $Z=1$ and
mass $m$ is the one-dimensional Brownian motion in the presence of an
external electric field, $\vec{E}\neq 0$, in the absence of a magnetic field $\vec{B}= 0$. In the following we restrict or analysis to the case of the constant electric field, $\vec{E}={\rm constant}$. The random motion of the particle is described by the Langevin equation
\begin{equation}  \label{Lang1}
\frac{dv}{dt} = eE- \nu v + \xi _A(t).
\end{equation}

To treat $\xi _A $ as a random acceleration we must consider an ensemble of
systems, and define the random acceleration through its ensemble averages \cite{Bal2},
\begin{equation}
\left \langle \xi _A (t)\right \rangle =0, \left \langle \xi _A
\left(t_1\right)\xi _A \left(t_2\right)\right \rangle = \frac{A}{dt}\delta
\left(t_1-t_2\right).
\end{equation}

Due to the ensemble interpretation, the velocity $v$ and the position $x$ of the particle
are stochastic variables. According to the central limit theorem, they
should both have Gaussian distributions in the steady state \cite{Coff}. The variances
of these distributions are independently known, for the $x$ distribution
should obey the diffusion law, and the velocity distribution should be
Maxwell-Boltzmann. Thus we must have $\left \langle x^2\right \rangle =2Dt$,
and $m\left \langle v^2\right \rangle =K_B T$, where $D$ is the diffusion
coefficient, and $T$ the absolute temperature \cite{Bal1,Coff}. By calculating these
variances via the Langevin equation, we can relate the parameters $c_0 =
A/dt $ and $\nu $ to physical properties. Hence we have $c_0/2m\nu =k_BT$,
and $\nu =k_BT/mD$ \cite{Bal1,Coff}.

\subsubsection{The energy balance equation}

We define the average kinetic energy of the particle as
\begin{equation}
E_K=\frac{m}{2}\langle v^2 \rangle.
\end{equation}
Multiplying both sides of the Langevin equation Eq.~(\ref{Lang1}) by $mv$ we
obtain
\begin{equation}
\frac{m}{2}\frac{d}{dt}v^2+m\nu v^2=mevE+mv\xi _A.
\end{equation}
Taking the average of the above equation we obtain
\begin{equation}  \label{endis1}
\frac{d}{dt}E_K=\langle mv\xi _A\rangle -2\nu E_K,
\end{equation}
where we have assumed that $E={\rm constant}$, and $<v>=0$.
In Eq.~(\ref{endis1}) the term $\langle mv\xi _A\rangle$ represents the
average work done on the system, while the term $-2\nu E_K$ gives the rate
of the energy dissipation, which, for a charged particle, is due to the
electromagnetic radiation. Therefore we obtain the relation
\begin{equation}
\langle P\rangle=\frac{2}{3}\frac{e^2}{c^3}\langle a^2 \rangle=2\nu E_K.
\end{equation}
The average work done on the system can be obtained as $\langle mv\xi
_A\rangle=mc_0/2$, and therefore the energy balance equation becomes
\begin{equation}
\frac{d}{dt}E_K=\frac{mc_0}{2} -2\nu E_K,
\end{equation}
with the general solution given by
\begin{equation}
E_K(t)=\frac{mc_0}{4\nu}\left(1-e^{-2\nu t}\right).
\end{equation}
In the limit of large times, $\lim_{t\rightarrow \infty}E_K(t)=mc_0/4\nu=%
\mathrm{constant}$, and therefore we obtain for the average of the emitted
electromagnetic radiation the expression
\begin{equation}
\langle P\rangle =\frac{mc_0}{2}=m^2\nu k_BT.
\end{equation}

\subsubsection{Numerical approach and simulated light curves\label{sect:Anumerical}}

The differential equations are brought to a dimensionless form, by the following transformations

\begin{itemize}
\item dimensionless time: $\theta = \nu t$; $\nu= 1/\tau$, where $\nu$ is
the collision frequency in Brownian Motion;

\item dimensionless displacement: $q = x \left ( A \tau ^3 \right ) ^{-1/2}$;

\item dimensionless velocity $V = v/v_T$, where $v_T = \sqrt{A\tau}$;

\item dimensionless acceleration $a_d = \sqrt{\tau/c_0} \equiv dV/d\theta$;

\item dimensionless emitted power $\bar{P} = \frac{3}{2}\frac{\tau c^3}{(Ze)^2 c_0} P \equiv a_d^2 $;

\item dimensionless electric field $\bar{E} = Ee \sqrt{\tau / c _0}$.

\end{itemize}

The dimensionless equation describing the motion of a Brownian particle in a constant electric field is
\begin{equation}
\frac{dV (\theta)}{d\theta} = - V (\theta) + \bar{E} + \bar{\xi}_A (\theta),\label{eq:velA}
\end{equation}
where
\begin{equation}
\left \langle \bar{\xi}_A(\theta)\right \rangle =0, \left \langle \bar{\xi}
_A \left(\theta_1\right)\bar{\xi} _A \left(\theta_2\right)\right \rangle=%
\frac{1}{d\theta}\delta \left(\theta_1-\theta_2\right).\label{eq:noiseA}
\end{equation}

We shall continue to denote the dimensionless emitted power by $L(\theta)$ and interpret it as the luminosity produced by a charged particle in Brownian Motion.

To produce the numerical solution for the displacement, velocity and radiation pattern, Eqs.~\eqref{eq:velA}-\eqref{eq:noiseA} and
\begin{equation}
V(\theta) = \frac{dq}{d\theta}; \quad a_d (\theta) = \frac{dV}{d\theta}; \quad L(\theta) = a_d^2(\theta),
\end{equation}
are used, together with a first order Euler scheme. The variables are discretised in the usual manner, the dimensionless timestep $\theta$ is discretised in units of $h$, such that $\theta _n = n h$, where $n$ is an integer number. Accordingly, all the other variables depending on $\theta$ become, e.g., $V(\theta) = V(n h) \equiv V_n$.

For example, for the velocity, the discretised equation
\begin{equation}
\frac{V_{n+1}-V_n }{h} = -V_n+\bar{E}+\psi _n
\end{equation}
becomes, in update form
\begin{equation}
V_{n+1} = V_n + h(-V_n+\bar{E}+\psi _n),
\end{equation}
where $\psi _n$ is a number drawn at each timestep from $\mathcal{N} (0,h^{-1})$, where $\mathcal{N} (\mu, \sigma ^2)$ is a normal distribution of mean $\mu$ and variance $\sigma ^2$.

The initial conditions are $q(0) = a_d(0) = 0$ and $V(0) = V_0$. For the purpose of studying the case when one injects a high energetic electron into a distribution of plasma versus what happens the case when one introduces a thermalized electron, two regimes of initial velocities will be considered: $V_0 \sim 10^3$ and $V_0\sim 1$ respectively.

To summarize, we obtain numerical solutions for the variables $\{q, V, L\}$, where the parameter space is given by $\{\bar{E},  V_0\}$. For consistency, $d\theta = h = 0.01$ throughout.

\begin{figure*}[tbp]
\begin{center}
\includegraphics[width=8cm,angle=0]{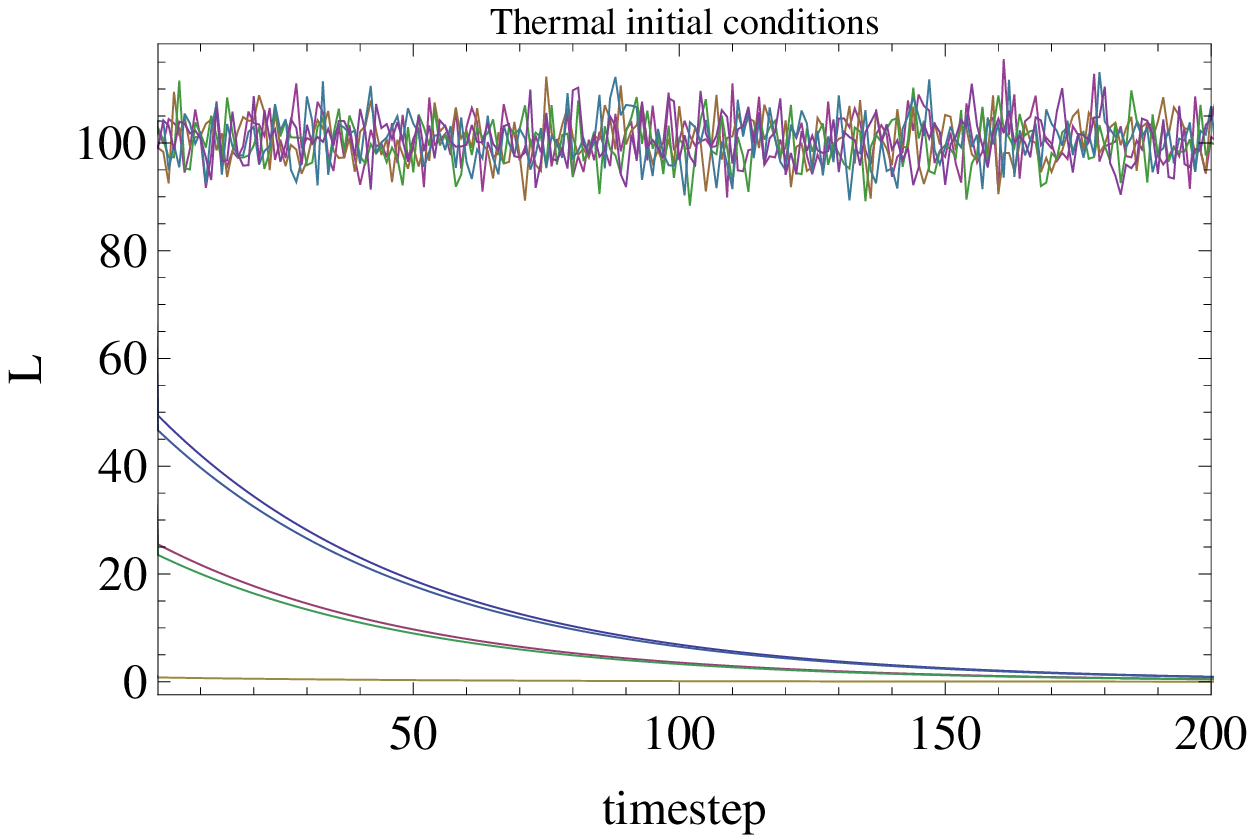} %
\includegraphics[width=8cm,angle=0]{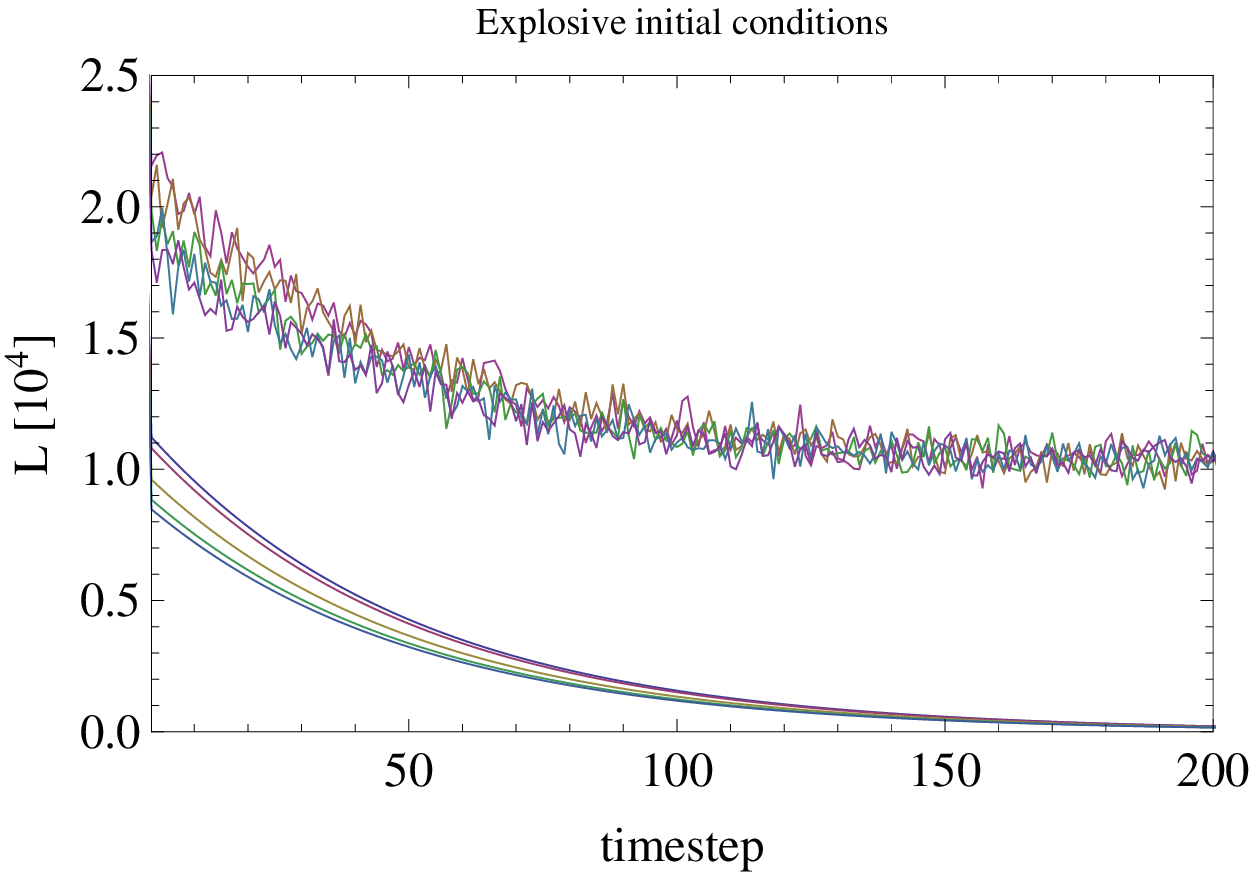}
\end{center}
\caption{Dimensionless power $L$ emitted by a charged particle in stochastic motion in a constant electric field. The solution of the noiseless differential equation is included. The colors are for different values of $\bar{E} \in \{\pm 7, \pm 5, 1\}$, corresponding to $e>0$ and $e<0$, respectively. The terms "thermal" and "explosive" refer to initial conditions of $V_0 = 0.1$ and $V_0=100$ respectively. For the thermal case with noise, the curves have been multiplied by $10^{-2}$ for presentation purposes. For the thermal case without noise, the lower curve is for $\bar{E}=1$, the next two are for $\bar{E}=\pm 5$ and the following two for $\bar{E}=\pm 7$.}
\label{fig:A-L}
\end{figure*}

The power emitted by a charged particle in stochastic one-dimensional motion in a constant electric field is represented, for two distinct sets of initial conditions, in Fig.~\ref{fig:A-L}; \textbf{for comparison purposes, also included is the solution to Eq.~\ref{eq:velA} without noise (i.e., $\bar{\xi} _A = 0$)}.

For the motion of a charged particle in a constant electric field, subjected to a friction force proportional to the velocity, we expect that the amount of radiated energy is and increasing function of the input energy; the influence of the input energy is considered here either by modifying $\vert E \vert$, or by modifying the initial conditions. The expected behavior is indeed recovered by our simulations, as can be seen both in Fig.~\ref{fig:A-L} and by consulting the column 3 of Table~\ref{tab:statA}. For the parameter space considered in this work, it can be seen that adding a noise component is enough not only to balance loss by friction, but in fact to dramatically increase the energy output of the system (Fig.~\ref{fig:A-L}). It thus becomes apparent that the energy radiated in this context is due to the energy received by the electron as random kicks in its Brownian motion. So, when it comes to mean values of the LC vector, they are higher than for the noiseless case and it is not as easy to discriminate between those corresponding to different values of the electric field. The random kicks play a very efficient role in homogenizing the behavior of the charged particle.

\subsubsection{Statistical characteristics of results}

The statistical characteristics (here and elsewhere in the paper) are based on a baseline of $10^3$ timesteps, even if for presentation purposes we show only portions of the data vector. Table~\ref{tab:statA} contains numerical values for the points in the parameter space covered by our simulations. These numerical values may be used to discriminate between different light curves as follows: it is noticed that the mean value increases with the value of the electric field (consequence of increasing energy injection), i.e. both values for $E \in \{-7,7\}$ are higher than for $E=1$, but the value for which the charge has the same sign as the applied electric field is larger. Also, the (overall) mean is always higher by approximately $6\%$ for the explosive initial conditions (ICs) than for the thermal ICs. However, ICs have a dramatic effect on the dispersion, as expected. The standard deviation increases by $300\%$ for the explosive ICs. Also dramatic is the effect on the skewness and kurtosis, as can be noticed from the values in the last two columns. Note that for the thermal case, the skewness and kurtosis have values very close to their theoretical values for a gaussian process, i.e. $0$ and $3$ respectively.

At this point a comment is in order regarding the appropriateness of studying curves such as the LC for the explosive case by means of these statistical characteristics. When studying the behavior of systems, one usually uses data which has been cleaned for the transitory part, i.e., data produced by the system in equilibrium. This is unfeasible in astronomy and astrophysics from at least two points of view: astronomers do not have the luxury of observing many different realisations of an identically prepared system and some of the most interesting physics in astrophysics is the one pertaining to the transitory state. Characteristics such as the skewness and kurtosis can clearly offer at least an idea on the nature of the source.

\begin{figure*}[tbp]
\begin{center}
\includegraphics[width=8cm,angle=0]{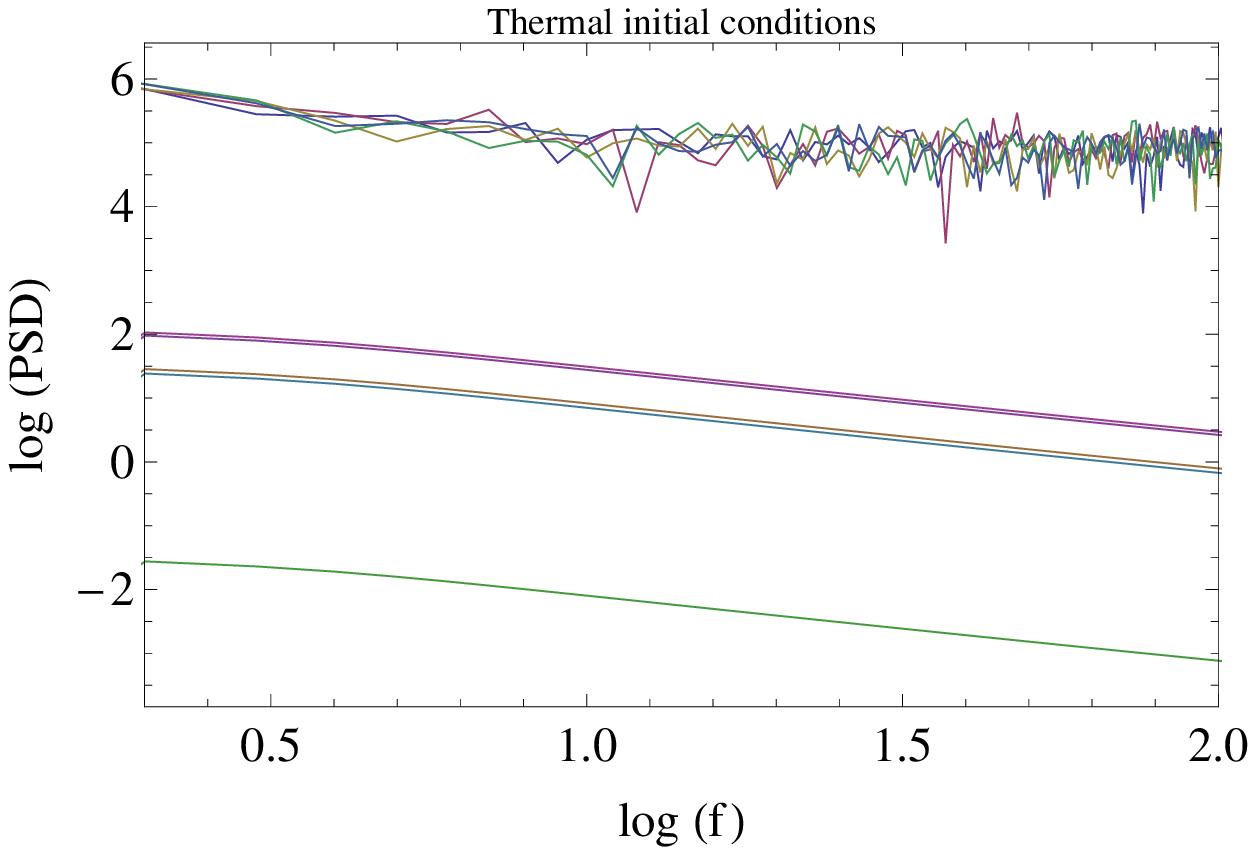} %
\includegraphics[width=8cm,angle=0]{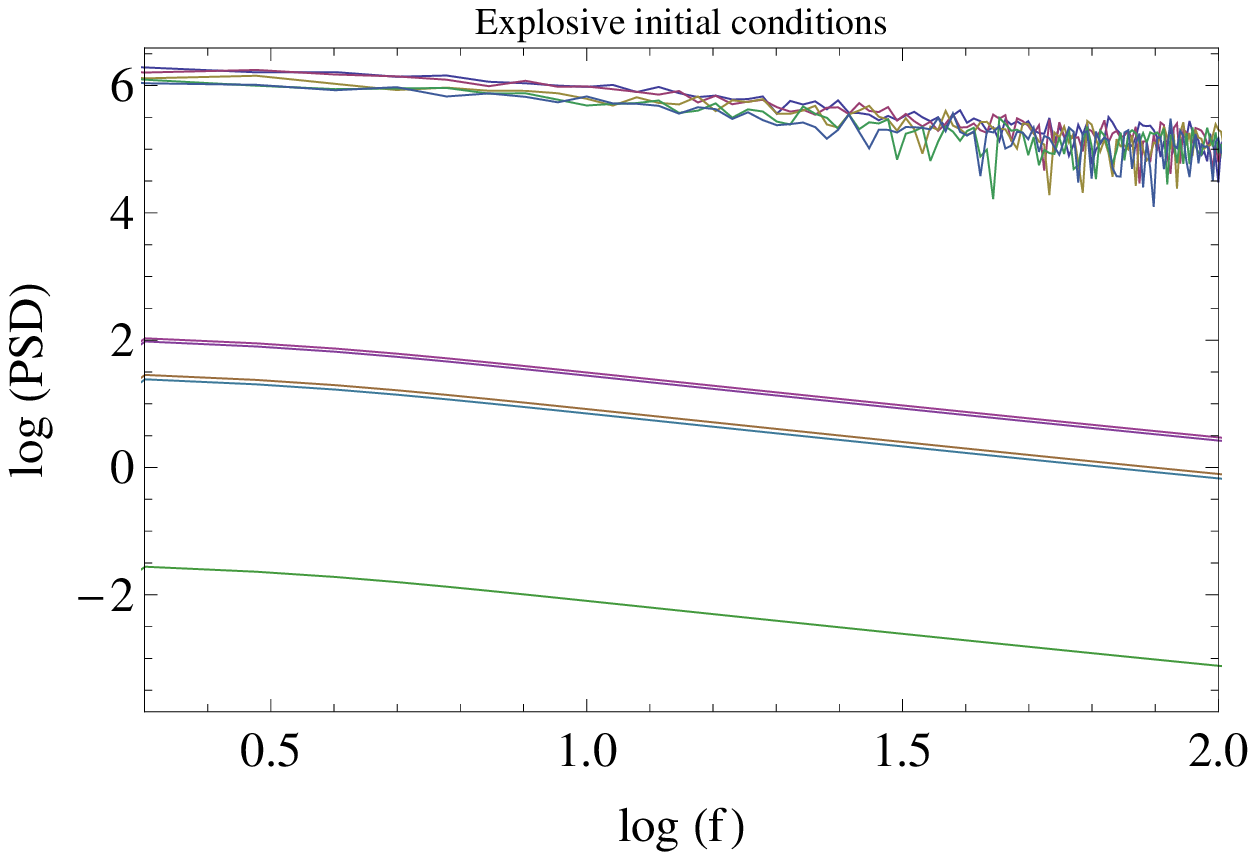}
\end{center}
\caption{Log-Log representation of the PSD corresponding to the stochastic motion of a charged particle in a constant electric field.  The colors are for different values of $\bar{E} \in \{\pm 7, \pm 5, 1\}$, corresponding to $e>0$ and $e<0$, respectively. The terms "thermal" and "explosive" refer to initial conditions of $V_0 = 0.1$ and $V_0=100$ respectively. For the cases without noise, the lower curve is for $\bar{E}=1$, the next two are for $\bar{E}\pm 5$ and the following two for $\bar{E}\pm 7$.}
\label{fig:A-PSD}
\end{figure*}

For the case of a charge moving in a constant electric field while subjected to relatively small Brownian kicks (recall that $\bar{E}$ may be interpreted as the ratio between the influence of the electric energy with respect to the energy injected by Brownian motion) the PSDs are generally flat curves (Fig.~\ref{fig:A-PSD}), reflecting the fact that the noise content is trivial. \textbf{Also included is the PSD of the noiseless counterpart}. The actual value of the $\bar{E}$ free parameter has little importance for both the qualitative and quantitative look of the PSD in both initial condition cases. There is no feature in the PSD which indicates a characteristic frequency nor features that look like QPOs, but this is what we would expect for this case.

\subsection{Radiation of a charged particle in a harmonic external potential
\label{sect:B1}}

\subsubsection{Equations and physics}

The Langevin equation for the one dimensional motion of a charged particle with mass $m$
and charge $e$ in a harmonic potential with natural frequency $\omega _{0}$
is given by
\begin{equation}
\frac{d^{2}x}{dt^{2}}+\nu \frac{dx}{dt}+\omega _{0}^{2}x=\xi _{B},
\label{l1}
\end{equation}%
where the stochastic force $\xi _{B}$ has the properties
\begin{equation}
\left\langle \xi _{B}(t)\right\rangle =0,\left\langle \xi _{B}\left(
t_{1}\right) \xi _{B}\left( t_{2}\right) \right\rangle =\frac{B}{dt}\delta
\left( t_{1}-t_{2}\right) .
\end{equation}

The total energy per unit mass of the charged particle in the harmonic
potential is given by
\begin{equation}
E=\frac{1}{2}v^{2}+\frac{1}{2}\omega _{0}^{2}x^{2}.
\end{equation}

For the variation of the energy of the particle we obtain, with the use of
Eq.~(\ref{l1}), the expression
\begin{equation}
\frac{dE}{dt}=P=\xi _{B}\frac{dx}{dt}-\nu \left( \frac{dx}{dt}\right) ^{2}=%
\frac{2}{3}\frac{e^{2}}{c^{3}}a^{2}.
\end{equation}

\subsubsection{Numerical approach and light curve\label{sect:Bnumerical}}

The same dimensionless variables are again used, together with
\begin{itemize}

\item {} dimensionless frequency: $W=\omega _0 \tau$, where $\omega _0$
is the frequency of an external harmonic potential.

\end{itemize}

The dimensionless form of the Langevin differential equation Eq.~(\ref{l1}) becomes
\begin{equation}\label{harmosc}
\frac{d^2q (\theta)}{d\theta ^2} + \frac{dq(\theta)}{d\theta} + W^2 q
(\theta)= \bar{\xi} _B (\theta),
\end{equation}
where
\begin{equation}
\left \langle \bar{\xi}_B(\theta)\right \rangle =0, \left \langle \bar{\xi}
_B \left(\theta_1\right)\bar{\xi} _B \left(\theta_2\right)\right \rangle=%
\frac{1}{d\theta}\delta \left(\theta_1-\theta_2\right).
\end{equation}

In the absence of the external stochastic force, $\bar{\xi} _B (\theta)\equiv 0$, the general solution of Eq.~(\ref{harmosc}) with initial conditions $q(0)=q_0$ and $\left.dq/d\theta\right|_{\theta =0}=\dot{q}_0$ is given by
\bea
q(\theta)& =&\frac{e^{-\theta /2}}{\sqrt{1-4 W^2}}\Bigg[ (2 \dot{q}_0+q_0) \sinh \left(\frac{1}{2} \theta \sqrt{1-4 W^2}\right)+\nonumber\\
&&\sqrt{1-4 W^2} q_0 \cosh \left(\frac{1}{2} \theta  \sqrt{1-4
   W^2}\right)\Bigg].
\eea
The physical characteristics of the particle motion depends on the sign of the quantity  $1-4W^2$. If $1-4W^2<0$, the particle will oscillate at the natural damped frequency $\omega =\sqrt{4W^2-1}/2$.

The same numerical algorithm as in Section~\ref{sect:Anumerical} is employed, i.e., we discretise the equations and implement an Euler scheme for the update equations. We obtain numerical solutions for the variables $\{q, V, P\}$, where the parameter space is given by $\{\ W,  V_0\}$.

\begin{figure*}[tbp]
\begin{center}
\includegraphics[width=8cm,angle=0]{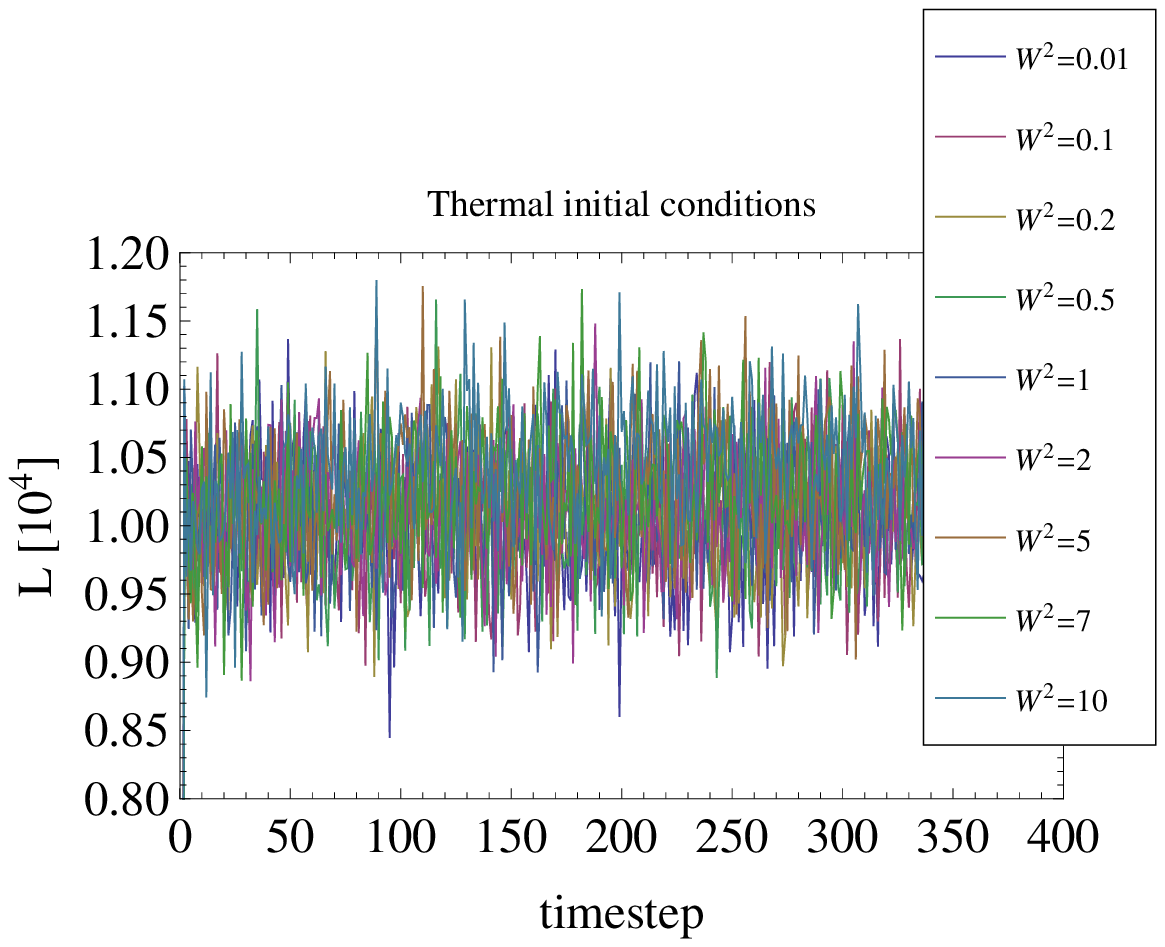} %
\includegraphics[width=8cm,angle=0]{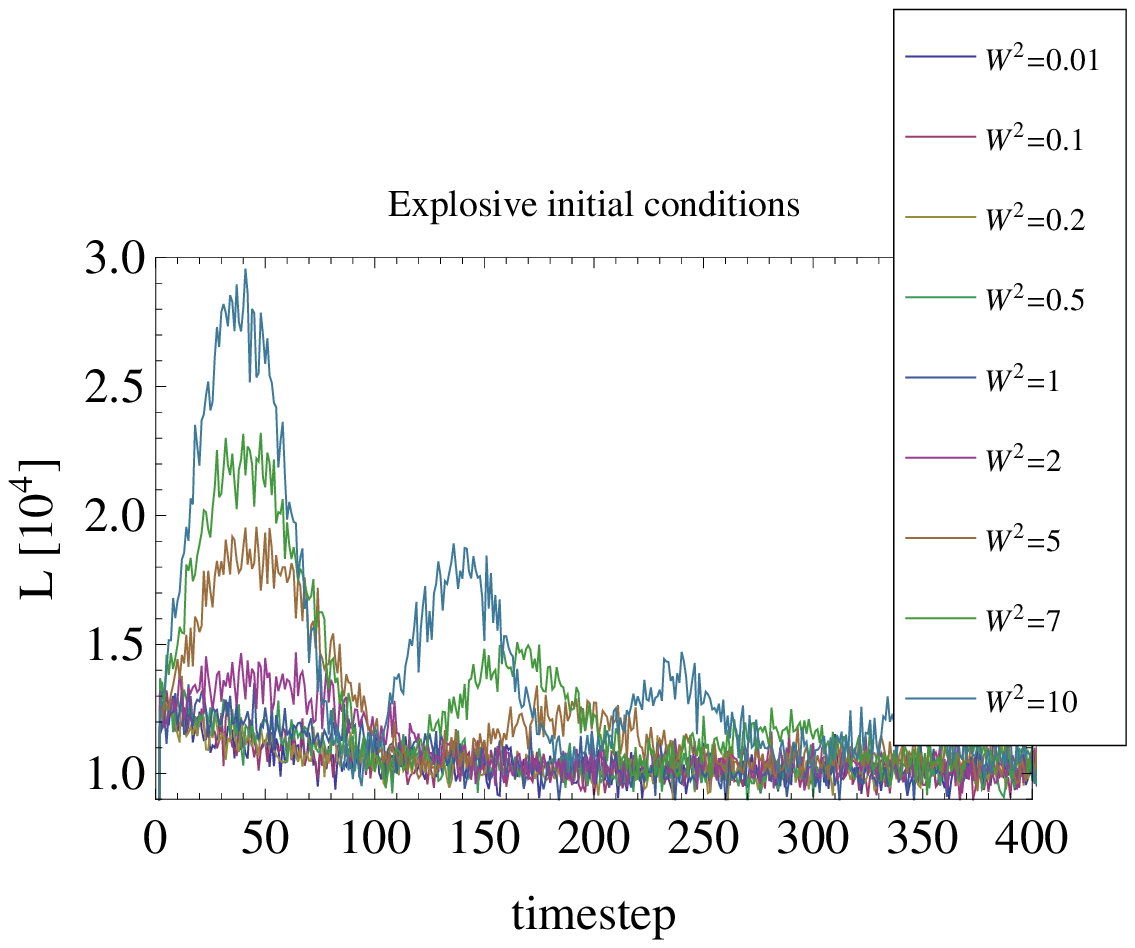}
\end{center}
\caption{Dimensionless power $L$ emitted by a charged particle in stochastic motion in a harmonic external potential of dimensionless frequency $W^2$. The terms "thermal" and "explosive" refer to initial conditions of $V_0 = 1$ and $V_0=50$ respectively.}
\label{fig:B-L}
\end{figure*}

\begin{figure*}[tbp]
\begin{center}
\includegraphics[width=8cm,angle=0]{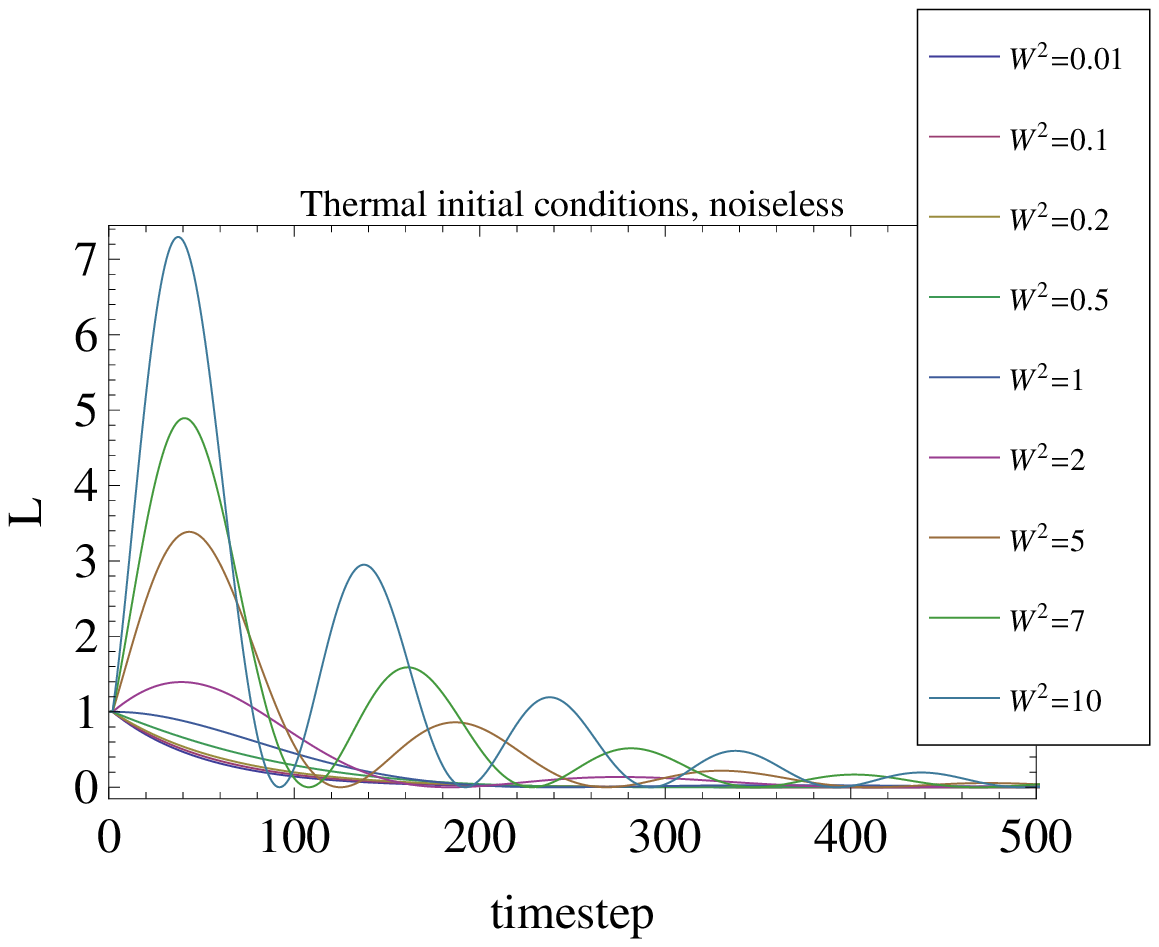} %
\includegraphics[width=8cm,angle=0]{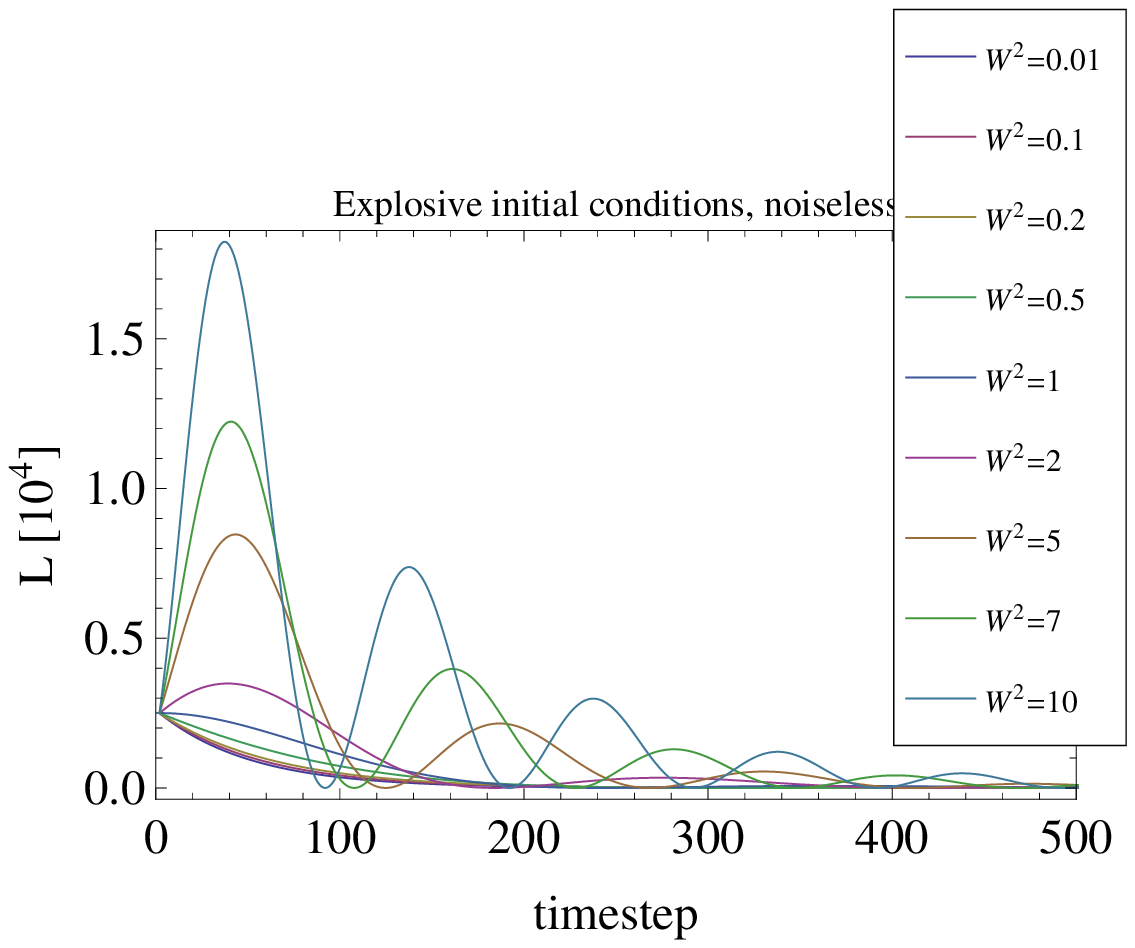}
\end{center}
\caption{Dimensionless power $L$ emitted by a charged particle in motion in a harmonic external potential of dimensionless frequency $W^2$. The terms "thermal" and "explosive" refer to initial conditions of $V_0 = 1$ and $V_0=50$ respectively.}
\label{fig:B-L-nn}
\end{figure*}

The time variation of the electromagnetic radiation emitted by a charged particle in stochastic motion in a harmonic potential is represented in Fig.~\ref{fig:B-L}; the noiseless counterpart is given in Fig.~\ref{fig:B-L-nn}.

For the motion of a charged particle, undergoing friction in an external potential, we would expect that the light curve is a sinusoidal-like curve, with the amplitude decreasing in time. This indeed can be seen in Fig.~\ref{fig:B-L-nn}. Naturally, since the friction remains constant, the position of the peaks depends on the value of $W$; also, the mean value of the LC is an increasing function of $W$, as more energetic electrons emit more energy when subjected to otherwise identic conditions; this can be seen in column 3 of Table~\ref{tab:statB}. When the noise is turned on (Fig.~\ref{fig:B-L}), the random kicks are very efficient at homogeneizing the behavior of the charged particle for the thermal case: there is little difference between the LCs for $W^2$ varying across $3$ orders of magnitude. However, in the explosive case there is a clear difference between  LCs produces by various $W^2$ values. The two sources of energy injection other than the noise are in this case large enough to leave a visible signature in the LC.

\subsubsection{Statistical analysis of results}

Table~\ref{tab:statB} contains the analysis of the statistical characteristics for the case of a charged Brownian particle, with friction, in a harmonic potential of dimensionless equivalent frequency $W$. As expected, the mean value of the power output increases as $W^2$ increases and is larger consistently for a larger energy input through the different ICs. In the case of the dispersion, while for the thermal case the change is of $\approx 3.4 \%$ for different values of $W$, in the explosive case, the dispersion changes by $\approx 140 \%$ for different values of $W$. The skewness is pronounced for the thermal case, but rather constant with varying $W$. For the explosive case, the skewness is almost zero for $W^2 = 1$, but increases by two order of magnitude for $W^2 = 5$.

\begin{figure*}[tbp]
\begin{center}
\includegraphics[width=8cm,angle=0]{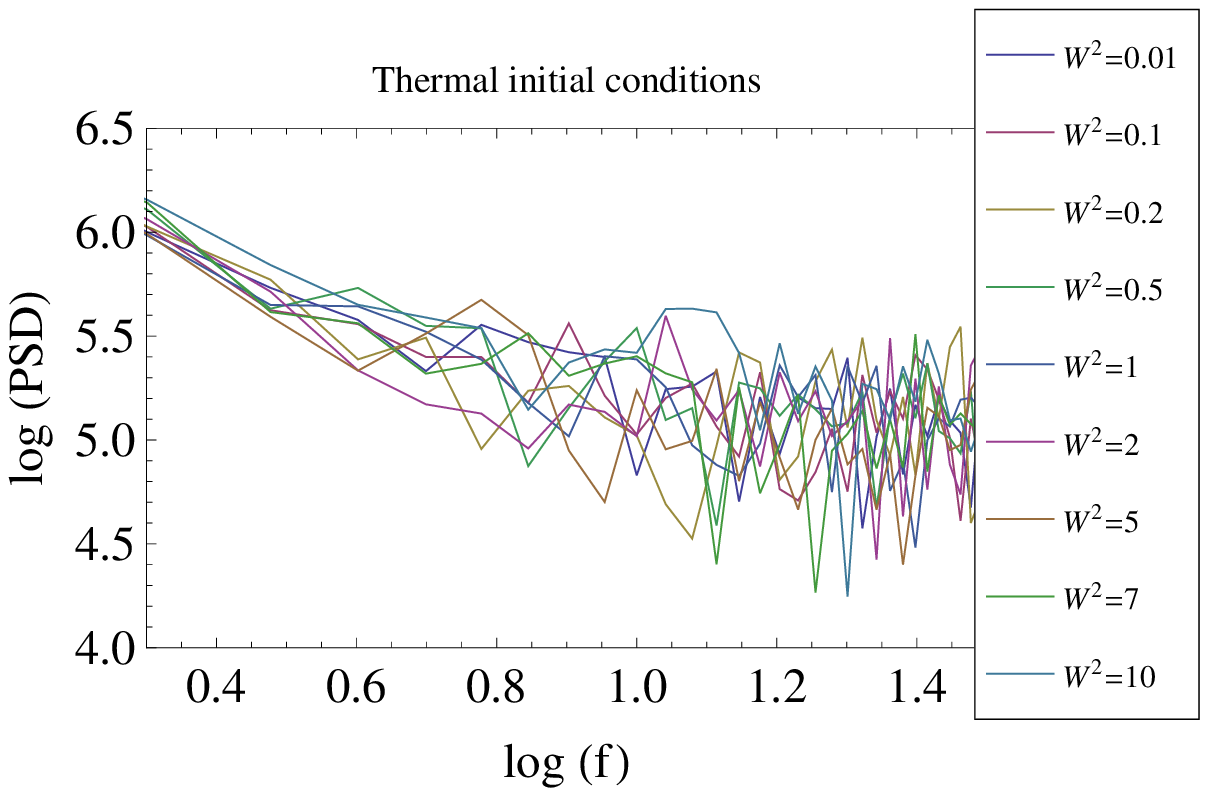} %
\includegraphics[width=8cm,angle=0]{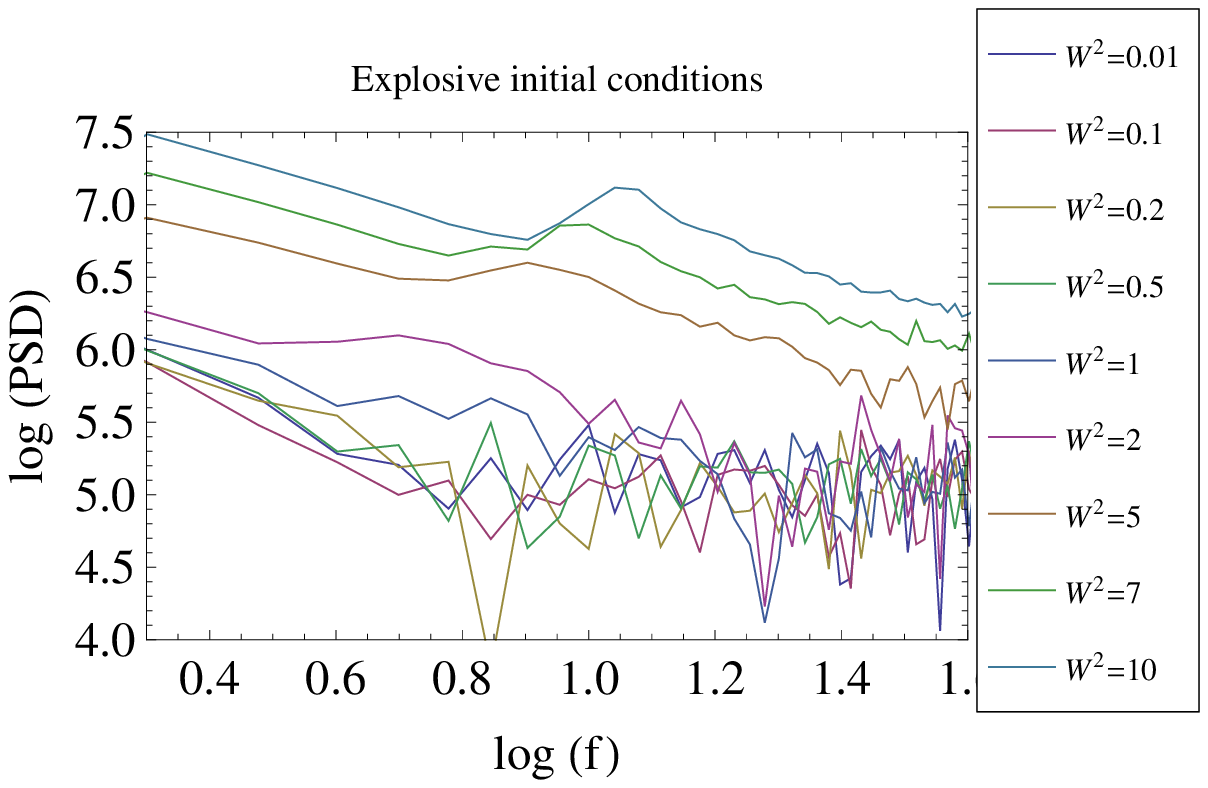}
\end{center}
\caption{Log-Log representation of the PSD as a function of frequency of the radiation emitted by a charged particle in stochastic motion in a harmonic potential of dimensionless frequency $W^2$. The terms "thermal" and "explosive" refer to initial conditions of $V_0 = 1$ and $V_0=50$ respectively.}
\label{fig:B-PSD}
\end{figure*}

\begin{figure*}[tbp]
\begin{center}
\includegraphics[width=8cm,angle=0]{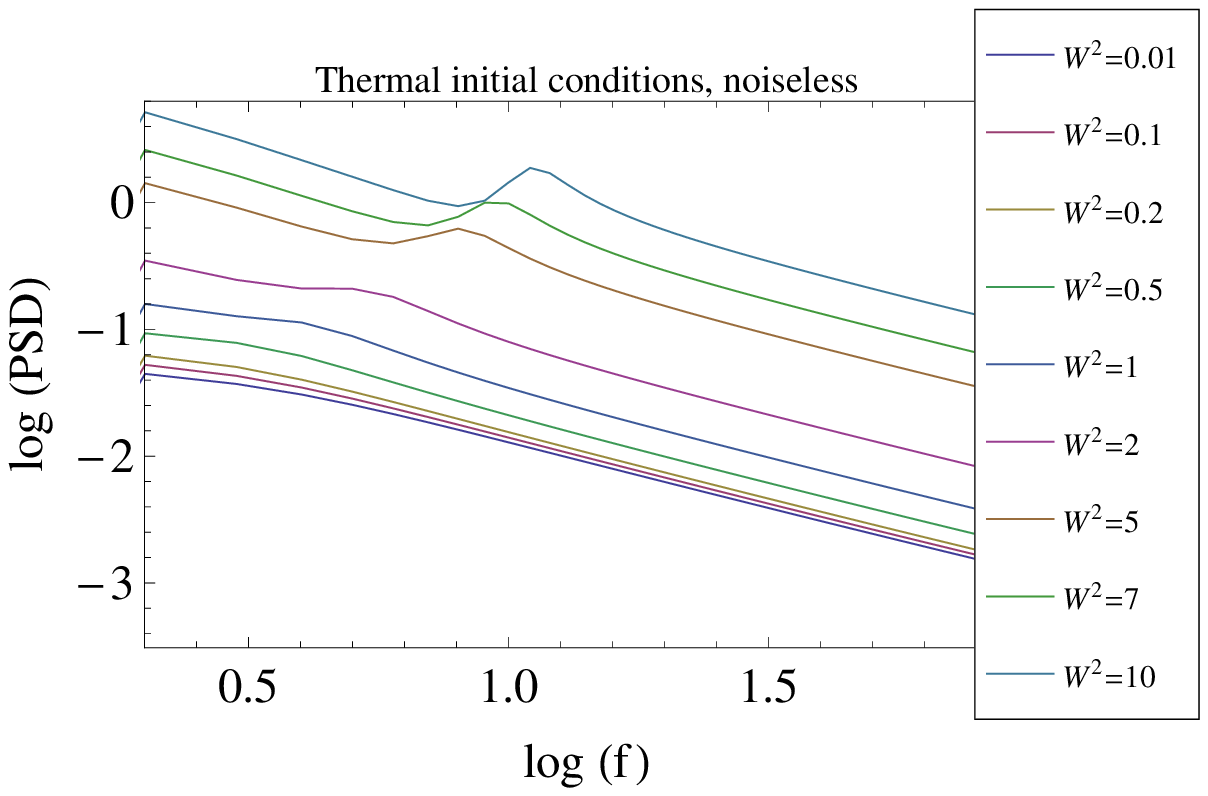} %
\includegraphics[width=8cm,angle=0]{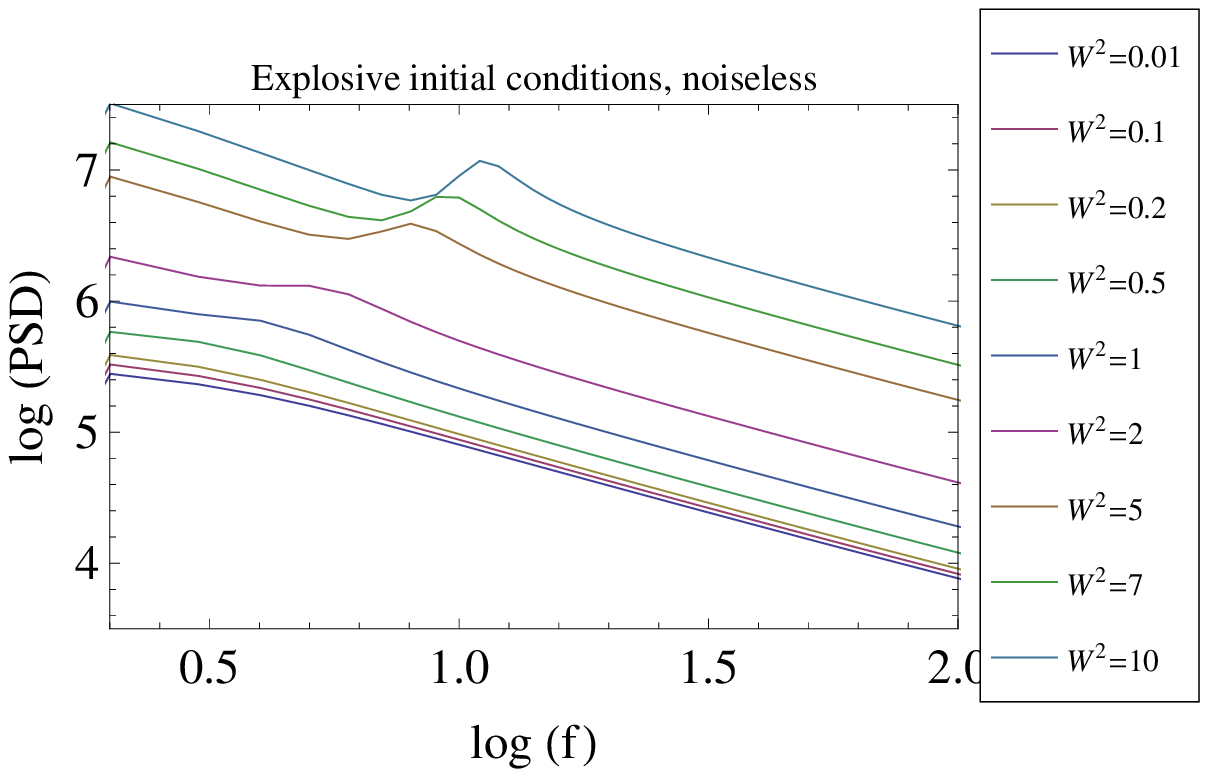}
\end{center}
\caption{Log-Log representation of the PSD as a function of frequency of the radiation emitted by a charged particle in motion in a harmonic potential of dimensionless frequency $W^2$. The terms "thermal" and "explosive" refer to initial conditions of $V_0 = 1$ and $V_0=50$ respectively.}
\label{fig:B-PSD-nn}
\end{figure*}

A technical note regarding the reading of the PSD, namely the clarification of the meaning of a feature appearing at $\log f = x$; the calculation of the PSD is based on taking the correlation between the values of the LC $j$ timesteps apart. When calculating the associated frequency, one has to take into account that we meshed the temporal axis in quanta of $h=0.01$, such that in fact the value $x$ in the PSD corresponds to a period $T = 10^{2-x}$. Conversely, if one expects periodicity with some period $\tau = 2\pi/\omega$, then the connection between $\omega$ and $x$ is given by $x = 2 - \log (2\pi/\omega)$.

Since the motion is periodic, we expect that the PSD will display well defined peaks at the frequencies corresponding to $W$. This can be seen in Fig.~\ref{fig:B-PSD-nn}; for $W^2 \in (0.01, 10)$, the quantity $\log f = \in ( 0.2,1.7 )$. One can see that the theoretical behavior is indeed recovered by the simulations. When the noise is turned on, these general features characterizing the LC are recovered in the PSD as well. For both IC cases, the power is greater than for the noiseless case and generally greater for the explosive conditions; different values of $W^2$ produce clear signatures in the PSD of the explosive LC. The PSD contains no new feature brought on by noise, but the existing peaks are widened, as one would expect from the interaction with a heath bath.

\subsection{The case of the generalized Langevin equation}

\subsubsection{Equations and physics}

In the presence of a colored noise, which accounts for the general memory and retarded effects of the frictional force, and on the  fluctuation-dissipation theorems, the motion of a charged particle in a harmonic potential is described by the generalized Langevin equation, which in the one-dimensional case can be written as
\begin{equation}\label{l2}
\frac{d^2 x}{dt^2}+\int_0^t{\gamma (t-t^{\prime }) \frac{dx\left(t'\right)}{dt^{\prime }}}%
dt^{\prime  }+\omega _0^2x=\xi _C (t),
\end{equation}
where
\begin{equation}
\gamma (t) = \frac{\alpha}{\tau _d} \exp \left \{ -t/\tau\right \},
\end{equation}
\begin{equation}
\langle \xi _C (t) \xi _C (t^{\prime }) \rangle = \frac{1}{\beta} \gamma
(t-t^{\prime }).
\end{equation}

The time variation of the total energy of the harmonically oscillating charged particle with motion described by the generalized Langevin equation is given by
\be
\frac{dE}{dt}=P=\xi _C\frac{dx}{dt}-\frac{dx}{dt}\int_0^t{\gamma (t-t^{\prime }) \frac{dx\left(t'\right)}{dt^{\prime }}}%
dt^{\prime  }=\frac{2}{3}\frac{e^2}{c^3}a^2.
\ee

\subsubsection{Numerical approach and light curve}

For the case of the generalized Langevin equation Eq.~(\ref{l2}), in addition to the other dimensionless quantities discussed in Sections~\ref{sect:Anumerical} and~\ref{sect:Bnumerical}, we define
\begin{itemize}
\item {} dimensionless friction amplitude: $\bar{\alpha} = \alpha \tau$;

\item {} dimensionless friction kernel: $\bar{\gamma} (\theta) = \bar{\alpha}
e^{-\theta}$;

\item {}dimensionless correlation amplitude for the stochastic force $C_1 =
\alpha \nu/\beta (\nu v_T)^{-2}$.
\end{itemize}

The dimensionless equation becomes
\begin{equation}
\frac{d^2q}{d\theta ^2} + \int _0 ^\theta \bar{\gamma} (\theta - \theta
^{\prime }) \frac{dq}{d\theta ^{\prime }}d\theta ^{\prime 2 } + W^2 q = \bar{\xi}_C
(\theta),
\end{equation}
where
\begin{equation}
\langle \bar{\xi} _C (\theta) \bar{\xi} _C (\theta ^{\prime }) \rangle = C_1
e^{\theta ^{\prime }- \theta}.
\end{equation}

This type of equation is solved by rewriting it as a set of two coupled
equations, with the help of an auxiliary dimensionless variable $Z(\theta)$
\begin{equation}
\frac{d^2q}{d\theta ^2} + W^2 q = Z(\theta),
\end{equation}
\begin{equation}
\frac{dZ}{d\theta} = -Z-\bar{\alpha}\frac{dq}{d\theta} + \eta _C (\theta),
\end{equation}
where
\begin{equation}
\langle \eta _C (\theta) \eta _C (\theta ^{\prime }) \rangle = \bar{C}
\delta (\theta - \theta^{\prime }),
\end{equation}
$\bar{C} = 2 \bar{\alpha} v_T^{-2}\beta ^{-1}$and $Z(0)$ is drawn from
a distribution with
\begin{equation}
\langle Z^2(0) \rangle = 0.5 \bar{C},
\end{equation}
for each realization of the stochastic process.

The equations for $Z$ and $q$ are solved by a second order Runge-Kutta
procedure based on the algorithm developed in~\cite{hershkowitz1998}. The update form of the equations is

\begin{equation}
q_{n+1} = q_n + V_n h + \frac{1}{2}h^2 (-\Omega ^2 q_n + Z_n),
\end{equation}
\begin{equation}
V_{n+1} = V_n + h (-\Omega ^2 q_n + Z_n)+ \frac{1}{2}h^2 (-\Omega ^2 V_n -Z_n -\bar{\alpha} V_n) + c_2,
\end{equation}
\begin{eqnarray}
Z_{n+1} &=& Z_n  + h(-Z_n -\bar{\alpha} V_n) + \\ && \frac{h^2}{2} \left ( Z_n + \bar{\alpha} V_n + \bar{\alpha} \Omega ^2 q_n -  \bar{\alpha} Z_n \right ) + c_1 - c_2  \nonumber,
\end{eqnarray}
where $c_1 = x$ and $c_2 = h (x/2 + y/(2\sqrt{3}))$, with $x, y \in \mathcal{N}(0,\bar{C})$, drawn at each timestep.

We obtain numerical solutions for the variables $\{q, V, P\}$, where the parameter space is given by $\{\ W, V_0\}$ and we set $\bar{\alpha}$ and $\bar{C}$ to some fixed values.

\begin{figure*}[tbp]
\begin{center}
\includegraphics[width=8cm,angle=0]{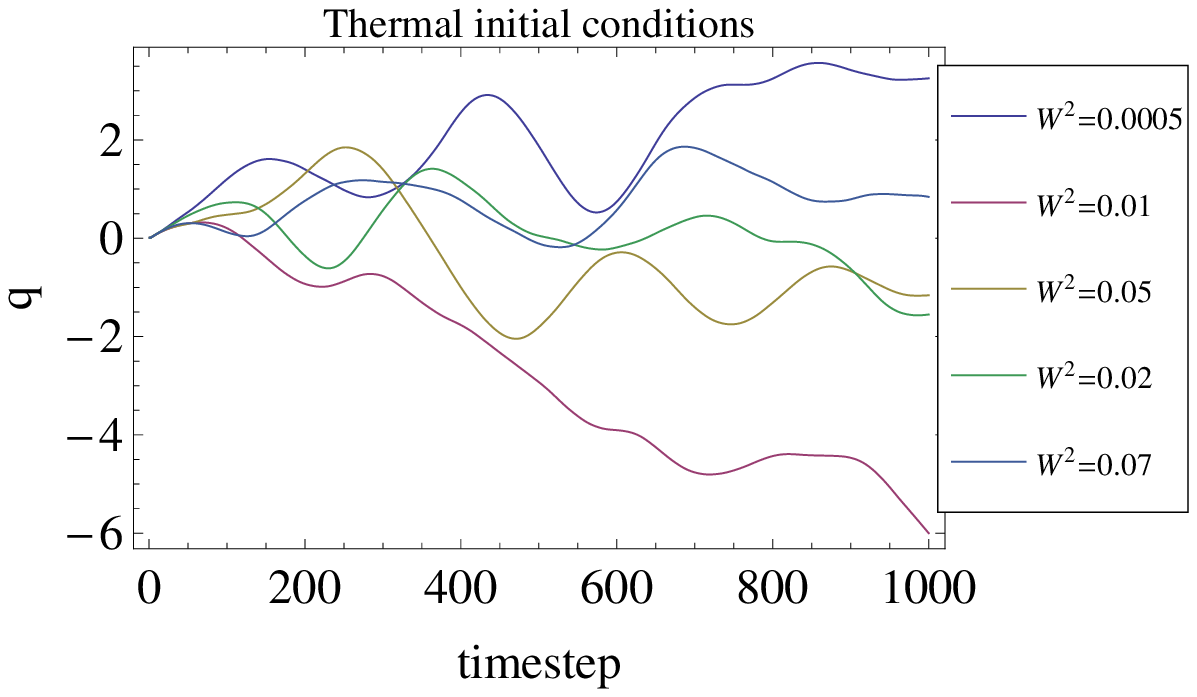} %
\includegraphics[width=8cm,angle=0]{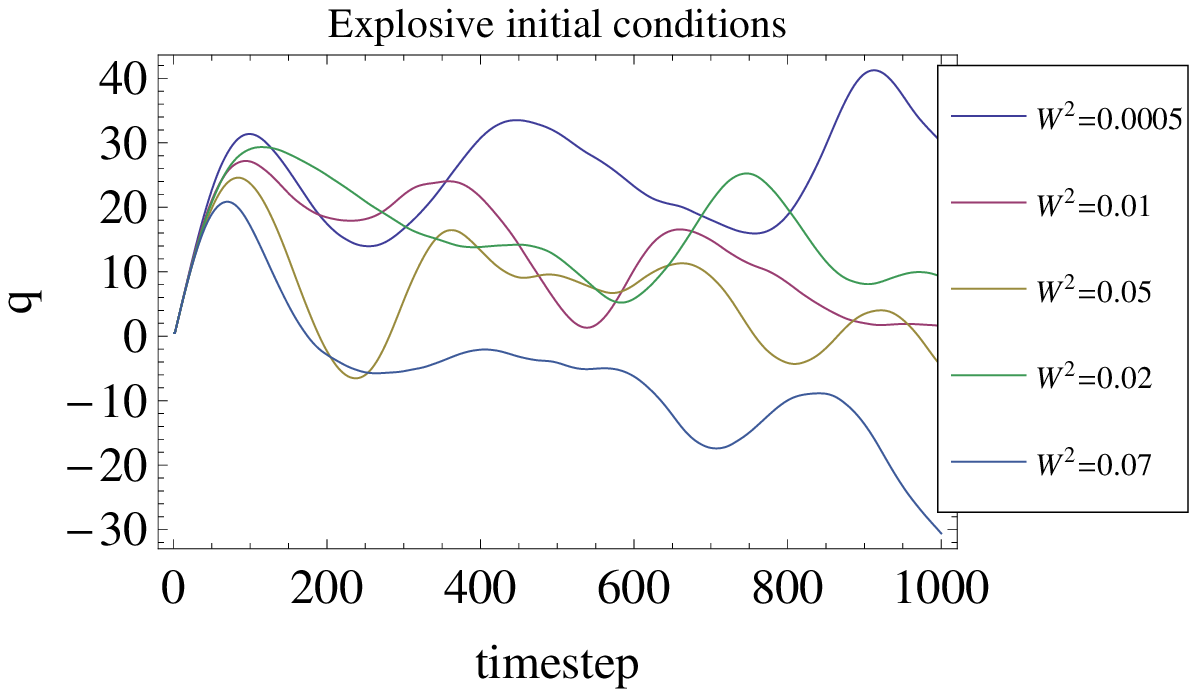}
\end{center}
\caption{ Dimensionless displacement $q$ of a charged particle in stochastic motion described by the generalized Langevin equation with exponential friction kernel, with $\bar{\alpha} = 5$. The terms "thermal" and "explosive" refer to initial conditions of $V_0 = 1$,  $\bar{C}=10$ and $V_0=50$,  $\bar{C}=100$  respectively.}
\label{fig:C-q}
\end{figure*}

\begin{figure*}[tbp]
\begin{center}
\includegraphics[width=8cm,angle=0]{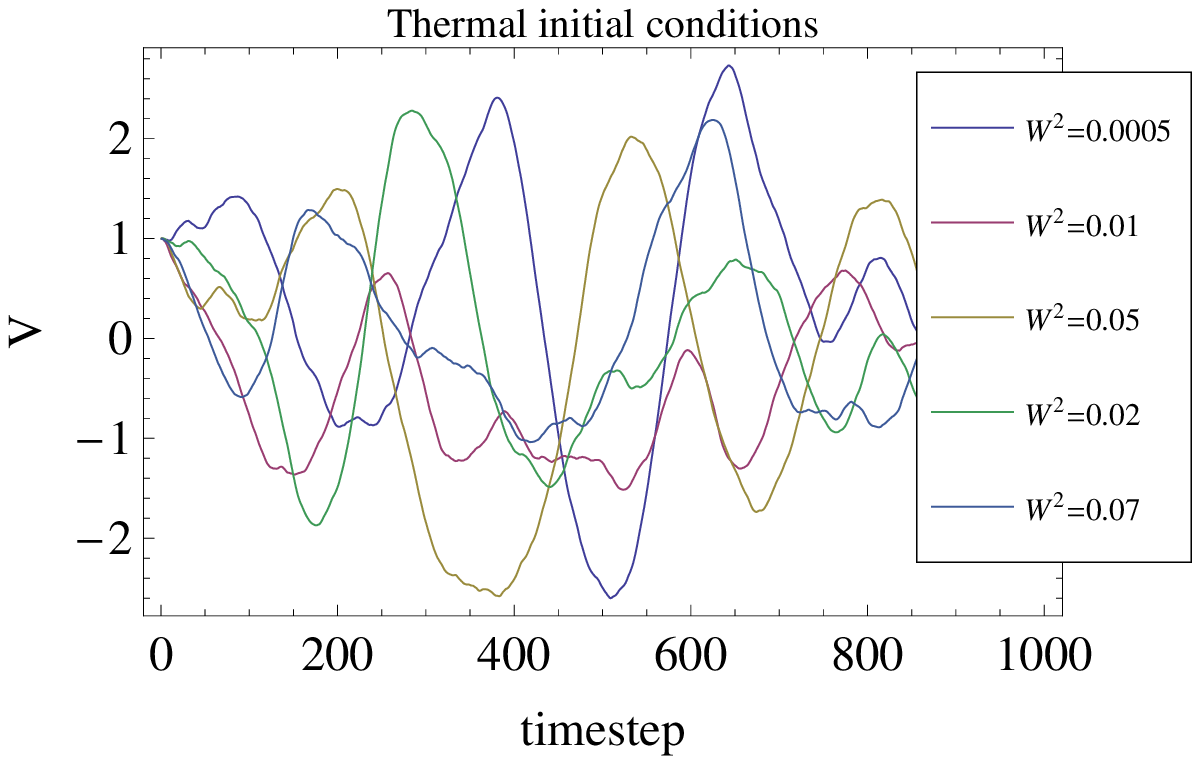} %
\includegraphics[width=8cm,angle=0]{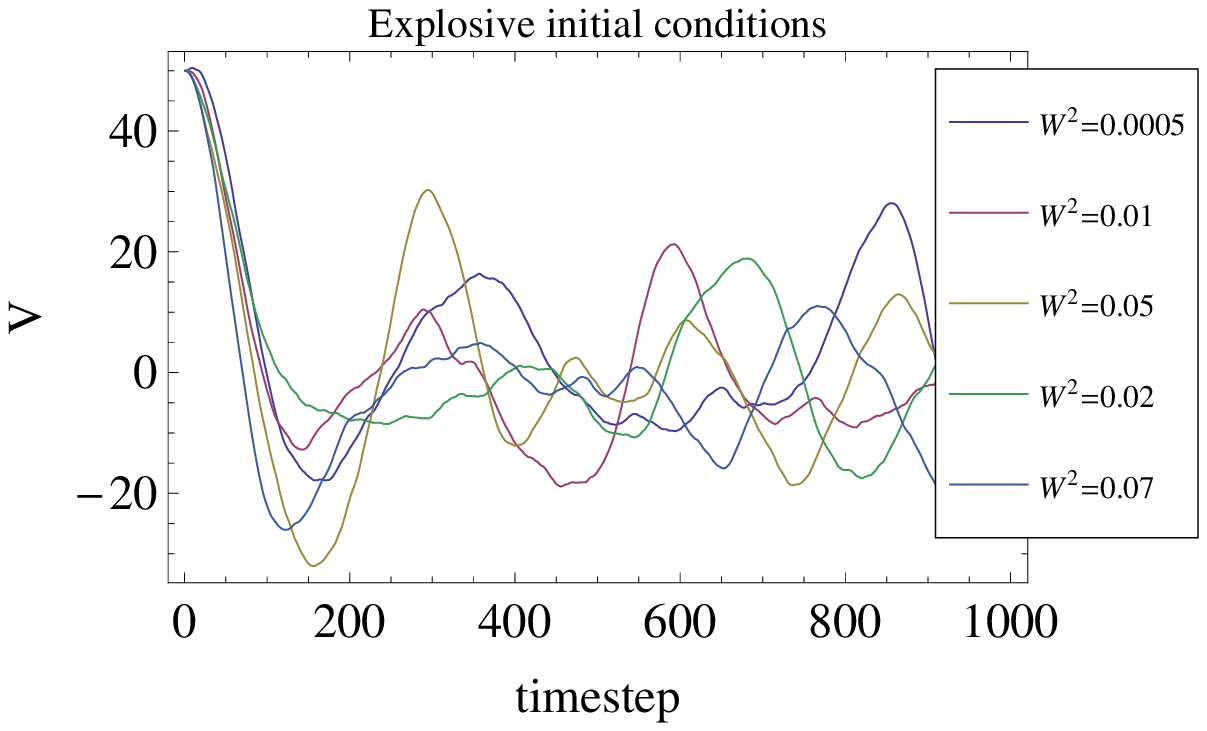}
\end{center}
\caption{Dimensionless velocity of a charged particle in stochastic motion described by the generalized Langevin equation with exponential friction kernel, with $\bar{\alpha} = 5$. The terms "thermal" and "explosive" refer to initial conditions of  $V_0 = 1$,  $\bar{C}=10$ and $V_0=50$,  $\bar{C}=100$ respectively.}
\label{fig:C-V}
\end{figure*}

\begin{figure*}[tbp]
\begin{center}
\includegraphics[width=8cm,angle=0]{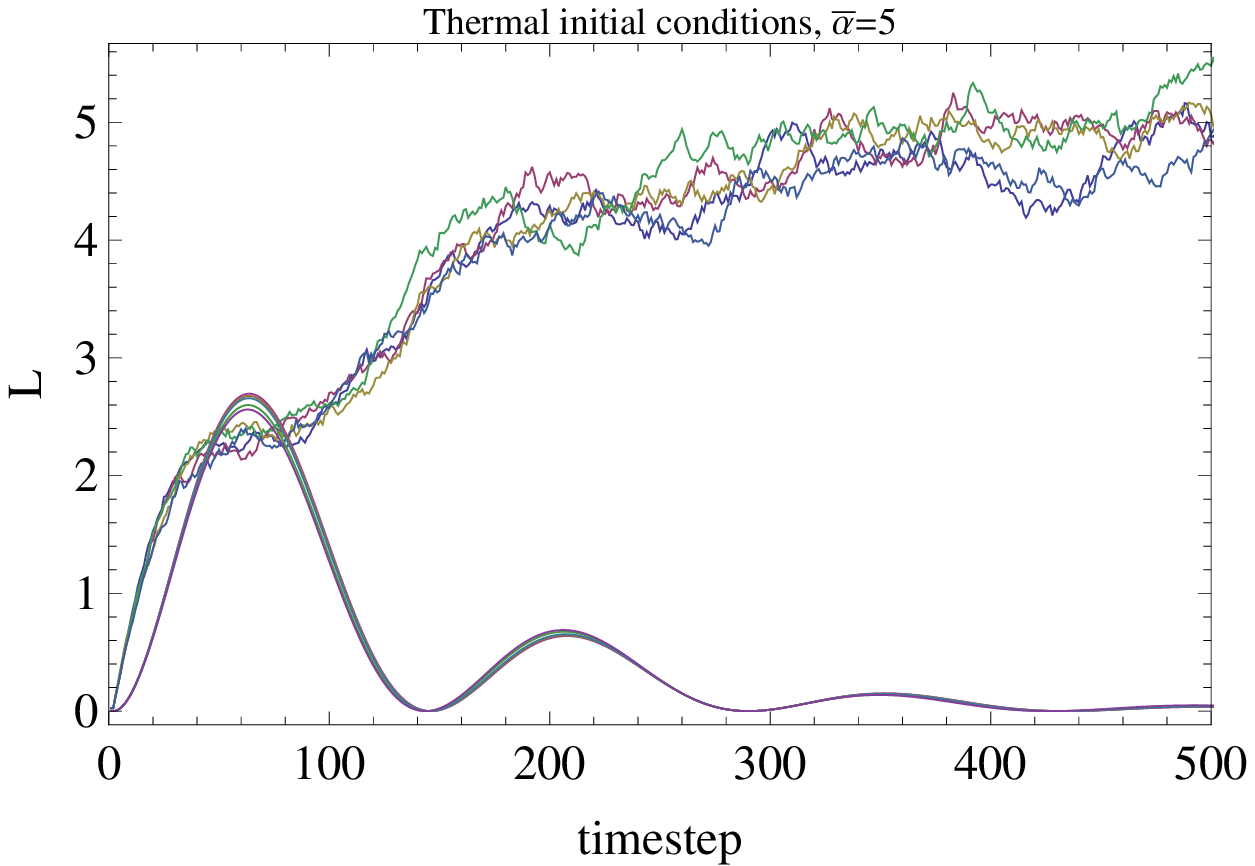} %
\includegraphics[width=8cm,angle=0]{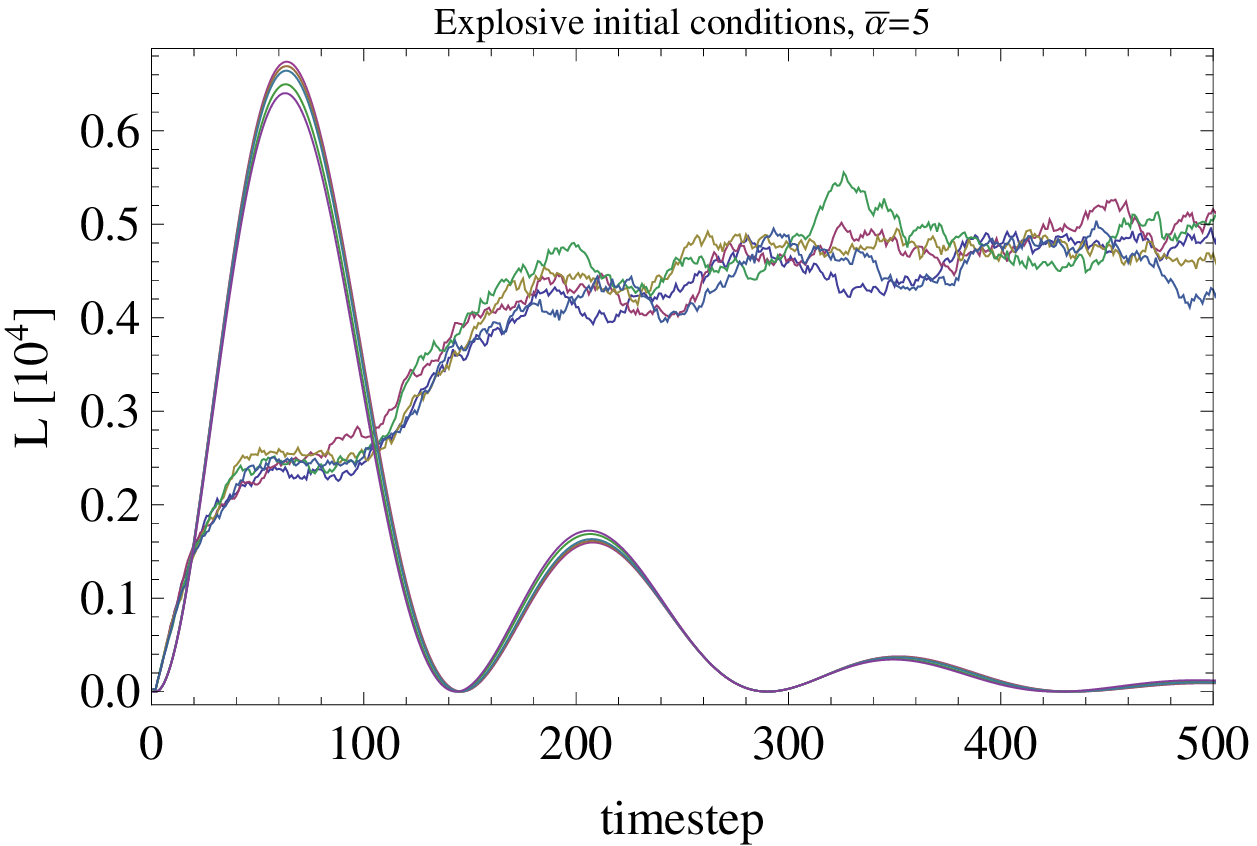}
\end{center}
\caption{Dimensionless electromagnetic power $L$ emitted by a charged particle in stochastic motion described by the generalized Langevin equation with exponential friction kernel, with $\bar{\alpha} = 5$. The terms "thermal" and "explosive" refer to initial conditions of  $V_0 = 1$,  $\bar{C}=10$ and $V_0=50$,  $\bar{C}=100$ respectively; the variable parameter is $W^2 \in \{ 0.0005, 0.01, 0.02, 0.05, 0.07 \}$, but it is seen not to have a big effect on the energy output. Both noise and noiseless cases included. For presentation purposes, the light curves for the case with noise were multiplied by $10^{-3}$ (thermal) and by $10^{-2}$ (explosive).}
\label{fig:C-L-l}
\end{figure*}

\begin{figure*}[tbp]
\begin{center}
\includegraphics[width=8cm,angle=0]{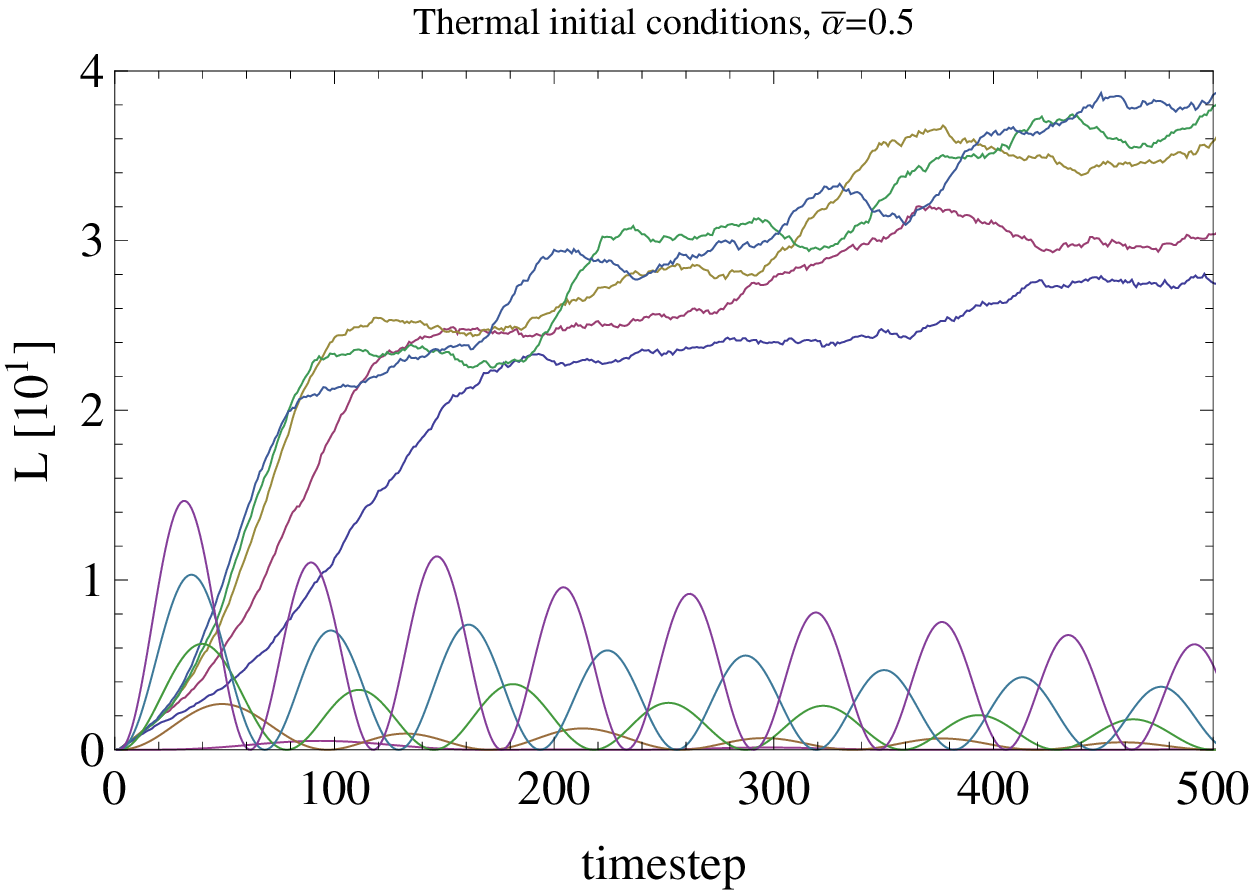} %
\includegraphics[width=8cm,angle=0]{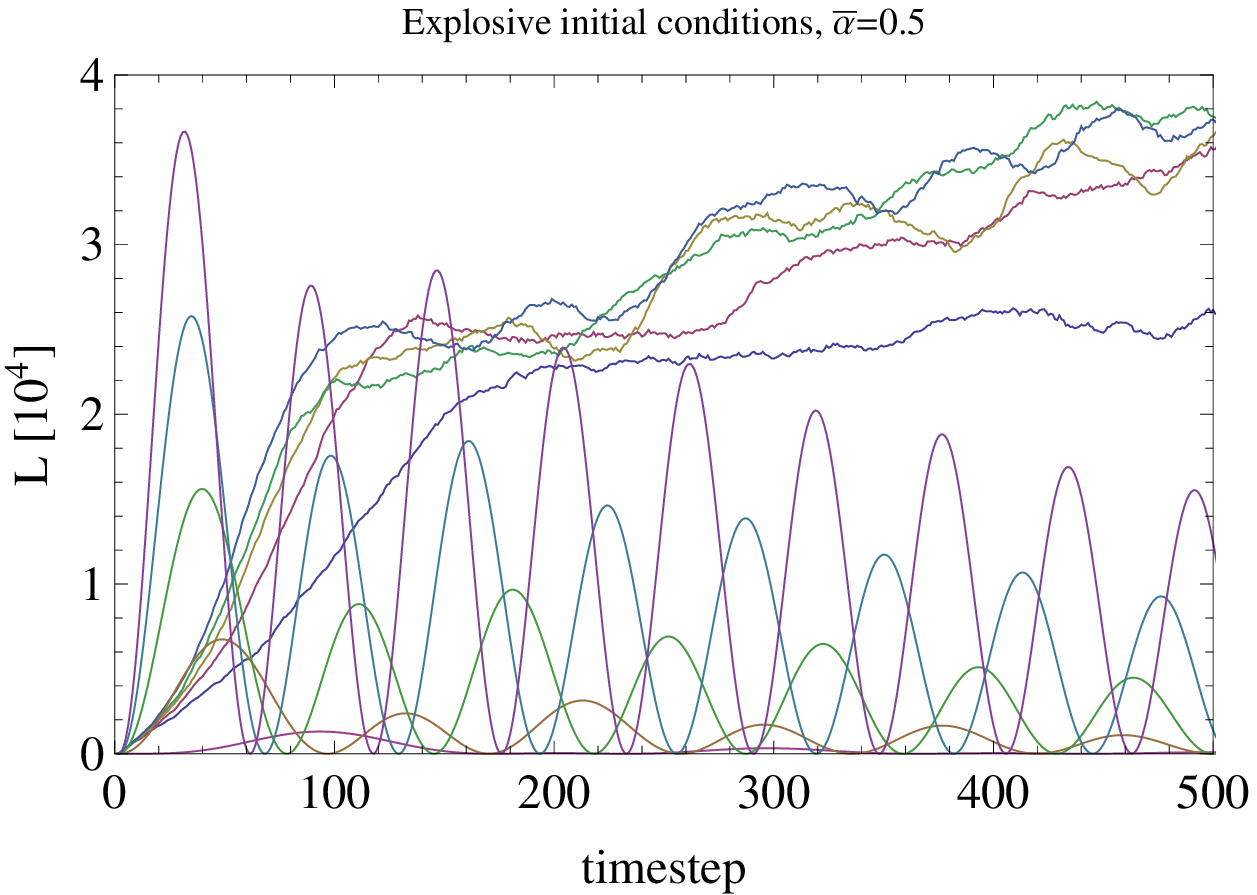}
\end{center}
\caption{Dimensionless electromagnetic power $L$ emitted by a charged particle in stochastic motion described by the generalized Langevin equation with exponential friction kernel, with $\bar{\alpha} = 0.5$. The terms "thermal" and "explosive" refer to initial conditions of  $V_0 = 1$,  $\bar{C}=10$ and $V_0=50$,  $\bar{C}=100$ respectively; the variable parameter is $W^2 \in \{ 5,10,15,20,25 \}$; the amplitude is an increasing function of $W$. Both noise and noiseless cases included. For presentation purposes, the light curves for the case with noise were multiplied by $10^{-3}$ (thermal) and by $10^{-2}$ (explosive).}
\label{fig:C-L-g}
\end{figure*}

\begin{figure*}[tbp]
\begin{center}
\includegraphics[width=8cm,angle=0]{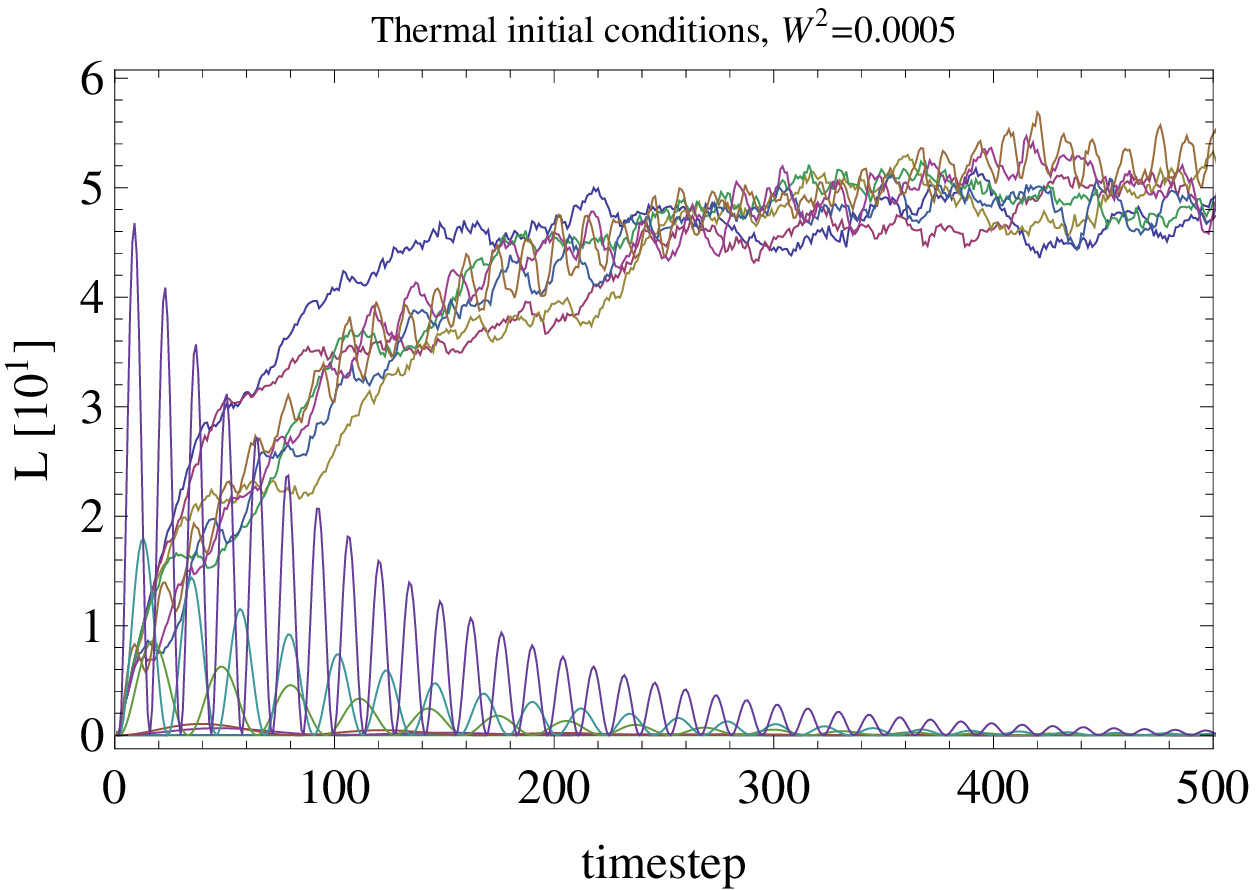} %
\includegraphics[width=8cm,angle=0]{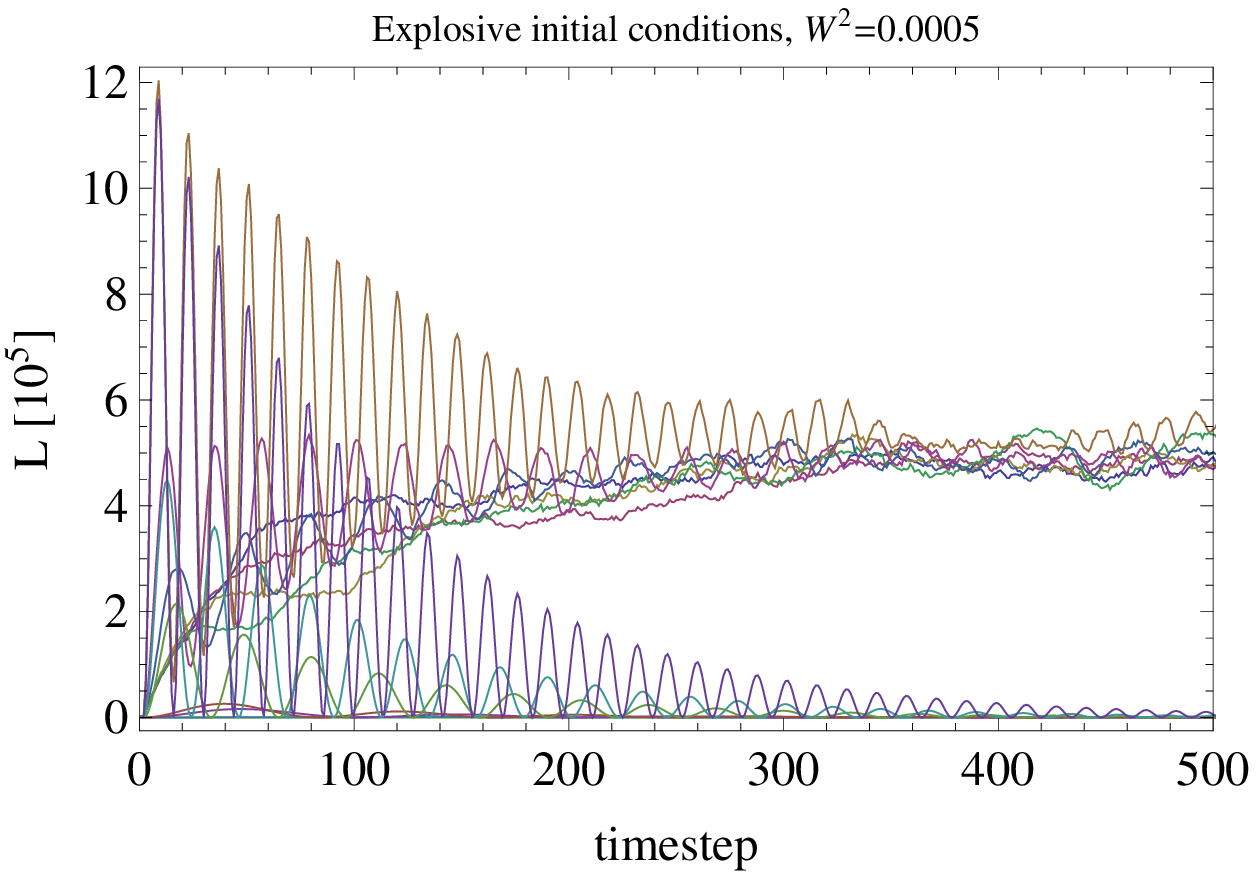}
\end{center}
\caption{Dimensionless electromagnetic power $L$ emitted by a charged particle in stochastic motion described by the generalized Langevin equation with exponential friction kernel, with $W^2 = 0.0005$. The terms "thermal" and "explosive" refer to initial conditions of  $V_0 = 1$,  $\bar{C}=10$ and $V_0=50$,  $\bar{C}=100$ respectively; the variable parameter is $\bar{\alpha} \in \{ 0.1,1,10,15,100,200,500 \}$; the amplitude is an increasing function of $\bar{\alpha}$. Both noise and noiseless cases included. For presentation purposes, the light curves for the case with noise were multiplied by $10^{-1}$ (thermal).}
\label{fig:C-L-f}
\end{figure*}

The dimensionless displacement and velocities of the stochastic motion with memory of the charged particle are presented in Figs.~\ref{fig:C-q} and \ref{fig:C-V}. The electromagnetic power emitted by the particle is shown in Figs.~\ref{fig:C-L-l}-\ref{fig:C-L-f}; \textbf{each of the figures also includes the noiseless counterpart.}

For the case of a charged particle moving in an external potential, while undergoing friction with memory, we expect LCs with a sinusoidal-like appearance while the amplitude is decreasing in time; also, the means and period depend on the $W^2$ parameter. When the friction is constant, we see that for $W^2 <1$ the emission is friction dominated, while for $W^2>1$ the emission is oscillation dominated. If $W^2$ is kept fixed and the friction kernel amplitude $\bar{\alpha}$ is varied, it is seen (Fig.~\ref{fig:C-L-f} and Table~\ref{tab:statC-III}) that the mean value of the LC is an increasing function of $\bar{\alpha}$. This might be a consequence of the fact that a larger memory kernel means that more energy is stored in the system.

When the noise is turned on, the overall energy injected, and thus emitted by the charged particle is increased. When $W^2$ is kept constant, there is a point in parameter space in which the system switches from friction dominated to noise dominated, as can be seen in Fig.~\ref{fig:C-L-f} right, upper group of LCs.

\subsubsection{Statistical analysis of results}

Tables~\ref{tab:statC} and~\ref{tab:statC-III} contain the analysis of the statistical characteristics for the case of a charged Brownian particle, with a memory friction kernel, in a harmonic potential of dimensionless equivalent frequency $W$. As for the other cases, the mean value of the emitted power is larger in the case of explosive conditions, in this case by two orders of magnitude; it varies, within the same ICs, by approximately $3\%$ for the thermal case and by approximately $4\%$ for the explosive case. The same comment can be made about the dispersion, The novelty here is that the skewness and kurtosis vary very little with the change of ICs; they significantly depart from their corresponding Gaussian values.

\begin{figure*}[tbp]
\begin{center}
\includegraphics[width=8cm,angle=0]{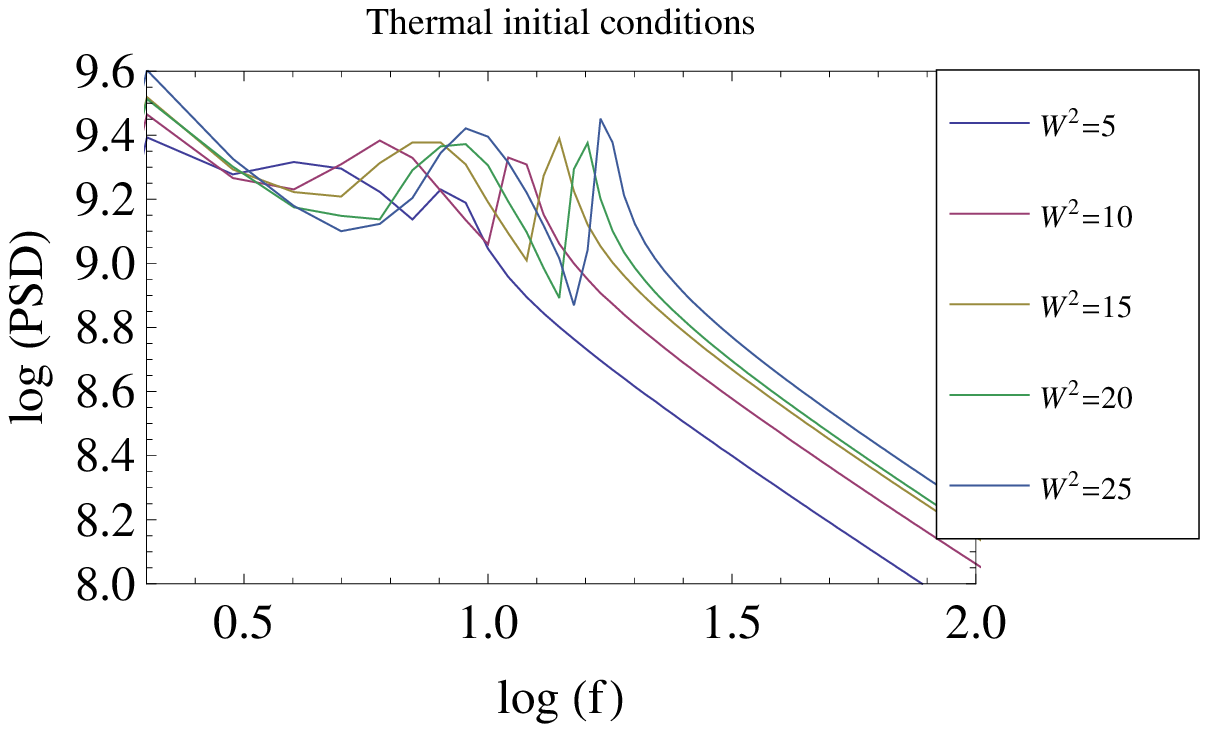} %
\includegraphics[width=8cm,angle=0]{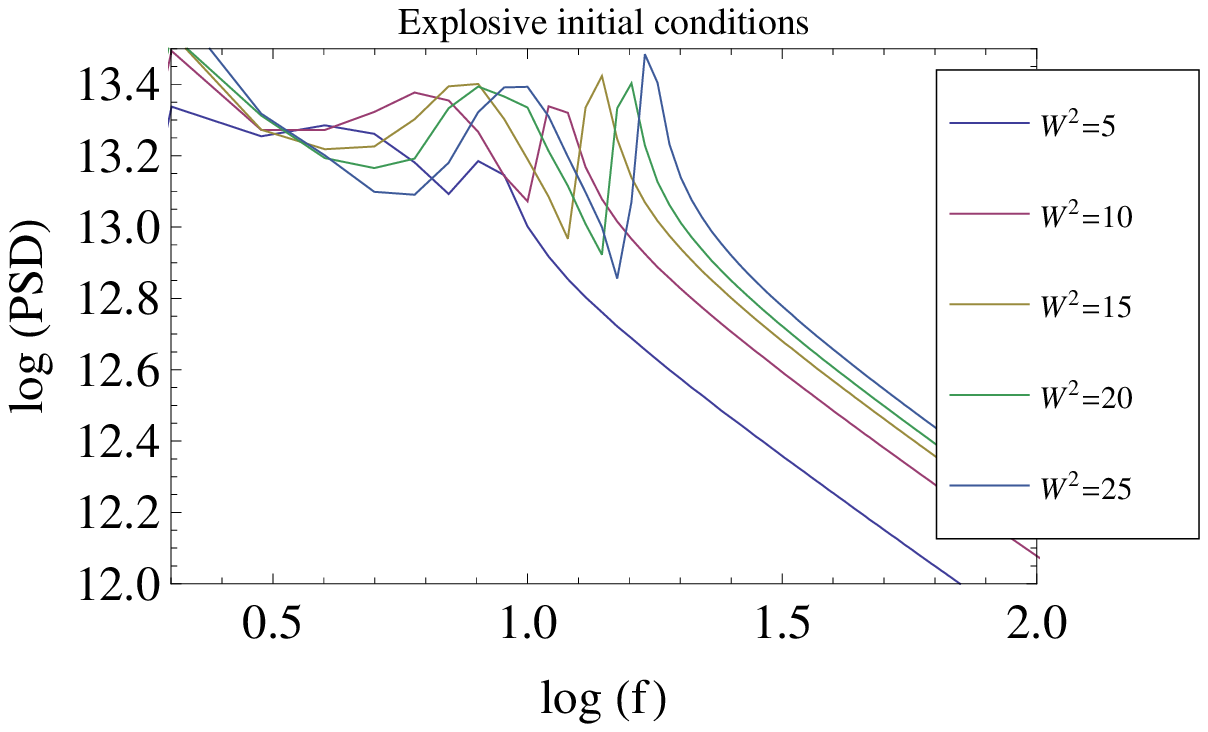}
\end{center}
\caption{Log-Log representation of the PSD of a charged particle in stochastic motion described by the generalized Langevin equation with exponential friction kernel, with $\bar{\alpha} = 0.5$. The terms "thermal" and "explosive" refer to initial conditions of  $V_0 = 1$,  $\bar{C}=10$ and $V_0=50$,  $\bar{C}=100$ respectively.}
\label{fig:C-PSD-II}
\end{figure*}

\begin{figure*}[tbp]
\begin{center}
\includegraphics[width=8cm,angle=0]{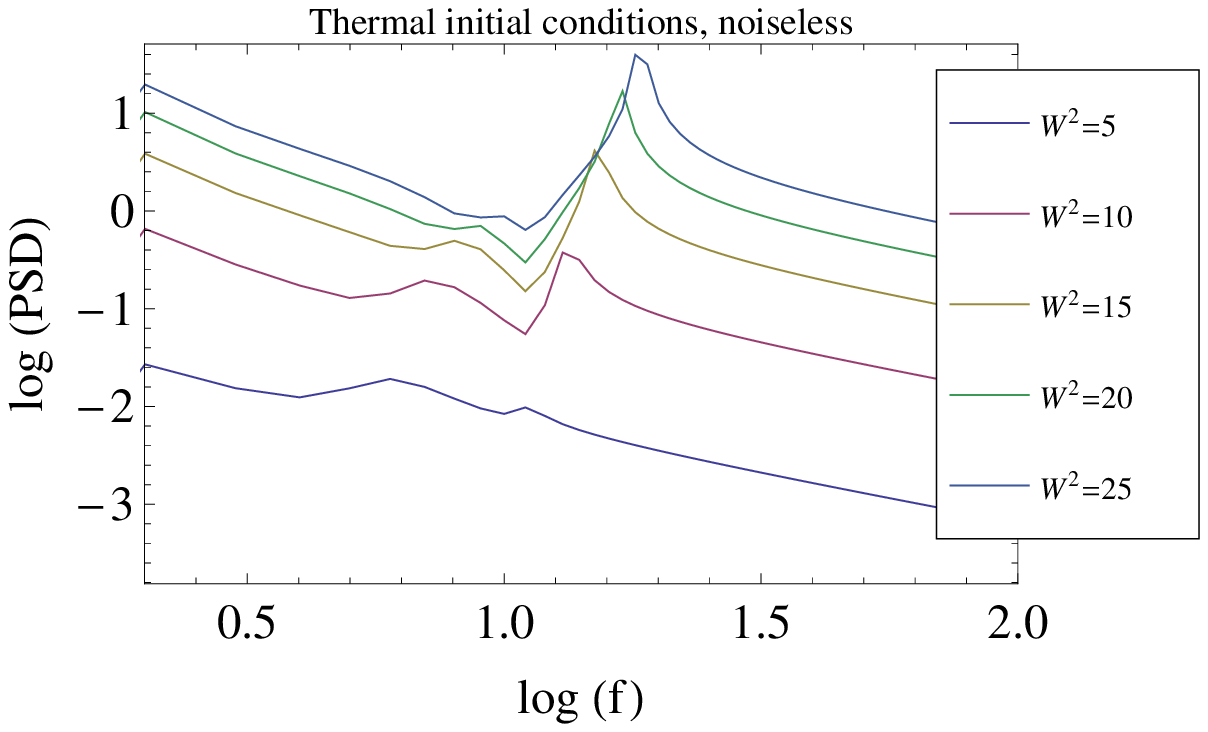} %
\includegraphics[width=8cm,angle=0]{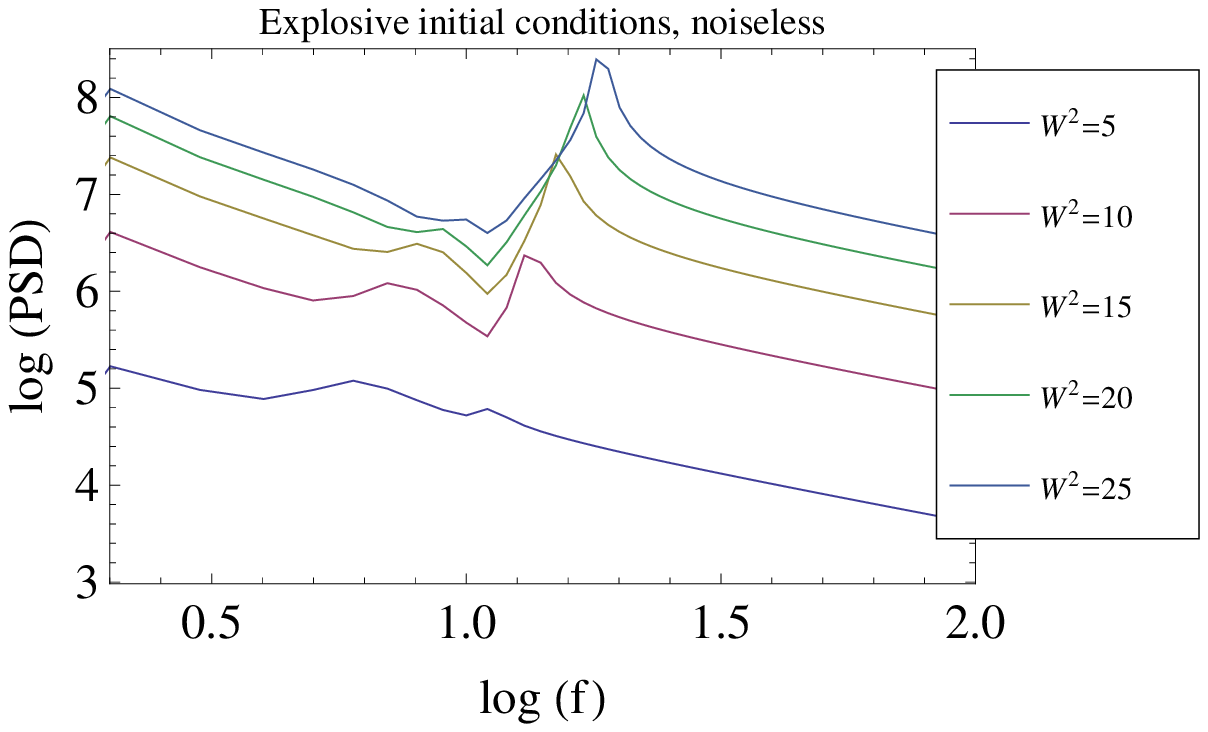}
\end{center}
\caption{Log-Log representation of the PSD of a charged particle in motion described by the generalized Langevin equation with exponential friction kernel, with $\bar{\alpha} = 0.5$. The terms "thermal" and "explosive" refer to initial conditions of  $V_0 = 1$,  $\bar{C}=10$ and $V_0=50$,  $\bar{C}=100$ respectively.}
\label{fig:C-PSD-II-nn}
\end{figure*}

\begin{figure*}[tbp]
\begin{center}
\includegraphics[width=8cm,angle=0]{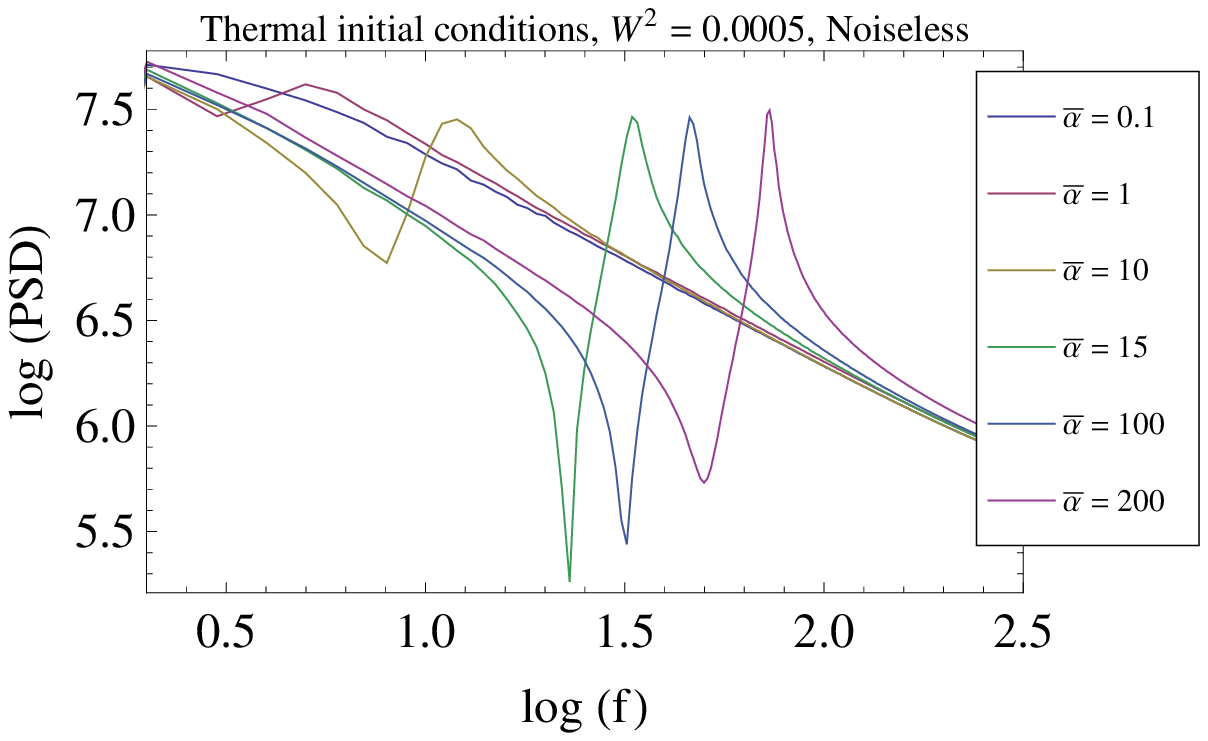} %
\includegraphics[width=8cm,angle=0]{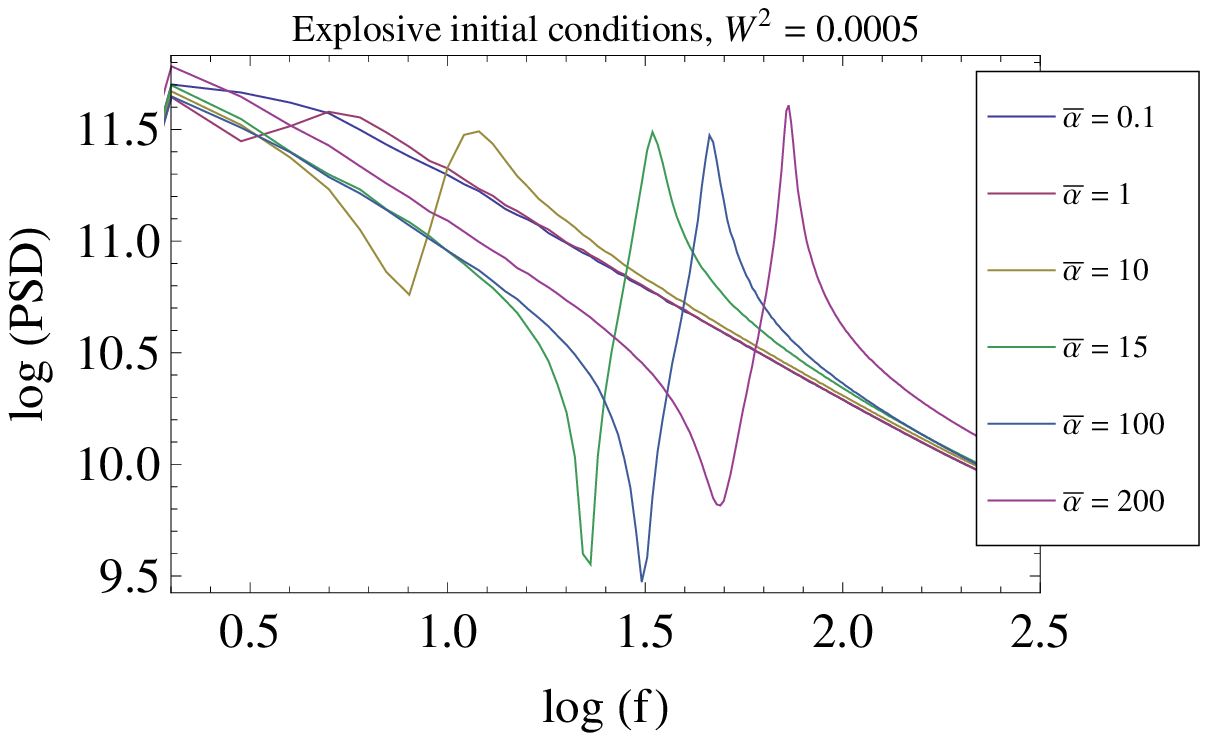}
\end{center}
\caption{Log-Log representation of the PSD of a charged particle in stochastic motion described by the generalized Langevin equation with variable exponential friction kernel $\bar{\alpha}$ and $W^2 = 0.0005$; left: initial conditions of  $V_0 = 1$,  $\bar{C}=10$; right: initial conditions of  $V_0 = 50$,  $\bar{C}=100$.}
\label{fig:C-PSD-III}
\end{figure*}

\begin{figure*}[tbp]
\begin{center}
\includegraphics[width=8cm,angle=0]{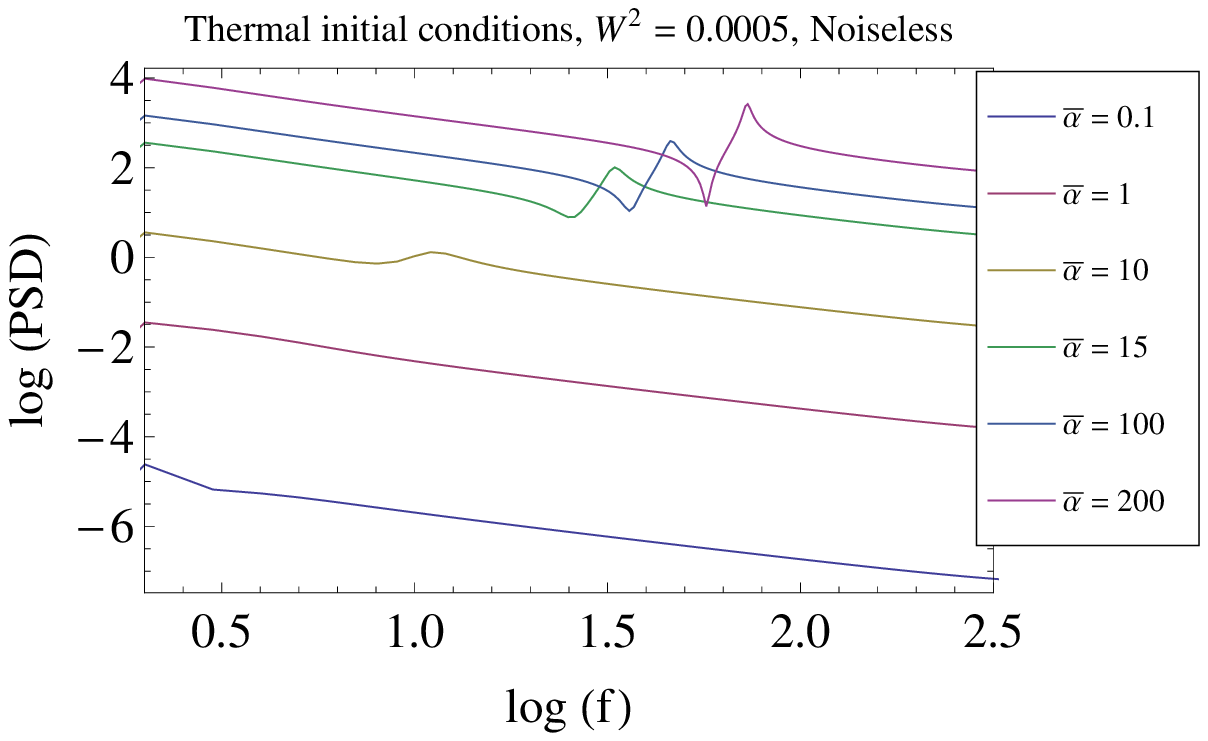} %
\includegraphics[width=8cm,angle=0]{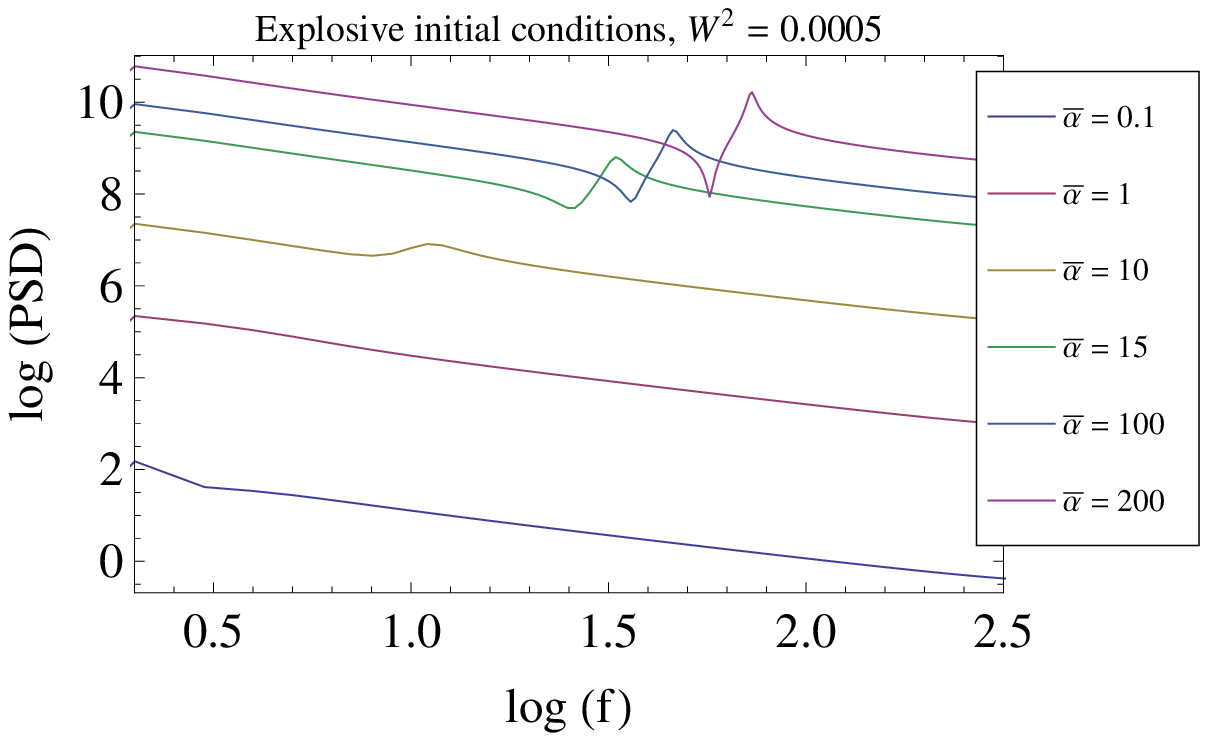}
\end{center}
\caption{Log-Log representation of the PSD of a charged particle in motion described by the generalized Langevin equation with variable exponential friction kernel $\bar{\alpha}$ and $W^2 = 0.0005$; left: initial conditions of  $V_0 = 1$,  $\bar{C}=10$; right: initial conditions of  $V_0 = 50$,  $\bar{C}=100$.}
\label{fig:C-PSD-III-nn}
\end{figure*}

The very interesting consequence is seen clearer in the PSD: since for $W^2 =0.0005$ the value of $\log f =-0.44$, than the new feature appearing in the PSD (Fig.~\ref{fig:C-PSD-III-nn}) is a signature of the memory existent in the system, due to the friction kernel. The position of the peak shifts as $\bar{\alpha}$ increases, thus as the amplitude of the memory increases, the motion becomes correlated over longer timespans. When the noise is turned on, the PSDs for $W^2>1$ (Fig.~\ref{fig:C-PSD-II} compared with Fig.~\ref{fig:C-PSD-II-nn}) are adequate to fully see the effects of the various influences present in the system; for this set of $W^2$, the value of $\log f \in {1.55,1.90}$, which cannot be seen as such in either of the plots. However, there is a feature for smaller values of $\log f$, and the position of this feature moves to the right as $W^2$ increases. When $W^2$ is constant and the friction varies (Fig.~\ref{fig:C-PSD-III}) the position of the additional peak stays roughly the same with respect to the noiseless case, however the height of the peak becomes larger. We infer that this happens because, due to the memory of the friction, the system is not efficient in instant radiation of its energy, and a correlation appears, overwriting the expected periodicity and producing a new one.

\begin{figure*}[tbp]
\begin{center}
\includegraphics[width=8cm,angle=0]{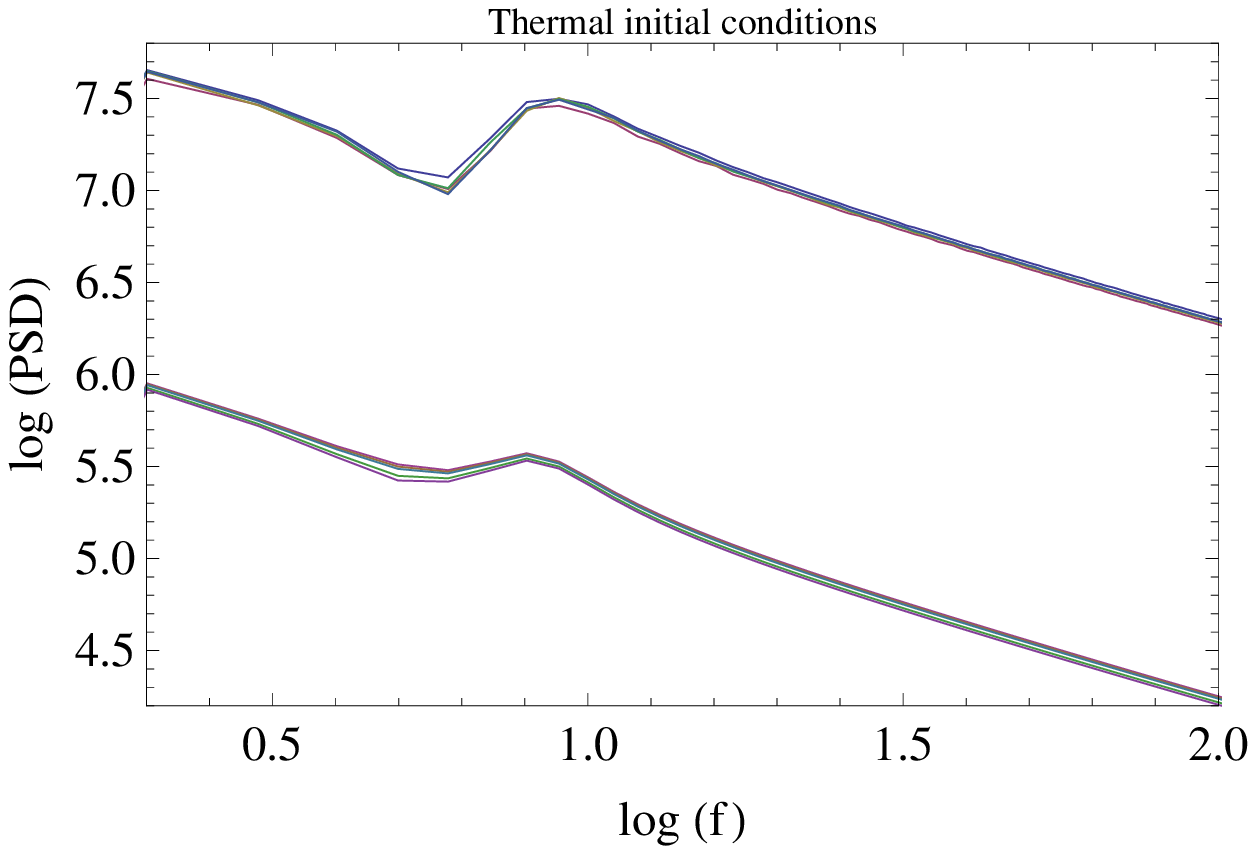} %
\includegraphics[width=8cm,angle=0]{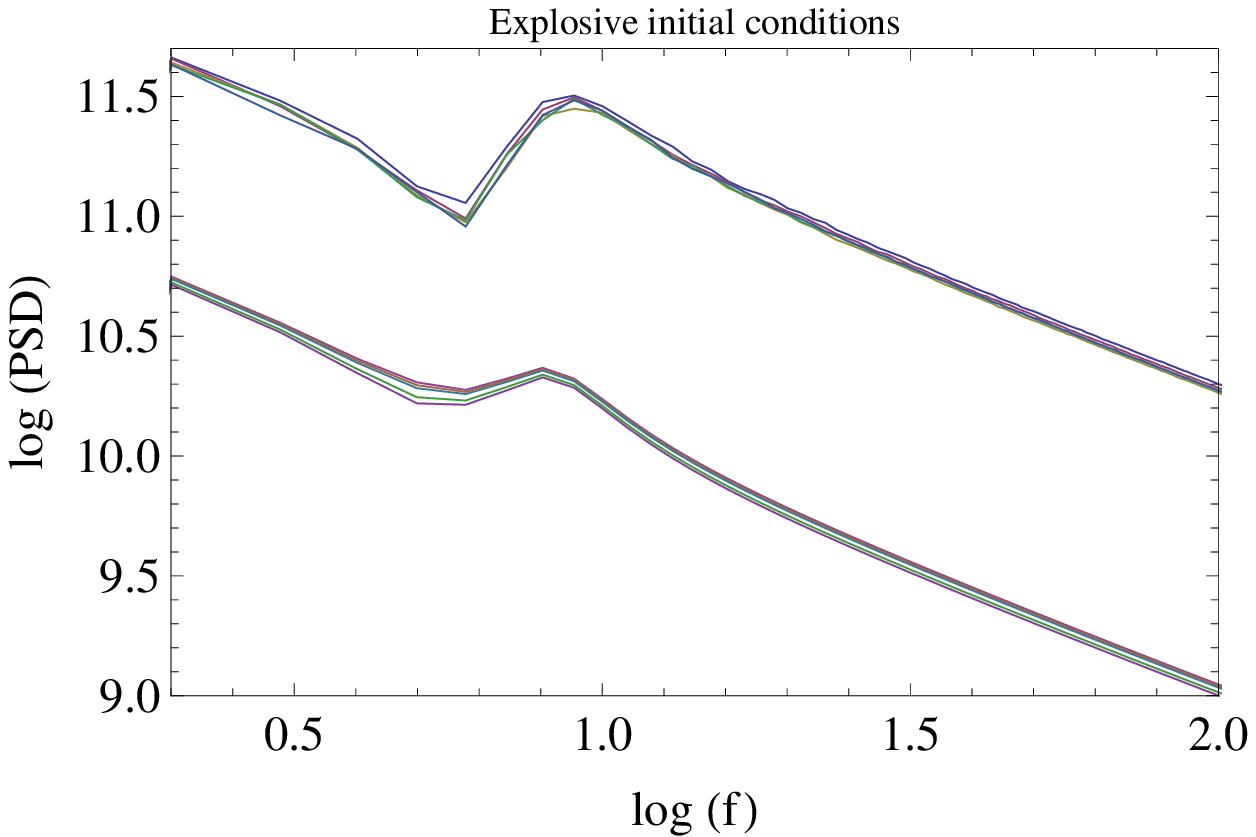}
\end{center}
\caption{Log-Log representation of the PSD of a charged particle in stochastic motion described by the generalized Langevin equation with exponential friction kernel, with $\bar{\alpha} = 5$. The terms "thermal" and "explosive" refer to initial conditions of  $V_0 = 1$,  $\bar{C}=10$ and $V_0=50$,  $\bar{C}=100$ respectively; the variable parameter is $W^2 \in \{ 0.0005, 0.01, 0.02, 0.05, 0.07 \}$. Both noise and noiseless cases included; for presentation purposes, the noiseless group has been moved up the vertical axis by 6 units (thermal) and 3 units (explosive).}
\label{fig:C-PSD-I}
\end{figure*}

\subsection{Radiation pattern from stochastic particle motion in a constant
magnetic field\label{sect:D}}

\subsubsection{Equations and physics}

The equation of motion of a charged particle in a magnetic field in the presence of a stochastic force  $\vec{\xi} ^D(t)$ and of interparticle collisions, generating a force proportional to the particle velocity, is given by the Langevin type equation \cite{b4}
\begin{equation}\label{l4}
\frac{d\vec{v}}{dt} = \frac{Ze}{mc}\left [ \vec{v}(t)\times \vec{B}\right ]
- \nu \vec{v}(t) + \vec{\xi} ^D(t)
\end{equation}
where
\begin{equation}
\left \langle \xi ^D_i \left(t\right)\xi ^D _j \left(t^{\prime
}\right)\right \rangle = \frac{D}{dt} \delta _{ij} \left(t-t^{\prime
}\right).
\end{equation}
where $i,j=x,y$ and, the  constant magnetic field is oriented along the $z$ direction,
\begin{equation}
\vec{B} = B\hat{z},B={\rm constant}.
\end{equation}

\subsubsection{Numerical approach and simulated light curves}

In addition to the other dimensionless quantities discussed in Sections~\ref{sect:Anumerical}, we define a
\begin{itemize}
\item dimensionless magnetic frequency: $\bar{\Omega} = \Omega \tau = B_0%
\frac{Ze}{mc}\tau$.
\end{itemize}

For the case of stochastic motion in a constant magnetic field described by Eqs.~(\ref{l4}), the equation is split into components and afterwards
made dimensionless as
\begin{equation}
\frac{dX}{d\theta} = V_x; \quad \frac{dV_x}{d\theta}=\bar{\Omega}V_y - V_x + \bar{\xi}_x (\theta),\label{eq:DX}
\end{equation}
\begin{equation}
\frac{dY}{d\theta} = V_y; \quad\frac{dV_y}{d\theta}=-\bar{\Omega}V_x - V_y + \bar{\xi}_y (\theta),\label{eq:DY}
\end{equation}
\begin{equation}
\frac{dZ}{d\theta} = V_z; \quad\frac{dV_z}{d\theta}=- V_z + \bar{\xi}_z (\theta),
\end{equation}
where
\begin{equation}
\left \langle \bar{\xi} _i \left(\theta \right)\bar{\xi} _j \left(\theta
^{\prime }\right)\right \rangle =\frac{1}{d\theta}\delta _{ij} \left(\theta
- \theta ^{\prime }\right).
\end{equation}

In the case of the motion in a deterministic magnetic field, with $\bar{\xi}_x (\theta)=\bar{\xi}_y (\theta)\equiv 0$, the solution of the equation of motion of the charged particle in the constant magnetic field is given by
\be
V_x(\theta)=e^{-\theta }\left[ V_x(0)\cos \left(\bar{\Omega }\theta
   \right)+V_y(0) \sin \left(\bar {\Omega} \theta \right)\right],\label{eq:dvx}
   \ee
   \be
V_y(\theta)=e^{-\theta}\left[
   V_y(0) \cos \left(\bar{\Omega} \theta \right)-V_x(0) \sin \left(\bar{\Omega }\theta \right)\right],\label{eq:dvy}
\ee
\bea
X(\theta)&=&X(0)+\frac{e^{-\theta }}{\bar{\Omega }^2+1}\times \nonumber\\
&& \Bigg\{\left[V_x(0) \bar{\Omega }-V_y(0)\right]\sin \left(\bar{\Omega}\theta \right) -
   \Bigg[V_x(0)+\nonumber\\
 &&  V_y(0) \bar{\Omega }\Bigg]\cos \left(\bar{\Omega }\theta \right)\Bigg\},
\eea
\bea
Y(\theta)&=&Y(0)+\frac{e^{-t}}{\bar{\Omega }^2+1} \Bigg\{\left[V_x(0)+V_y(0)\bar{\Omega }\right]\sin \left(\bar{\Omega}\theta  \right) +\nonumber\\
&&\left[V_x(0) \bar{\Omega }-V_y(0)\right]\cos \left( \bar{\Omega}\theta \right)
   \Bigg\}.
\eea

The solution $\{ X,Y,V_x,V_y \}$ to Eqs.~\eqref{eq:DX}-\eqref{eq:DY} in the presence of the stochastic noise was obtained by employing the method of~\cite{b4}. More precisely, we used their equations (16) suitably adapted for the dimensionless case. For $\{Z,V_z \}$ an Euler scheme as the one described in Section~\ref{sect:Anumerical} was used.

\begin{figure*}[tbp]
\begin{center}
\includegraphics[width=7.5cm,angle=0]{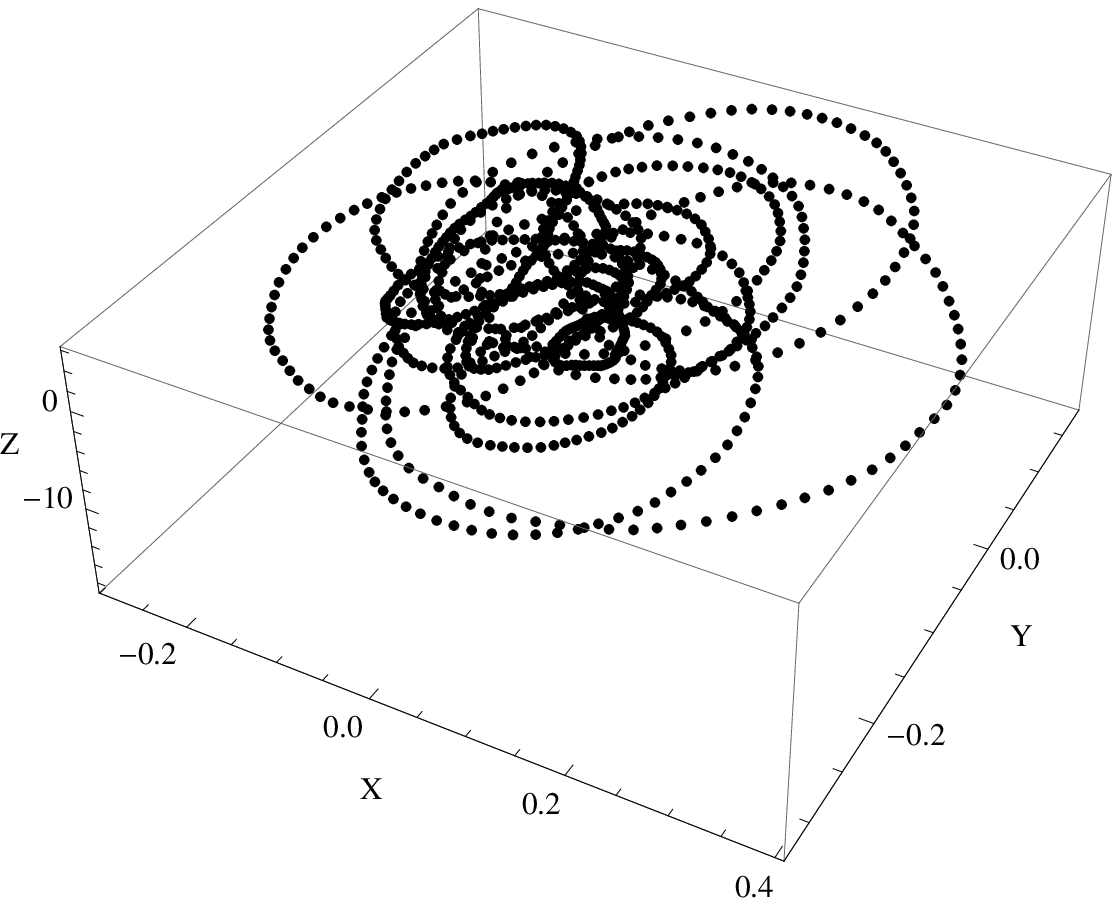}
\includegraphics[width=7.5cm,angle=0]{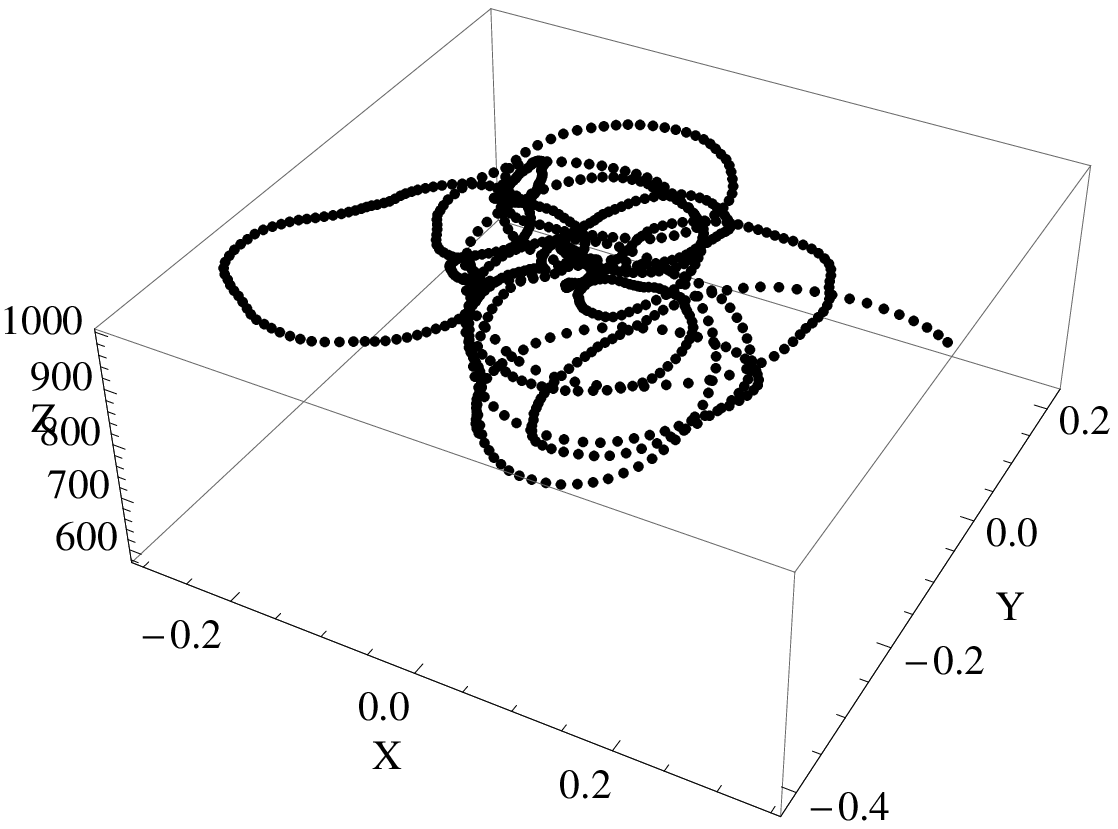}
\end{center}
\caption{Displacement of a charged particle in Brownian motion in a constant magnetic field with dimensionless Larmor frequency $\bar{\Omega} = 10$, for
thermal initial conditions $V_{0x}=V_{0y}=V_{0z}=0.2$ (left figure), and for
explosive initial conditions, $V_{0z}=10^3$, $V_{0x}=V_{0y}=0$ (right figure).}
\label{fig:D-Q}
\end{figure*}

\begin{figure*}[tbp]
\begin{center}
\includegraphics[width=7.5cm,angle=0]{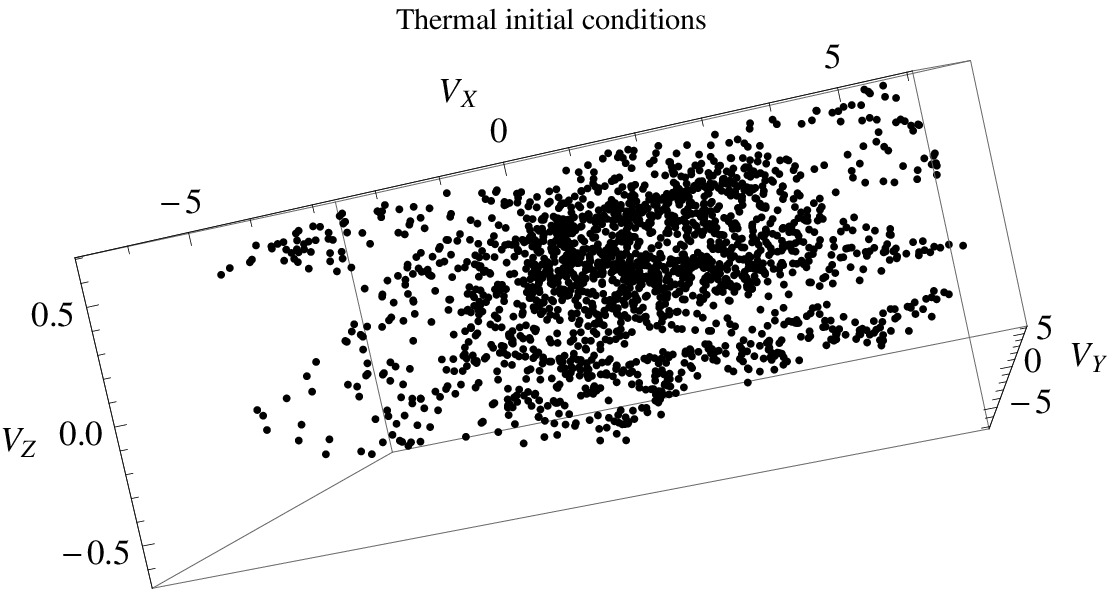}
\includegraphics[width=7.5cm,angle=0]{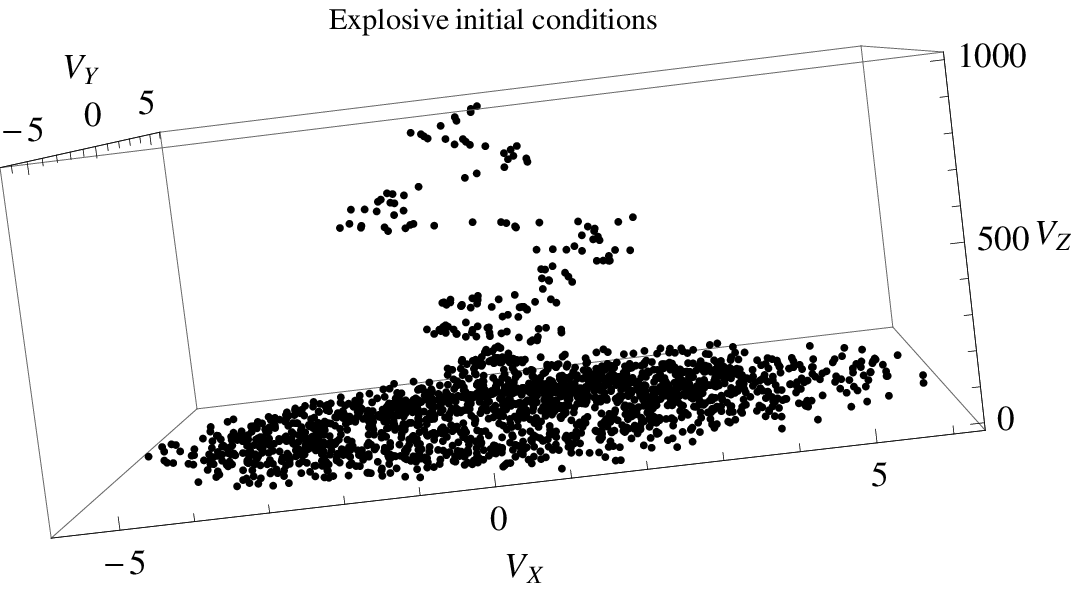}
\end{center}
\caption{Velocity of a charged particle in Brownian motion in a magnetic field with dimensionless frequency $\bar{\Omega} = 10$, for
thermal initial conditions $V_{0x}=V_{0y}=V_{0z}=0.2$ (left figure), and for
explosive initial conditions, $V_{0z}=10^3$, $V_{0x}=V_{0y}=0$ (right figure).}
\label{fig:D-V}
\end{figure*}

\begin{figure*}[tbp]
\begin{center}
\includegraphics[width=8cm,angle=0]{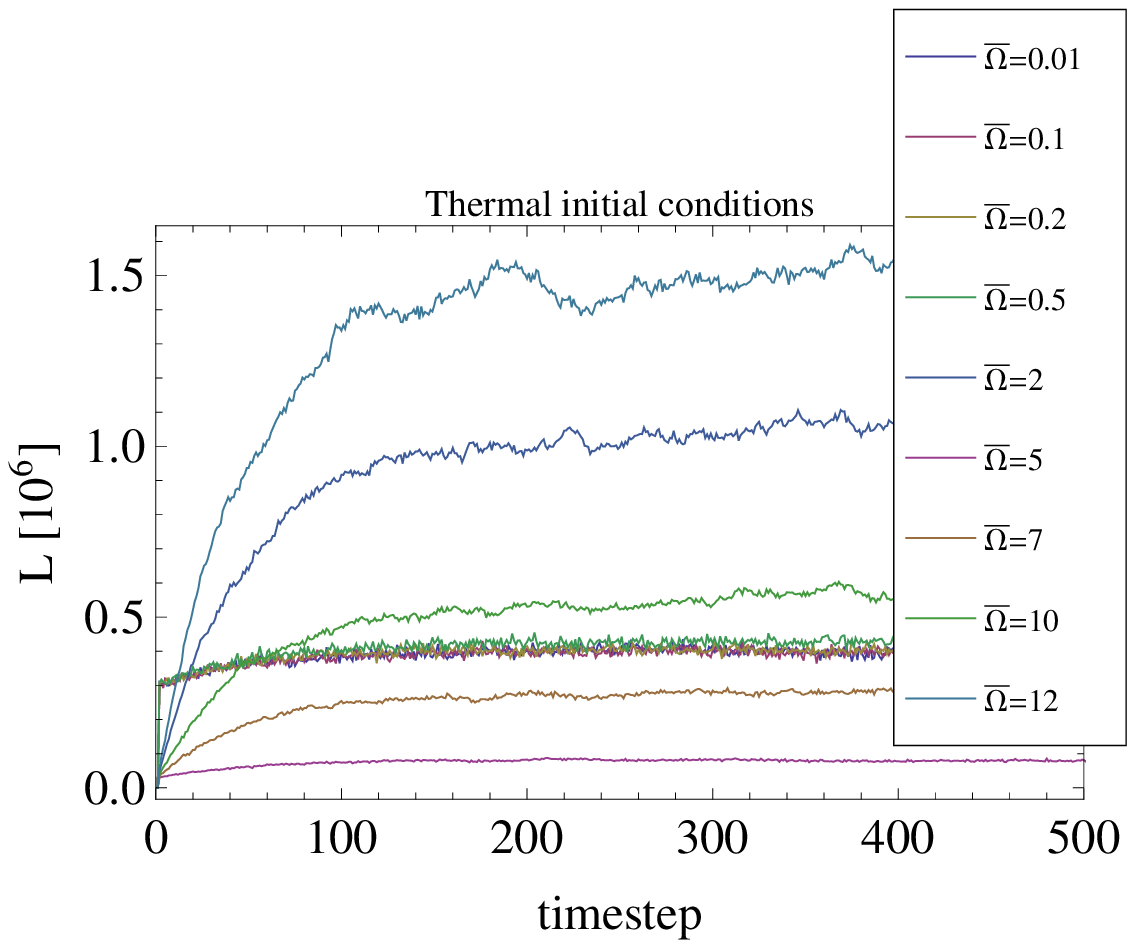} %
\includegraphics[width=8cm,angle=0]{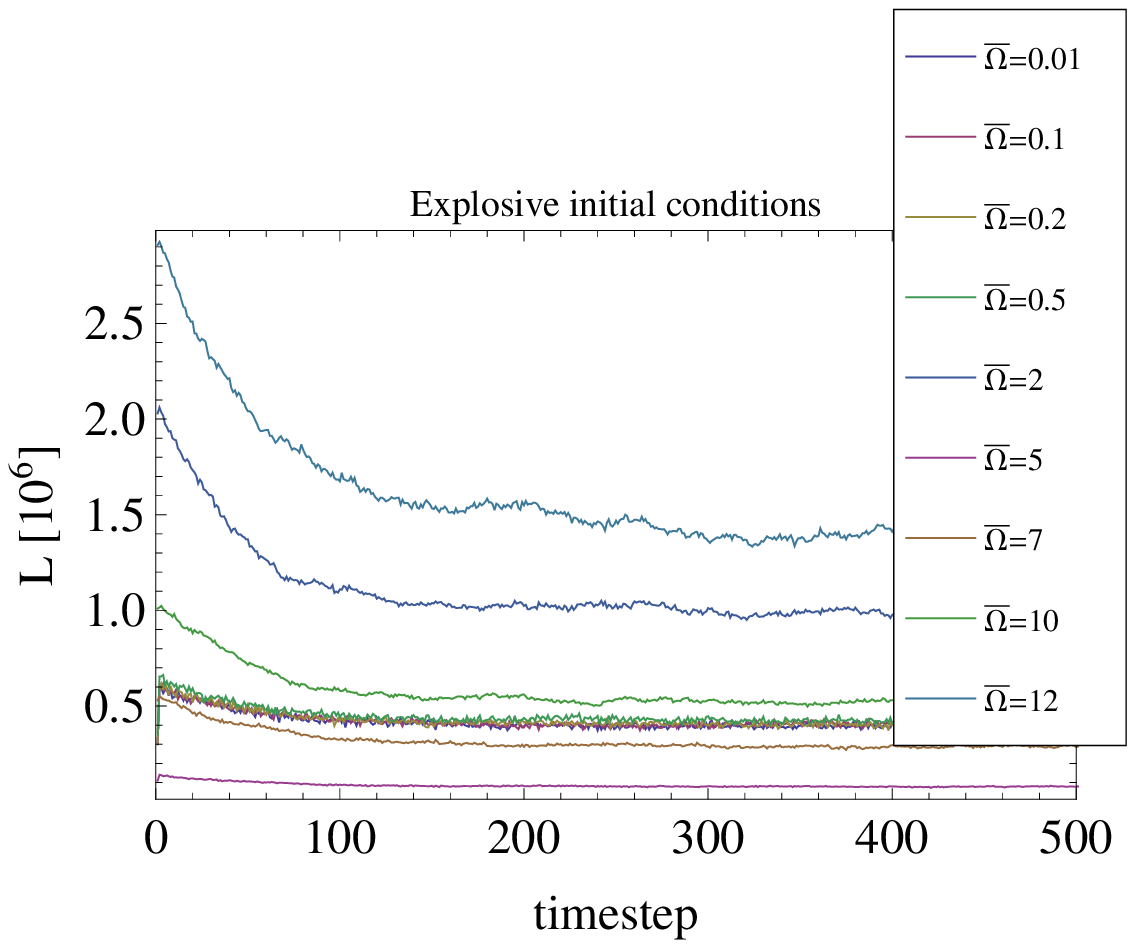}
\end{center}
\caption{Dimensionless power $L$ emitted by a charged particle in Brownian motion in a constant magnetic field. The terms "thermal" and "explosive" refer to initial conditions  $V_{0x} = V_{0y} = V_{0z} = 0.2$, and $V_{0x} = V_{0y} = V_{0z} = 100$, respectively. For presentation purposes, the curves corresponding to $\bar{\Omega} >1$ were multiplied by $10$.}
\label{fig:D-L}
\end{figure*}

\begin{figure*}[tbp]
\begin{center}
\includegraphics[width=8cm,angle=0]{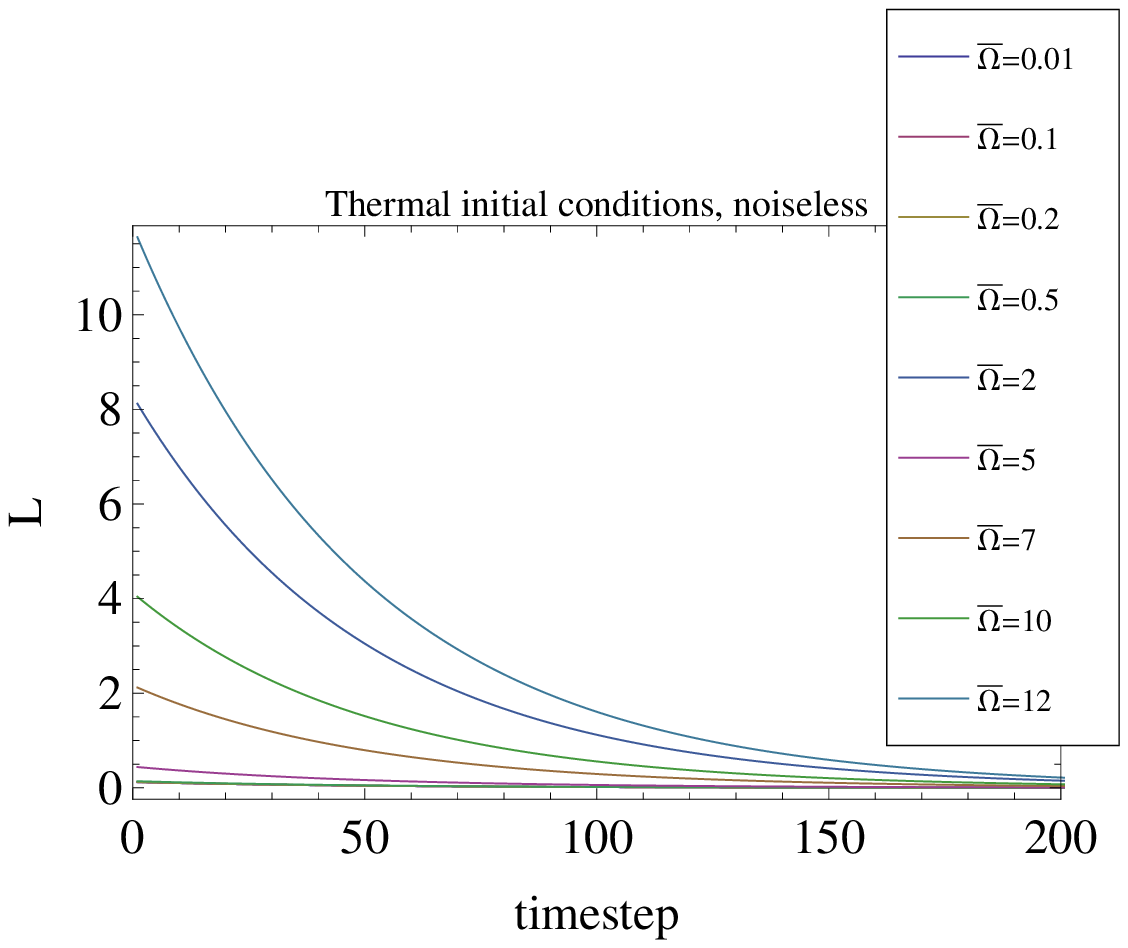} %
\includegraphics[width=8cm,angle=0]{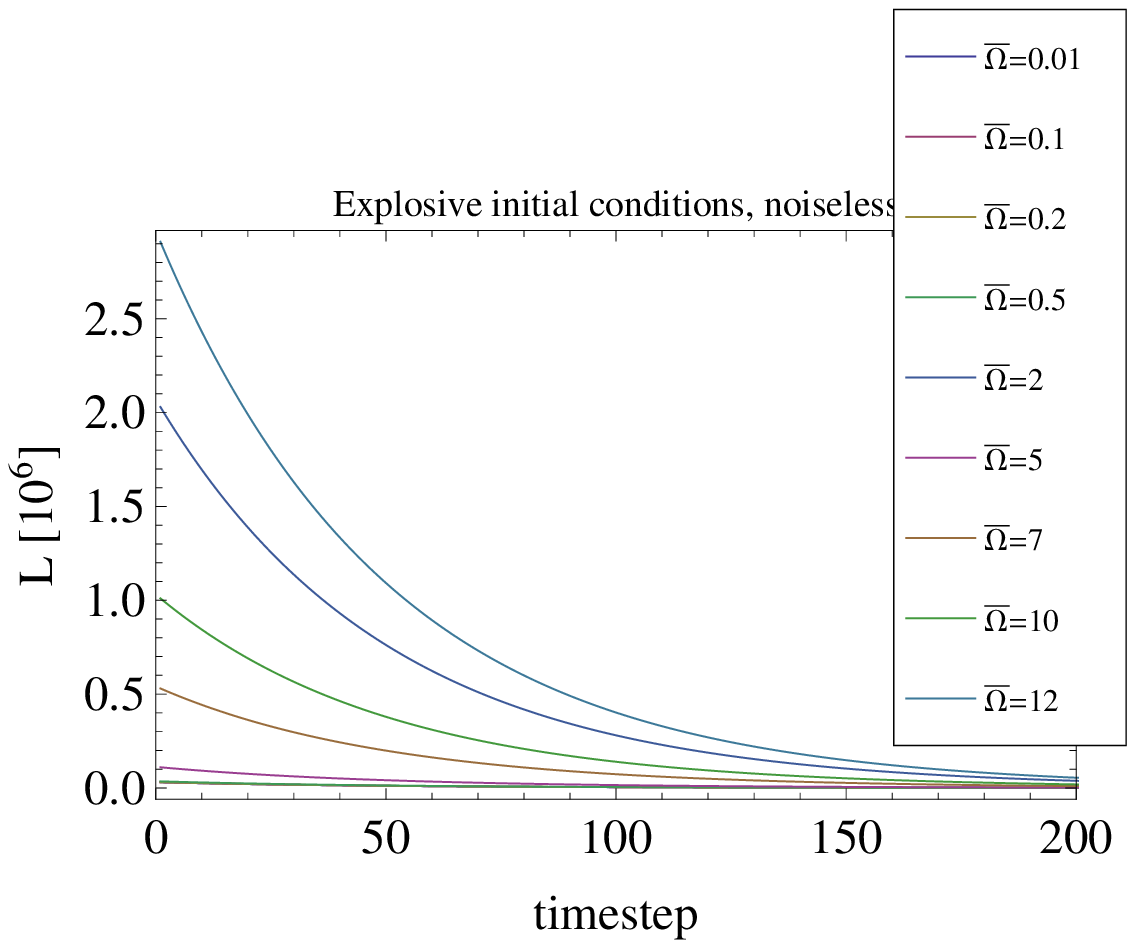}
\end{center}
\caption{Dimensionless power $L$ emitted by a charged particle in motion in a constant magnetic field. The terms "thermal" and "explosive" refer to initial conditions  $V_{0x} = V_{0y} = V_{0z} = 0.2$, and $V_{0x} = V_{0y} = V_{0z} = 100$, respectively.}
\label{fig:D-L-nn}
\end{figure*}

The acceleration producing the radiation
pattern was obtained as
\begin{equation}
a_{d,n}^2 = a_{x,n}^2 + a_{n,y}^2 + a_{z,n}^2,
\end{equation}
where
\begin{equation}
a_{x,n+1} = a_{x,n} + h (\bar{\Omega}V_{y,n} - V_{x,n} + \psi _x),
\end{equation}
\begin{equation}
a_{y,n+1}=a_{y,n}+h(-\bar{\Omega}V_{x,n} - V_{y,n} + \psi_y),
\end{equation}
\begin{equation}
a_{z,n+1}=a_{z,n}+h (- V_{z,n} + \psi_z),
\end{equation}
with $\psi _j \in \mathcal{N} (0, h^{-1})$ at each timestep. The displacement of the charged particle in stochastic motion is presented in Fig.~\ref{fig:D-Q}, while the 3D velocity is shown in Fig.~\ref{fig:D-V}. The power emitted during the random motion in the constant magnetic field is depicted in Fig.~\ref{fig:D-L} and noiseless counterpart is shown in Fig.~\ref{fig:D-L-nn}.


Although the motion of the particle is periodic, the observational signature will not exhibit periodicity. A simple calculation for the noiseless case (Eqs.~\eqref{eq:dvx}-\eqref{eq:dvy}) shows that the contribution to the LC from the $x$ and $y$ directions, $a_x^2 + a_y^2 = e^{-2\theta}(1+\bar{\Omega}^2)$ is obviously not periodic. So there is a very clear difference between the observational signature of a charged particle moving in an external harmonic potential (previous two cases) and that of a charged particle moving in a constant magnetic field.

The motion of a charged particle in a constant magnetic field while undergoing friction is expected to produce a light curve with an intensity which is decaying in time, with mean value depending on the energy injected in the system, through the IC and the value of the magnetic field. This is indeed recovered in Fig.~\ref{fig:D-L-nn} and column 3 of Table~\ref{tab:statD}.  When a noise component is turned on, the energy level is generally enhanced and thus in this parameter space it seems that the energy radiated from charged particles generally comes from the random kicks these particles are subjected to.

\subsubsection{Statistical analysis of results}

Table~\ref{tab:statD} contains the analysis of the statistical characteristics for the case of a charged Brownian particle undergoing friction in a magnetic field. Even in the case of thermal injection, there is a dramatic increase in the mean power output with little variation of the dimensionless Larmor frequency $\bar{\Omega}$. In the explored parameter space, for fixed magnetic energy content, the power output does not significantly change even if the emission occurs following and explosion. This insensitivity on ICs for fixed $\bar{\Omega}$ can also be seen for the dispersion and for the kurtosis. The skewness, however, changes sign, although generally keeping the same absolute value to within a few percent.

\begin{figure*}[tbp]
\begin{center}
\includegraphics[width=8cm,angle=0]{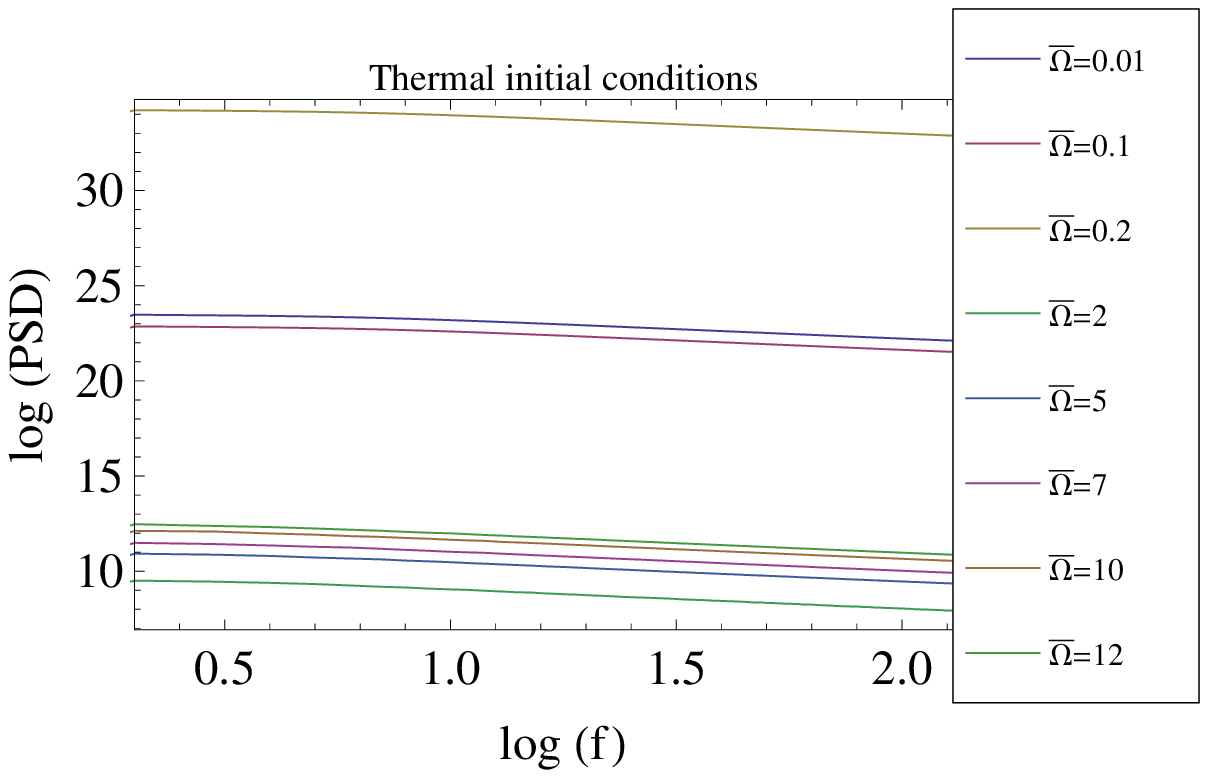} %
\includegraphics[width=8cm,angle=0]{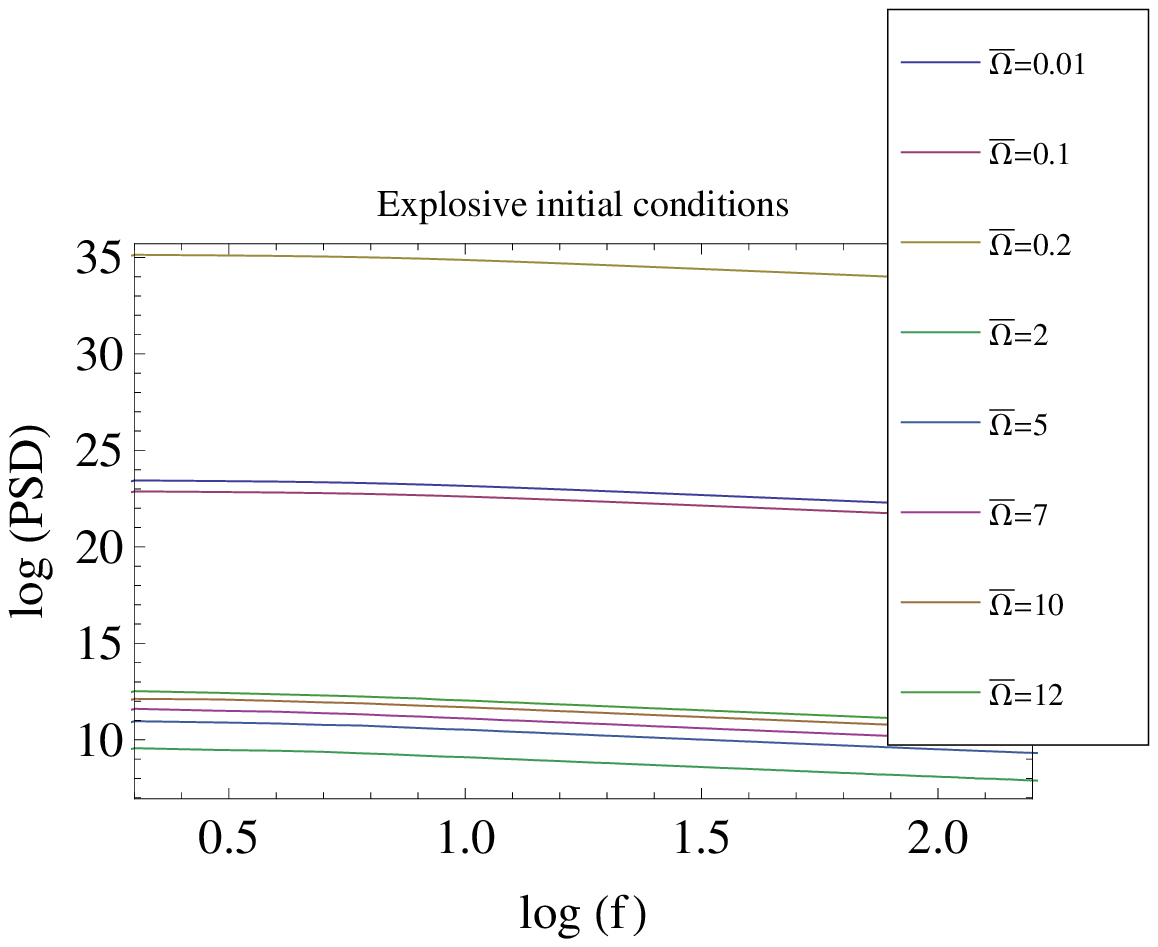}
\end{center}
\caption{Log-Log representation of the PSD of the electromagnetic power emitted by a charged particle in Brownian motion in a constant magnetic field. The terms "thermal" and "explosive" refer to initial conditions of $V_{0x} = V_{0y} = V_{0z} = 0.2$ and $V_{0x} = V_{0y} = V_{0z} = 100$ respectively.}
\label{fig:D-PSD}
\end{figure*}

\begin{figure*}[tbp]
\begin{center}
\includegraphics[width=8cm,angle=0]{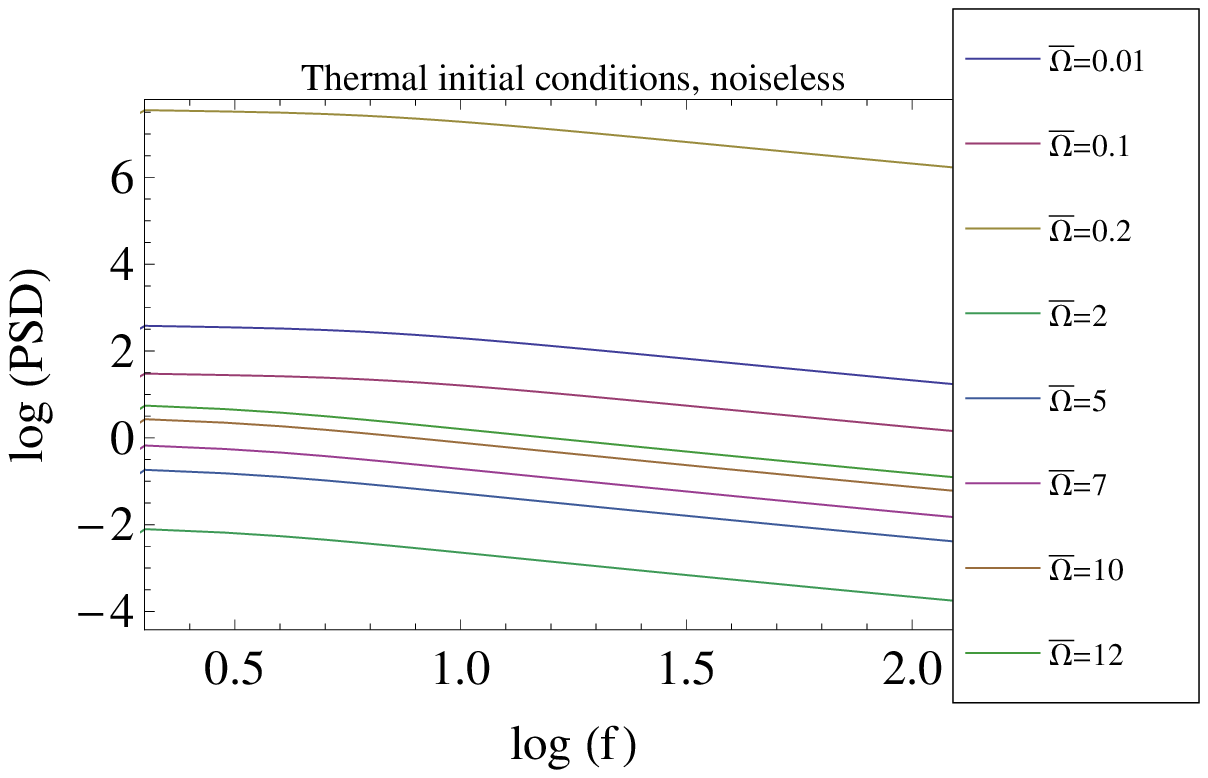} %
\includegraphics[width=8cm,angle=0]{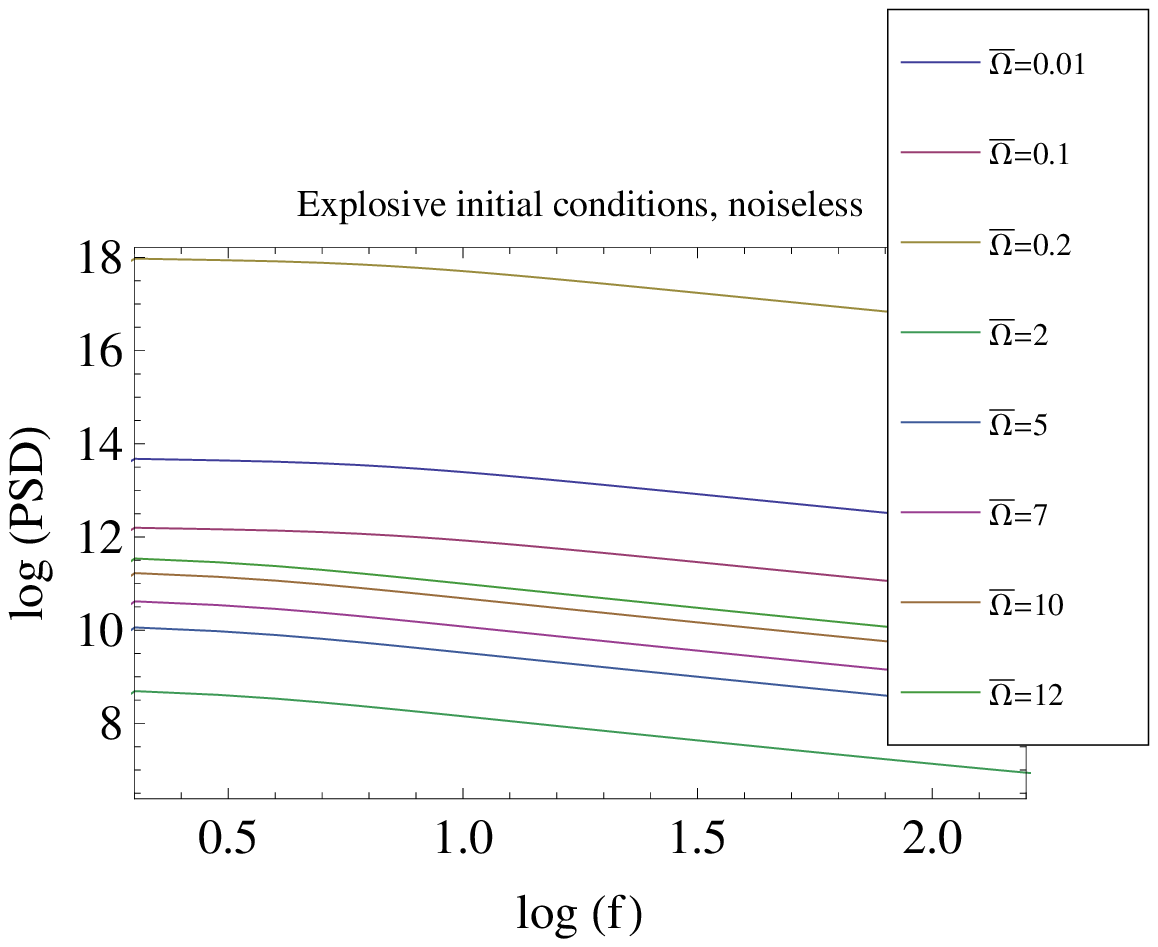}
\end{center}
\caption{Log-Log representation of the PSD of the electromagnetic power emitted by a charged particle in motion in a constant magnetic field. The terms "thermal" and "explosive" refer to initial conditions of $V_{0x} = V_{0y} = V_{0z} = 0.2$ and $V_{0x} = V_{0y} = V_{0z} = 100$ respectively.}
\label{fig:D-PSD-nn}
\end{figure*}

Since, as argued above, no periodicity is expected in the LC, the PSD for the noiseless case is featureless (Fig.~\ref{fig:D-PSD-nn}). When the noise is turned on, the PSD is still generally smooth and the effect of the noise is seen in the overall increase of power at all frequencies (Fig.~\ref{fig:D-PSD}). The numerical value of $\bar{\Omega}$ influences the spectral power, i.e., the relative influence of the power for a fixed frequency grows with growing $\bar{\Omega}$.

\section{Discussions and final remarks}\label{sect:C}

In the present paper we have considered the radiation properties of charged particles in Brownian motion. To model the electromagnetic emissivity properties of the particles we have adopted the Langevin and the generalized Langevin equation, respectively, which give a full description of the particle-external environment system. In order to solve the Langevin equations we have adopted a numerical approach, based on the use of some numerical integrators.  Due to the stochastic, random distribution of the physical parameters of the particles, complex radiation patterns can be generated in the presence of some random forces generated by the particle environment. In particular we have found that for specific oscillation frequencies some peaks are present in the PSD curves of the emitted electromagnetic radiation for charged Brownian particles moving in external harmonic potentials.  The presence of resonant stochastic peaks in the PSD of the energy emission can be used to explain astrophysical observations. For example, in \cite{sr4,sr5} it was shown that the simulated PSD curves of luminosity for stochastically oscillating general relativistic disks \cite{M1}  have the same profile as the observed PSD of black hole X-ray binaries in the low-hard state. Hence resonant effect in the accretion disk oscillations may provide an alternative interpretation of the persistent low-frequency quasi-periodic oscillations in astrophysical systems. The stochastic electromagnetic radiation model, representing an interplay between deterministic and stochastic processes, including resonance phenomena, could also be used to explain strongly peaked astrophysical effects in black hole-binary systems.

Quasi Periodic Oscillations (QPOs) are a commonly occurring phenomenon in astrophysical observations, with many kHz QPOs  detected in the light curves of neutron stars and  black hole sources \cite{QPO1}. Since the orbital frequencies of high frequency QPOs  lie in the range of orbital frequencies
of geodesics just few Schwarzschild radii outside the central massive source \cite{QPO2}. This observation strongly suggest that the QPOs could be related to the orbital motion of particles in an accretion disk \cite{QPO3}. An interesting theoretical model explaining the QPO properties is the Abramowicz-Kluzniak resonance model \cite{QPO4}, which starts from pointing out the importance of the observed 3:2 frequency ratio, and
that the commensurability of frequencies may be a clear signature of the existence of resonance in the astrophysical system. From observational point of view the fact that for kHz QPOs  the frequencies scale with $1/M$, where $M$ is the mass of the central object, provide a strong support to the idea that they
are due to orbital oscillations \cite{QPO2}. By adopting the above approach it follows that from a mathematical point of view QPOs can be modelled by analyzing  the time evolution of perturbed nearby Keplerian geodesics, which can be described by the equations \cite{QPO2}
\be\label{qpo1}
\ddot{z}(t) + \omega _{\theta } z(t) = f \left[\rho (t), z(t), r_0, \theta _0\right],
\ee
\be\label{qpo2}
\ddot{\rho }(t) + \omega _{r } \rho (t) = g \left[\rho (t), z(t), r_0, \theta _0\right],
\ee
where $z(t)$ and $\rho (t)$ denotes the small deviations from the circular orbit $\left(r_0,\theta _0\right)$, $\omega _{\theta }$ and $\omega _r$ are the epicyclic frequencies, while the functions $f$ and $g$ describe the couplings between particle motion and external perturbations. However, the astrophysical model described by Eqs.~(\ref{qpo1}) and (\ref{qpo2}) does not take into account the complexity of the processes taking place in the central regions of accretion disks, like, for example, the Magneto-Rotational Instability induced turbulence. Therefore a {\it stochasticized} version of Eqs.~(\ref{qpo1}) and (\ref{qpo2}) was proposed in \cite{QPO2}, with the evolution of the perturbations described by the stochastic equations
\be\label{qpo3}
\ddot{z}(t) + \omega _{\theta } z(t) - f \left[\rho (t), z(t), r_0, \theta _0\right]=\sigma _z\beta (t),
\ee
\be\label{qpo4}
\ddot{\rho }(t) + \omega _{r } \rho (t) = g \left[\rho (t), z(t), r_0, \theta _0\right],
\ee
where $\sigma _z$ is a constant, while $\beta (t)$ is a continuous Gaussian white-noise process, with zero mean, and unit variance, respectively. In this model the noise term acts only along the vertical direction. By using the above equations it is possible to show that the presence of the stochastic noise can trigger the appearance of resonances in the epicyclic oscillations of nearly Keplerian disks \cite{QPO2}. From a mathematical point of view Eqs.~(\ref{qpo3}) and (\ref{qpo4}) are similar to the oscillation equations considered in the present paper, without the dissipation term included. Hence the present approach may allow the study of more general QPO models, in which the complex physical behavior of the central region of the accretion disks may be modelled by stochastic differential equations involving dissipative and memory effects.

For cases II and III, preliminary calculations show that aside from the qualitative similarity to observational QPO signatures, some of the peaks have a Q factor which is larger than 2, but not much; however, the "industry" of accurate and un-debatable calculation of Q factors from nontrivial, multi-peaks PSD is beyond the purpose of this paper and we reserve it for future work.

We point out that in the parameter space considered in this paper, there was no additional feature in the PSDs calculated for cases where there was no reason to expect periodicity, i.e. the first and last. When there was a periodic behavior embedded in the system, the complex interplay between noise, friction, memory and a harmonic potential lead to the appearance of \textbf{a new feature in the PSD, with QPO characteristics}.

Another possible astrophysical application of the models developed in the present paper can be related to the study of the properties of the Gamma Ray Bursts (GRBs). GRBs are very powerful, sudden,  and short cosmic gamma-ray emissions, from astrophysical sources situated at cosmological distances.   The typical energy fluxes in GRBs range from $10^{-5}$ to $5 \times 10^{-4}$ erg cm${}^{-2}$,  while the time intervals during which the gamma ray emissions take place are of the order of
$10^{-2}$ to $10^{3}$ s \cite{Pi04,Kum}. Non-thermal photons are also observed in GRB explosions, and their presence is usually explained
by either synchrotron emission, or inverse Compton  scattering by
relativistic electrons in strong magnetic fields \cite{Kum}. The electrons, obeying a power-law distribution, are assumed to have been accelerated to relativistic energies in the shocks generated in the optically thin regions of the matter outflowing from the explosion center.
A possible alternative to the synchrotron radiation model for explaining the non-thermal radiative properties of GRBs is represented by the jitter radiation model. In this model the radiation  originates from relativistic electrons accelerated in strong stochastic magnetic fields \cite{Fl06}-\cite{Fl07b}. On the other hand the physical characteristics of the radiation in stochastic magnetic fields, or a turbulent cosmic medium, can be studied by using the Langevin equation description, together with the numerical methods developed in the present approach to the problem of the radiation emission by charged particles in Brownian motion. Therefore the theoretical models introduced in the present paper may contribute to a better understanding of the astrophysical processes taking place during the main explosive phase, or during the afterglow of the GRB explosions.

The astrophysical implications of the methods and results obtained in our theoretical and numerical analysis will be explored in detail in a future publication. In the present paper we have only presented some basic theoretical tools that can be used for the in-depth modelling and comparison of the astrophysical observations of stochastically varying luminosity sources with the theoretical predictions of the physical models.

\section*{Acknowledgments}

We would like to thank to the anonymous referee for comments and suggestions that helped us to significantly improve our manuscript. GM is partially supported by a grant of the Romanian National Authority of Scientific Research, Program for research - Space Technology and Advanced Research - STAR, project number 72/29.11.2013.

\appendix

\section{Tables with the statistical characteristics of the stochastic Light Curves}

\newpage

\begin{table}\caption{Values of the statistical characteristics for the LC in case I as a function of the parameter space.}
  \begin{tabular}{| l | c | c| c | c| r | }
    \hline
    $V_0$ & $E$ & $\mu$ & $\sigma$ & $s$ & $\kappa$ \\ \hline
0.1&-7&10039.963&460.881&0.181&3.072\\ \hline
0.1&-5&10053.522&456.594&0.025&3.029\\ \hline
0.1&1&10020.254&437.080&0.148&3.341\\ \hline
0.1&5&10061.864&450.921&0.107&2.894\\ \hline
0.1&7&10073.099&438.762&0.104&3.103\\ \hline
100&-7&10654.364&1880.597&4.144&24.983\\ \hline
100&-7&10604.786&1817.379&3.991&24.165\\ \hline
100&-7&10550.766&1617.566&4.360&32.368\\ \hline
100&-7&10500.008&1523.748&4.689&39.015\\ \hline
100&-7&10508.306&1459.364&4.461&35.469\\ \hline

\multicolumn{5}{c}{ Noiseless }  \\ \hline
0.1&-7&2.581&8.060&4.750&33.520\\ \hline
0.1&-5&1.331&4.152&4.717&32.855\\ \hline
0.1&1&0.042&0.135&5.814&57.154\\ \hline
0.1&5&1.233&3.880&4.980&38.296\\ \hline
0.1&7&2.443&7.678&4.937&37.396\\ \hline
100&-7&583.976&1806.001&4.481&28.260\\ \hline
100&-5&563.045&1746.772&4.565&29.863\\ \hline
100&1&502.714&1578.482&4.907&36.750\\ \hline
100&5&464.543&1474.446&5.236&43.853\\ \hline
100&7&446.072&1424.988&5.442&48.475\\ \hline
  \end{tabular}\label{tab:statA}
\end{table}

\begin{table}\caption{Values of the statistical characteristics for the LC in case II as a function of the parameter space.}
  \begin{tabular}{| l | c | c| c | c| r | }
    \hline
    $V_0$ & $W^2$ & $\mu$    & $\sigma$ & $s$     & $\kappa$ \\ \hline
1&0.01&10042.336&546.597&-6.165&115.429\\ \hline
1&0.1&10054.237&564.972&-5.573&101.853\\ \hline
1&0.2&10035.968&549.911&-6.043&112.473\\ \hline
1&0.5&10072.870&549.593&-6.017&114.444\\ \hline
1&1&10097.764&560.838&-5.767&106.647\\ \hline
1&2&10131.502&552.469&-6.092&114.488\\ \hline
1&5&10279.330&579.003&-5.555&100.956\\ \hline
1&7&10363.076&568.170&-6.005&112.241\\ \hline
1&10&10512.427&582.914&-5.848&107.347\\ \hline
50&0.01&10187.129&625.712&-1.113&26.376\\ \hline
50&0.1&10190.245&648.480&-0.705&24.113\\ \hline
50&0.2&10212.970&632.853&-0.964&26.154\\ \hline
50&0.5&10280.704&737.837&-0.095&16.833\\ \hline
50&1&10332.067&768.806&0.035&15.000\\ \hline
50&2&10523.102&1029.668&1.294&9.830\\ \hline
50&5&11072.753&1869.612&2.621&10.729\\ \hline
50&7&11449.823&2499.379&2.830&11.413\\ \hline
50&10&12057.341&3478.019&3.056&12.653\\ \hline

\multicolumn{5}{c}{ Noiseless }  \\ \hline
1&0.01&0.052&0.155&3.884&18.647\\ \hline
1&0.1&0.056&0.161&3.686&16.945\\ \hline
1&0.2&0.061&0.168&3.481&15.245\\ \hline
1&0.5&0.076&0.194&2.999&11.451\\ \hline
1&1&0.102&0.244&2.613&8.574\\ \hline
1&2&0.153&0.358&2.559&8.089\\ \hline
1&5&0.314&0.738&2.952&10.930\\ \hline
1&7&0.428&1.004&3.092&12.130\\ \hline
1&10&0.607&1.415&3.192&13.120\\ \hline
50&0.01&129.378&387.101&3.884&18.647\\ \hline
50&0.1&140.609&402.425&3.686&16.945\\ \hline
50&0.2&153.130&420.912&3.481&15.245\\ \hline
50&0.5&190.936&484.290&2.999&11.451\\ \hline
50&1&254.382&608.915&2.613&8.574\\ \hline
50&2&383.218&894.795&2.559&8.089\\ \hline
50&5&786.035&1846.235&2.952&10.930\\ \hline
50&7&1069.036&2511.187&3.092&12.130\\ \hline
50&10&1517.157&3538.026&3.192&13.120\\ \hline
  \end{tabular}\label{tab:statB}
\end{table}

\begin{table}\caption{Values of the statistical characteristics for the LC in case III as a function of the parameter space.}
  \begin{tabular}{| l | c | c| c | c| r | }
    \hline
    $V_0$                 & $W^2$ &     $\mu$      & $\sigma$  & $s$     & $\kappa$ \\ \hline

    \multicolumn{5}{c}{$\bar{\alpha} = 5$}  \\ \hline
1, $\bar{C}=10$ &0.0005&4358.459&958.759&-1.989&6.657\\ \hline
1, $\bar{C}=10$&0.01&4465.579&988.686&-2.113&6.944\\ \hline
1, $\bar{C}=10$&0.02&4398.845&968.181&-2.096&6.886\\ \hline
1, $\bar{C}=10$&0.05&4493.401&978.447&-2.128&7.147\\ \hline
1, $\bar{C}=10$&0.07&4291.154&939.824&-2.083&7.055\\ \hline
50, $\bar{C}=100$&0.0005&443242.296&99867.551&-1.902&6.319\\ \hline
50, $\bar{C}=100$&0.01&459093.604&102252.455&-1.919&6.599\\ \hline
50, $\bar{C}=100$&0.02&436537.988&90562.257&-2.300&8.035\\ \hline
50, $\bar{C}=100$&0.05&444968.159&93650.477&-2.279&7.937\\ \hline
50, $\bar{C}=100$&0.07&434342.677&93998.582&-2.068&7.114\\ \hline
    \multicolumn{5}{c}{$\bar{\alpha} = 5$, Noiseless }  \\ \hline
1, $\bar{C}=10$&0.0005&0.249863&0.587454&2.96388&10.9935\\ \hline
1, $\bar{C}=10$&0.01&0.248641&0.583183&2.95912&10.976\\ \hline
1, $\bar{C}=10$&0.02&0.243824&0.565613&2.93701&10.8897\\ \hline
1, $\bar{C}=10$&0.05&0.24739&0.578728&2.95391&10.9564\\ \hline
1, $\bar{C}=10$&0.07&0.241584&0.557079&2.92469&10.8387\\ \hline
50, $\bar{C}=100$&0.0005&624.658&1468.63&2.96388&10.9935\\ \hline
50, $\bar{C}=100$&0.01&621.604&1457.96&2.95912&10.976\\ \hline
50, $\bar{C}=100$&0.02&609.559&1414.03&2.93701&10.8897\\ \hline
50, $\bar{C}=100$&0.05&618.474&1446.82&2.95391&10.9564\\ \hline
50, $\bar{C}=100$&0.07&603.96&1392.7&2.92469&10.8387\\ \hline

\multicolumn{5}{c}{$\bar{\alpha} = 0.5$}  \\ \hline

1, $\bar{C}=10$&5&28182.484&9477.725&-1.122&4.151\\ \hline
1, $\bar{C}=10$&10&35063.365&12108.226&-0.823&3.370\\ \hline
1, $\bar{C}=10$&15&39347.932&13571.939&-0.711&3.324\\ \hline
1, $\bar{C}=10$&20&41565.638&15284.847&-0.538&2.825\\ \hline
1, $\bar{C}=10$&25&41892.573&15589.830&-0.432&2.786\\ \hline
50, $\bar{C}=100$&5&2.826$\cdot$10$^6$&950773.090&-1.063&3.928\\ \hline
50, $\bar{C}=100$&10&3.595$\cdot$10$^6$&1.223$\cdot$10$^6$&-0.915&3.459\\ \hline
50, $\bar{C}=100$&15&3.940$\cdot$10$^6$&1.408$\cdot$10$^6$&-0.706&2.964\\ \hline
50, $\bar{C}=100$&20&4.041$\cdot$10$^6$&1.465$\cdot$10$^6$&-0.612&2.900\\ \hline
50, $\bar{C}=100$&25&4.084$\cdot$10$^6$&1.433$\cdot$10$^6$&-0.580&3.020\\ \hline

    \multicolumn{5}{c}{$\bar{\alpha} = 0.5$, Noiseless }  \\ \hline

1, $\bar{C}=10$&5&0.050&0.112&2.994&11.216\\ \hline
1, $\bar{C}=10$&10&0.337&0.506&2.761&11.279\\ \hline
1, $\bar{C}=10$&15&1.049&1.214&1.947&7.182\\ \hline
1, $\bar{C}=10$&20&2.110&2.108&1.471&5.094\\ \hline
1, $\bar{C}=10$&25&3.410&3.154&1.155&3.960\\ \hline
50, $\bar{C}=100$&5&124.436&280.137&2.994&11.216\\ \hline
50, $\bar{C}=100$&10&842.511&1265.895&2.761&11.279\\ \hline
50, $\bar{C}=100$&15&2623.710&3035.262&1.947&7.182\\ \hline
50, $\bar{C}=100$&20&5276.062&5269.579&1.471&5.094\\ \hline
50, $\bar{C}=100$&25&8524.133&7884.123&1.155&3.960\\ \hline

  \end{tabular}\label{tab:statC}
\end{table}

\begin{table}\caption{Values of the statistical characteristics for the LC in case III as a function of the parameter space.}
  \begin{tabular}{| l | c | c| c | c| r | }
    \hline
    $V_0, \bar{C}$                 & $\bar{\alpha}$ &     $\mu$      & $\sigma$  & $s$     & $\kappa$ \\ \hline
        \multicolumn{5}{c}{$W^2 = 0.0005$}  \\ \hline
1, $\bar{C}=10$&0.1&4452.572&718.951&-3.622&17.589\\ \hline
1, $\bar{C}=10$&1&4577.085&916.024&-2.086&8.367\\ \hline
1, $\bar{C}=10$&10&4435.546&1005.919&-1.968&6.368\\ \hline
1, $\bar{C}=10$&15&4599.580&1012.180&-2.366&8.127\\ \hline
1, $\bar{C}=10$&100&4424.343&995.141&-2.421&8.460\\ \hline
1, $\bar{C}=10$&200&4514.335&996.517&-2.477&8.899\\ \hline
1, $\bar{C}=10$&500&4695.090&1041.969&-2.245&7.958\\ \hline
50, $\bar{C}=100$&0.1&452685.721&75615.100&-3.432&16.364\\ \hline
50, $\bar{C}=100$&1&451849.363&90594.726&-2.111&8.218\\ \hline
50, $\bar{C}=100$&10&442090.501&93866.607&-2.187&7.421\\ \hline
50, $\bar{C}=100$&15&440222.821&98208.836&-2.127&7.217\\ \hline
50, $\bar{C}=100$&100&474725.014&88159.804&-2.350&9.286\\ \hline
50, $\bar{C}=100$&200&480350.123&73293.850&-3.041&14.721\\ \hline
50, $\bar{C}=100$&500&536340.269&110262.146&1.258&12.969\\ \hline

        \multicolumn{5}{c}{$W^2 = 0.0005$, noiseless}  \\ \hline
1, $\bar{C}=10$&0.1&0.004&0.002&-0.032&1.776\\ \hline
1, $\bar{C}=10$&1&0.050&0.090&1.732&4.473\\ \hline
1, $\bar{C}=10$&10&0.500&1.223&3.314&13.870\\ \hline
1, $\bar{C}=10$&15&0.749&1.859&3.456&15.229\\ \hline
1, $\bar{C}=10$&100&4.987&12.685&3.741&18.352\\ \hline
1, $\bar{C}=10$&200&10.000&25.454&3.764&18.644\\ \hline
1, $\bar{C}=10$&500&25.984&64.768&3.700&18.143   \\ \hline
50, $\bar{C}=100$&0.1&9.848&4.744&-0.032&1.776\\ \hline
50, $\bar{C}=100$&1&124.678&226.067&1.732&4.473\\ \hline
50, $\bar{C}=100$&10&1249.205&3056.254&3.314&13.870\\ \hline
50, $\bar{C}=100$&15&1873.456&4647.519&3.456&15.229\\ \hline
50, $\bar{C}=100$&100&12468.011&31713.427&3.741&18.352\\ \hline
50, $\bar{C}=100$&200&24999.333&63633.987&3.764&18.644\\ \hline
50, $\bar{C}=100$&500&64959.634&161919.880&3.700&18.143   \\ \hline
  \end{tabular}\label{tab:statC-III}
\end{table}

\begin{table}\caption{Values of the statistical characteristics for the LC in case IV as a function of the parameter space, with $V_0=V_{0x}=V_{0y}=V_{0z}$.}
  \begin{tabular}{| l | c | c| c | c| r | }
    \hline
    $V_0$   & $\bar{\Omega}$    & $\mu$      & $\sigma$  & $s$     & $\kappa$ \\ \hline

0.2&0.01&39538.069&2221.934&-6.763&104.863\\ \hline
0.2&0.1&39490.746&2221.552&-6.776&104.661\\ \hline
0.2&0.2&39882.646&2216.489&-7.045&109.728\\ \hline
0.2&0.5&41854.708&2555.013&-5.864&77.840\\ \hline
100&0.01&40903.815&3139.004&3.193&15.363\\ \hline
100&0.1&41111.709&3201.374&3.157&15.350\\ \hline
100&0.2&41480.206&3299.796&3.253&15.552\\ \hline
100&0.5&43466.002&3480.511&3.422&16.877\\ \hline

0.2&2&996437.802&156456.780&-3.687&17.663\\ \hline
0.2&5&77822.854&8091.585&-3.923&22.470\\ \hline
0.2&7&272634.688&40580.606&-3.515&16.681\\ \hline
0.2&10&511739.362&77716.362&-3.695&18.258\\ \hline
0.2&12&1.412$\cdot$10$^6$&220092.720&-3.711&18.108\\ \hline
100&2&1.072$\cdot$10$^6$&152658.316&4.066&20.932\\ \hline
100&5&82831.353&9374.825&3.464&15.763\\ \hline
100&7&303399.185&42014.278&3.465&15.890\\ \hline
100&10&555533.183&80155.312&3.730&17.473\\ \hline
100&12&1.520$\cdot$10$^6$&242972.915&3.470&15.893\\ \hline
     \multicolumn{5}{c}{ Noiseless }  \\ \hline
0.2&0.01&0.006&0.018&3.905&18.841\\ \hline
0.2&0.1&0.006&0.018&3.905&18.841\\ \hline
0.2&0.2&0.006&0.019&3.905&18.840\\ \hline
0.2&0.5&0.007&0.021&3.904&18.836\\ \hline
100&0.01&1512.647&4544.336&3.905&18.841\\ \hline
100&0.1&1522.646&4574.348&3.905&18.841\\ \hline
100&0.2&1552.947&4665.295&3.905&18.840\\ \hline
100&0.5&1765.054&5301.922&3.904&18.836\\ \hline
0.2&2&0.410&1.231&3.901&18.808\\ \hline
0.2&5&0.022&0.067&3.902&18.817\\ \hline
0.2&7&0.107&0.321&3.901&18.810\\ \hline
0.2&10&0.204&0.612&3.901&18.809\\ \hline
0.2&12&0.588&1.764&3.901&18.808\\ \hline
100&2&102515.879&307700.036&3.901&18.808\\ \hline
100&5&5552.679&16670.272&3.902&18.817\\ \hline
100&7&26763.379&80333.032&3.901&18.810\\ \hline
100&10&51004.179&153090.473&3.901&18.809\\ \hline
100&12&146957.346&441088.678&3.901&18.808\\ \hline
  \end{tabular}\label{tab:statD}
\end{table}


\begin{thebibliography}{99}

\bibitem{rev1} R. Friedrich, J. Peinke, M. Sahimi, and M. Reza Rahimi Tabar, Physics Reports {\bf 506},  87 (2011).

\bibitem{rev2} S. R\"udiger, Physics Reports {\bf 534}, 39 (2014).

\bibitem{San1} J. R. Martín-Solís and R. Sanchez, Physics of Plasmas {\bf 13}, 012508 (2006).

\bibitem{And} F. Anderson, P. Helander, and L.-G. Eriksson, Physics of Plasmas {\bf 8}, 5221 (2001).

\bibitem{Bak} M. Bakhtiari, G. J. Kramer, M. Takechi, H. Tamai, Y. Miura, Y. Kusama,
and Y. Yamada, Phys. Rev. Lett. {\bf 94}, 215003 (2005).

\bibitem{San2} J. R. Martín-Solís and R. Sanchez, Physics of Plasma {\bf 15}, 112505 (2008).

\bibitem{Bal1} R. Balescu, Hai-Da Wang, and J. H. Misguich, Physics of Plasmas {\bf 1}, 3826 (1994).

\bibitem{4} H.-D. Wang, M. Vlad, E. Vanden Eijnden, F. Spineanu, J. H.
Misguich, and R. Balescu, Phys. Rev. E 51, 4844 (1995).

\bibitem{Bal2} R. Balescu, Statistical dynamics: Matter out of equilibrium, Imperial College Press, London, 1997

\bibitem{Bal3}  R. Balescu, Aspects of Anomalous Transport in Plasmas (Series in
Plasma Physics), Institute of Physics Publishing, Bristol and Philadelphia
(2005).

\bibitem{Coff} W. T. Coffey, Yu. P. Kalmykov, and J. T. Waldron, The Langevin Equation, with Applications to Stochastic Problems in Physics, Chemistry, and Electrical Engineering, World Scientific, Singapore, 2005

\bibitem{Kubo} R. Kubo, Reports on Progress in Physics {\bf 29},  255 (1966).

\bibitem{disk1} T. Harko and G. Mocanu, Mon. Not. R. Astron. Soc., 421, 3102 (2012)

\bibitem{disk2} T. Harko, C. S. Leung, and G. Mocanu, Eur. Phys. J. C  74, 2900 (2014)

\bibitem{disk3} C. S. Leung, G. Mocanu, and T. Harko, Journal of Astrophysics and Astronomy 35, 449 (2014)

\bibitem{b1}  J. B. Taylor, Phys. Rev. Lett. {\bf 6}, 262 (1961).
\bibitem{b2} B. Kursunoglu, Ann. Phys. {\bf 17}, 259 (1962).
\bibitem{b3} B. Kursunoglu, Phys. Rev. {\bf 132}, 21 (1963).
\bibitem{b4} D. S. Lemons and D. L. Kaufman,  IEEE Trans. on Plasma Science {\bf 27}, 5 (1999).

\bibitem{b5} M. Neuer and K. H. Spatschek, Phys. Rev. {\bf E 74}, 036401 (2006).

\bibitem{b6} M. Khoury, A. M. Lacasta, J. M. Sancho, A. H. Romero, and K. Lindenberg, Phys. Rev. {\bf B 78}, 155433 (2008).

\bibitem{b7} F. N. C. Paraan, M. P. Solon, and J. P. Esguerra,  Phys. Rev. {\bf E 77}, 022101 (2008).

\bibitem{b8} L. J. Hou, Z. L. Miskovic, A. Piel, and P. K. Shukla, Physics of Plasmas {\bf 16}, 053705 (2009).

\bibitem{b9} V. Lisy and J. Tothova, Transport Theory and Statistical Physics {\bf 42}, 365 (2013).

\bibitem{Nik79} Iu. A. Nikolaev and V. N. Tsytovich,  Phys. Scripta {\bf 20}, 665 (1979).

\bibitem{Top87a} I. N. Toptygin  and G. D. Fleishman,  Astrophys. Space
Science {\bf 132}, 213 (1987).

\bibitem{Top87b} I. N. Toptygin, G. D.  Fleishman,  and D. V. Kleiner,
Radiophys. and Quant. Electr. {\bf 30}, 334 (1987).

\bibitem{Med00} M. V. Medvedev,  Astrophys. J. {\bf 540}, 704 (2000).

\bibitem{Fl10} G. D. Fleishman and F. A. Urtiev, Monthly Notices of the
Royal Astronomical Society {\bf 406}, 644 (2010).

\bibitem{Fl06} G. D. Fleishman, Astrophys. J. {\bf 638}, 348 (2006)

\bibitem{Fl07} G. D. Fleishman G. and M. F. Bietenholz,  MNRAS {\bf 376}, 625 (2007).

\bibitem{Fl07a} G. D. Fleishman and I. N. Toptygin,  MNRAS {\bf 381}, 1473 (2007).

\bibitem{Fl07b} G. D. Fleishman and I. N. Toptygin,  Phys. Rev. {\bf E 76}, 017401 (2007).

\bibitem{Hij} 	H. Hijar, Phys. Rev. {\bf E 91}, 022139 (2015).

\bibitem{sr3} Z.-Y. Wang, P.-J. Chen, and L.-Y. Zhang, Chinese Physics {\bf B 24},  059801 (2015).

\bibitem{sr4}  Z.-Y. Wang, P.-J. Chen, and L.-Y. Zhang, Chinese Physics Letters {\bf 30}, 099801 (2013).

\bibitem{sr5} 	Z. Y. Wang, P. J. Chen, D. X.  Wang, and L. Y.  Zhang, Journal of Astrophysics and Astronomy {\bf 34}, 33 (2013).

\bibitem{LaLi} L. D. Landau and E. M.  Lifshitz, The classical theory
of fields (Oxford: Pergamon Press) 1971

\bibitem{ermak1980} D. L. Ermak, H. Buckholz, Journal of Comp. Phys. {\bf 35},
169 (1980).

\bibitem{hershkowitz1998} E. Hershkowitz, J. Chem. Phys. {\bf 108}, 9253 (1998).

\bibitem{tem04} M. Templeton, JAAVSO {\bf 32}, 41 (2004).
\bibitem{vau13} S. Vaughan,  Phil. Trans. R. Soc. A. {\bf 371}, 20110549 (2013).

\bibitem{M1} C. S. Leung, J. Y. Wei, T. Harko, and Z. Kovacs, J. Astrophys. Astr. {\bf 32}, 189 (2011).

    \bibitem{QPO1} W. N. Alston, M. L. Parker, J. Markeviciute, A. C. Fabian, M. Middleton, A.Lohfink, E. Kara, and C. Pinto, Monthly Notices of the Royal Astronomical Society {\bf 449}, 467 (2015).
        \bibitem{QPO2} R. Vio, P. Rebusco, P. Andreani, H. Madsen, and R. V. Overgaard, Astron. Astrophys. {\bf 452}, 386 (2006).

        \bibitem{QPO3} L. Stella and M. Vietri,  Astrophys. J. {\bf 492}, L59 (1998).
        \bibitem{QPO4} M. A. Abramowicz and W. Kluzniak,  Astron. Astrophys. {\bf 374}, L19 (2001).

\bibitem{Pi04} T. Piran, Rev. Modern Phys. {\bf 76}, 1143 (2004).

\bibitem{Kum} P. Kumar and B. Zhang, Physics Reports {\bf 561}, 1 (2015).

\end{thebibliography}
\end{document}